\documentclass{article}
\usepackage{arxiv}
\usepackage{natbib}
\usepackage{booktabs}       
\usepackage{nicefrac}       
\usepackage{microtype}      
\usepackage{lipsum}		
\usepackage{graphicx}
\usepackage{amsmath,amsfonts,amssymb}
\usepackage{doi}
\usepackage{amsthm}%
\usepackage{mathrsfs}%
\usepackage[title]{appendix}%
\usepackage{xcolor}%
\usepackage{textcomp}%
\usepackage{manyfoot}%
\usepackage{booktabs}%
\usepackage{algorithm}%
\usepackage{algorithmicx}%
\usepackage{algpseudocode}%
\usepackage{listings}%
\usepackage{bbm}
\DeclareMathOperator{\Tr}{Tr}
\DeclareMathOperator{\BigO}{\mathcal{O}}
\DeclareMathOperator{\tildeO}{\tilde{\mathcal{O}}}
\usepackage{etoolbox}
\usepackage{algorithm}
\usepackage{algpseudocode}
\usepackage{mathrsfs}
\usepackage{siunitx}
\usepackage{amsthm}

\algnewcommand{\LineComment}[1]{\State $\triangleright$ #1}
\algnewcommand{\NoteComment}[1]{\State [NOTE: #1]}
\newcommand{\norm}[2]{\left\lVert #1\right\rVert_{#2}}
\newcommand{\refSI}[1]{Section~S#1 in Supplementary Information}
\usepackage{subfigure}

\DeclareMathOperator*{\argmin}{arg\,min}
\DeclareMathOperator*{\inc}{\mathrel{+}=}

\usepackage{stmaryrd}
\usepackage{array}
\newcolumntype{P}[1]{>{\raggedright\arraybackslash}p{#1}}

\usepackage{ifthen}
\usepackage{bm}
\usepackage{amsfonts}
\usepackage{stackengine}
\usepackage{cleveref}
\let\oldhref\href
\renewcommand{\href}[2]{\oldhref{#1}{\hbox{#2}}}

\newcommand{\ket}[1]{\vert #1 \rangle}

\newcommand{\trunc}[3]{\left[\left[#1\right]\right]_{#2}^{#3}}
\usepackage[british]{babel}
\usepackage{hyphenat}
\theoremstyle{thmstyleone}%
\newtheorem{theorem}{Theorem}[section]
\newtheorem{definition}{Definition}[section]
\newtheorem{corollary}{Corollary}[section]
\newtheorem{lemma}{Lemma}[section]

\newtheorem{example}{Example}
\usepackage{tikz}
\usepackage{pgfplots} 
\usepackage{fancyhdr}
\usepackage{pgfplots} 
\newcommand{\IHPC}{Institute of High Performance Computing (IHPC), Agency for Science, Technology and Research (A*STAR)}
\newcommand{\CFAR}{Centre for Frontier AI Research (CFAR), Agency for Science, Technology and Research (A*STAR)}
\newcommand{\CQuERE}{Centre for Quantum Engineering, Research and Education, TCG CREST} 
\title{Quantum Policy Gradient in Reproducing Kernel Hilbert Space}
\author{David M. Bossens\\ \IHPC \\ \CFAR \\ \url{david_bossens@cfar.a-star.edu.sg} \And Kishor Bharti \\ \IHPC \\ 
\CQuERE\\
\url{bharti_kishor@ihpc.a-star.edu.sg} \And Jayne Thompson \\ \IHPC \\ \url{jayne_thompson@ihpc.a-star.edu.sg} }
\begin{document}
\maketitle
\thispagestyle{fancy}

\maketitle

\begin{abstract}
Parametrised quantum circuits offer expressive and data-efficient representations for machine learning. Due to quantum states residing in a high-dimensional Hilbert space, parametrised quantum circuits have a natural interpretation in terms of kernel methods. The representation of quantum circuits in terms of quantum kernels has been studied widely in quantum supervised learning, but has been overlooked in the context of quantum RL. This paper proposes the use of kernel policies and quantum policy gradient algorithms for quantum-accessible environments. After discussing the properties of such policies and a demonstration of classical policy gradient on a coherent policy in a quantum environment, we propose parametric and non-parametric policy gradient and actor-critic algorithms with quantum kernel policies in quantum environments. This approach, implemented with both numerical and analytical quantum policy gradient techniques, allows exploiting the many advantages of kernel methods, including data-driven forms for functions (and their gradients) as well as tunable expressiveness. The proposed approach is suitable for vector-valued action spaces and each of the formulations demonstrates a quadratic reduction in query complexity compared to their classical counterparts. We propose actor-critic algorithms based on stochastic policy gradient, deterministic policy gradient, and natural policy gradient, and demonstrate additional query complexity reductions compared to quantum policy gradient algorithms under favourable conditions.
\end{abstract}

\section{Introduction}
Reinforcement learning (RL) is a technique for interactively learning from an environment from rewards which has been successful across a wide range of applications. Unfortunately, RL has high sample complexity, i.e. it requires many data samples before a high-performing policy is learned. With the aim of reducing the sample complexity, several works have proposed applying RL systems within quantum-accessible environments, where interactions with the environment occur within a quantum system allowing to make use of superpositions across state-action trajectories.  While exponential sample complexity improvements have only been shown for a special case environment formulated around Simon's problem \citep{Dunjko2017}, recent quantum policy gradient algorithms demonstrate benefits in terms of quadratic sample complexity improvements when applying a parametrised quantum circuit within a quantum environment due to the properties of quantum superpositions \citep{Jerbi2022}. Moreover, several quantum RL works demonstrate that by using parametrised quantum circuits (PQCs), the number of parameters can be reduced compared to using classical neural networks  \citep{Lan2021,Chen2023} -- although this line of investigation has primarily focused on classical environments. 

Despite the promise of quadratic or better improvements, limited work has been done in designing quantum algorithms and suitable PQCs for quantum-accessible environments. Previous work has introduced various PQCs for classical RL \citep{Jerbi2021}, namely Raw-PQC and Softmax-PQC, which were put to the test as RL policies in classical environments based on numerical  experiments in hardware-efficient PQCs \citep{Kandala2017} with data-reuploading \citep{Perez-Salinas2020}. These PQCs were then used in a theoretical study on the sample complexity of quantum policy gradient techniques, including analytical gradient estimation using quantum Monte Carlo techniques and a numerical quantum gradient estimation technique based on central differencing \citep{Jerbi2022}. PQCs have also been applied to the quantum control context, where the overall sample complexity is not mentioned \citep{Wu2020} or is without quadratic improvement \citep{Sequeira2023a}.

Due to quantum states residing in a high-dimensional Hilbert space, PQCs have a natural interpretation in terms of kernel methods. While so far, this property has been discussed widely for supervised learning \citep{Schuld2019,Schuld2021}, this has not yet been adopted in RL. 

Our work is inspired by streams of work in classical RL that use kernel-based formulations of the policy \citep{Lever2015,Bagnell2003}. We formulate Gaussian and softmax policies based on quantum kernels and analyse their efficiency across various optimisation schemes with quantum policy gradient algorithms. While maintaining quadratic query complexity speedups associated with QPG, the use of quantum kernels for the policy definition leads to advantages such as analytically available policy gradients, tunable expressiveness, and techniques for sparse non-parametric representations of the policy within the context of vector-valued state and action spaces. This also leads to a quantum actor-critic algorithm with an interpretation related to the natural gradient. Unlike quantum algorithms for natural policy gradient \citep{Meyer2023,Sequeira2024}, the proposed algorithm is formulated within the kernel method framework and is tailored to the quantum accessible environment where it can exploit a quadratic sample complexity improvement as well as a variance reduction as is often associated with actor-critic RL.

\subsection{Using quantum kernels for reinforcement learning policies}
Kernel methods have strong theoretical foundations for functional analysis and supervised learning (see e.g. \citep{Scholkopf2003} for an overview). We review some of these useful properties here and how they can be applied to formulate and learn efficient policies for quantum RL.

Each kernel corresponds to an expressible function space through its reproducing kernel Hilbert space (RKHS; see Section~\ref{sec: RKHS}). The choice of the kernel function thereby provides an opportunity to balance the expressiveness, training efficiency, and generalisation. For instance, reducing the bandwidth factor to $c < 1$ of the squared cosine kernel 
\begin{equation}
\label{eq: bandwidth squaredcosine}
\kappa(s,s') = \prod_{j=1}^{d} \cos^2(c (s_j - s'_j)/2)
\end{equation}
restricts features to parts of the Bloch sphere, allowing improved generalisation \citep{Canatar2022}, as well as expressiveness control, considering expressiveness can be measured based on the distance to the Haar distribution \citep{Nakaji2021}. Optimising or tuning a single parameter is significantly more convenient compared to redesigning the ansatz of a PQC. More generally, kernel methods provide a relatively interpretable framework to form particular functional forms.

Kernel functions inherently define a particular feature-map. This interpretation follows from Mercer's theorem, which states that every square-integrable kernel function can be written as 
\begin{align*}
\kappa(s,s') = \sum_{i=1}^{\infty} \lambda_i e_i(s) e_i(s') \,,
\end{align*}
where for all $i$, $e_i$ is an eigenfunction such that $\lambda_i e_i(s') = \mathcal{T}_K[e_i](s') = \int_{\mathcal{X}} \kappa(s,s') e_i(s) d\mu(s)$ with eigenvalue $\lambda_i$, where $\mu$ is a strictly positive Borel measure (e.g. the Lebesgue measure) for continuous $\mathcal{X}$ or the counting measure for discrete $\mathcal{X}$. Mercer's theorem leads to the kernel trick, 
\begin{align*}
\kappa(s,s') = \langle \phi(s), \phi(s') \rangle \,,
\end{align*}
which allows writing the kernel function as an inner product based on a feature-map $\phi$. For quantum kernels, this conveniently allows a definition of kernels in terms of the data encoding as a feature-map. For instance, the basis encoding corresponds to the Kronecker delta kernel, the amplitude encoding corresponds to the inner product quantum kernel, etc. \citep{Schuld2021}. We construct kernel-based policies which construct kernel computations in quantum circuits both in the explicit view and the implicit (i.e. inner product) view (see Section~\ref{sec: QKPs}).

Kernel regression can be done in a data-driven (non-parametric) manner, i.e. based on a representative set of input-output pairs. In the context of RL, the data are state-action pairs, which often have lower dimensionality compared to parameter vectors. The \textit{representer theorem} guarantees that the optimal function approximator in the RKHS can be written as a linear combination of kernel evaluations based on input-output samples, which leads to the formulations of support vector machines and kernel regression. In quantum supervised learning, one uses this property to evaluate the kernel in a quantum device and then compute the prediction in a quantum or a classical device \citep{Schuld2019,Jerbi2023}. In a quantum RL setting, we analogously consider the optimal deterministic policy $\mu$ as a linear combination of kernel computations with regard to a select subset of the states:
\begin{align}
\label{eq: representer}
\mu(s) = \sum_{i=1}^{N} \beta_i \kappa(s,c_i) \,,
\end{align}
where $s$ is the current state, and $\{c_i,\beta_i\}_{i=1}^{N}$ are state-action pairs as representative data points (``representers'' for short). Using this quantity as the mean of a Gaussian distribution allows the quantum analogue of the Gaussian policies that are popular in classical RL with vector-valued action spaces. We will formulate a special class of PQCs which form circuits with Eq.~\ref{eq: representer} as their expectation, allowing a novel way to form expressive and coherent policies in the context of quantum environments (see Section~\ref{sec: QKPs}). We will show (see e.g. Section~\ref{sec: compatible QRKHS-AC} and Section~\ref{sec: deterministic AC}) that this comes with analytical forms for the gradient and that it is suitable for various non-parametric optimisation schemes. 

By performing regularisation in the context of kernel ridge regression, we make use of the result that for every RKHS $\mathcal{H}_K$ with reproducing kernel $K$, and any $g \in \mathcal{H}_K$,
\begin{align*}
\vert \vert  g  \vert \vert_{\mathcal{H}_K}^2 = \langle g ,g  \rangle_{\mathcal{H}_K} = \int ( \mathcal{R} g(x))^2 dx
\end{align*}
where the operator $\mathcal{R}: \mathcal{H}_K \to \mathcal{D}$ can be interpreted as extracting information from the function value which gets penalised during optimisation \citep{Scholkopf2003}. For instance, it can penalise large higher or lower order derivatives, large function values, or still other properties, leading to smoother optimisation landscapes and therefore improved convergence to the global optimum. This property contributes to an improved query complexity when considering actor-critic algorithms with smooth critic functions (see Section~\ref{sec: deterministic AC}) and can also be exploited when directly optimising the kernel (see Section~\ref{sec: parametric-kernel}).

\subsection{Overview of the contributions}
Motivated by the potential benefits of kernel policies, this work contributes the following theoretical results to the field of quantum RL:
\begin{itemize}
    \item In Section~\ref{sec: QKPs}, we propose two classes of quantum kernel policies (QKPs) for learning in quantum-accessible environments. First, we propose Representer PQCs, which incorporate representer theorem based formalisms directly within a quantum circuit and which are suitable for both analytical and numerical gradient based optimisation. Second, we propose Gaussian kernel-based policies based on a classically known mean function and covariance, which due to the mean and covariance being parametrised classically has a known analytical gradient, thereby removing the need for expensive estimation procedures required for analytical quantum policy gradient with traditional PQCs. Via Lemma~\ref{lem: setting N}, we also provide a formula to scale the number of representers based on kernel matching pursuit in vector-valued output spaces. Finally, using numerical policy gradient with classical sampling-based estimates, an empirical demonstration (see Section~\ref{sec: learnability RPQC}) shows that the proposed Representer PQC policies are learnable when applied coherently in a quantum circuit.
    \item In Section~\ref{sec: numerical policy gradient}, we use a central differencing approach on phase oracles of the value function for a numerical quantum policy gradient algorithm \citep{Jerbi2022} based on Representer PQCs. We report a query complexity comparable to \cite{Jerbi2022} but note the potentially lower number of parameters.
    \item In Section~\ref{sec: analytical policy gradient}, we use analytical quantum policy gradient algorithms which perform quantum multivariate Monte Carlo on binary oracles of the policy gradient. In Section~\ref{sec: REINFORCE}, we confirm that applying quantum analytical policy gradient to kernel-based policies yields a query complexity that is comparable to \cite{Jerbi2022} and gives quadratic improvements over classical policies.
    \item Section~\ref{sec: stochastic AC} proposes two further improvements in an algorithm we call Compatible Quantum RKHS Actor-Critic (CQRAC). First, the parameter dimensionality of the policy is reduced by using vector-valued kernel matching pursuit. Second, we formulate a quantum oracle, which we call the state-action occupancy oracle, which computes the policy gradient samples based on the critic's prediction on a particular state-action pair rather than on the cumulative reward of the trajectory, thereby reducing the variance of the estimate produced by analytical quantum policy gradient. Theorem~\ref{th: actor-critic query complexity}\textbf{a} demonstrates that the resulting query complexity depends on the maximal deviation from a baseline estimate, rather than on the maximal cumulative reward. Theorem~\ref{th: actor-critic query complexity}\textbf{b} provides an improved result which exploits an upper bound on the variance of the gradient of the log-policy, and thereby demonstrates how smooth policies such as the Gaussian kernel-based policy can give  additional query complexity benefits.
    \item Section~\ref{sec: deterministic AC} proposes Deterministic Compatible Quantum RKHS Actor-Critic (DCQRAC), which is based on the deterministic policy gradient theorem \citep{Silverb}. The approach makes use of similar formalisms as its non-deterministic counterpart, though with the key differences that it is based on state occupancy rather than state-action occupancy, and that the policy gradient takes a different form, leading to a different query complexity result. In particular, Theorem~\ref{th: dpg query complexity} demonstrates that the resulting query complexity depends on the norm of kernel features and the gradient norm of the critic, which illustrates the importance of techniques such as kernel matching pursuit and regularisation. 
    \item Section~\ref{sec: natural AC} proposes Compatible Quantum RKHS Natural Actor-Critic (CQRNAC), which is based on natural actor-critic \citep{Peters2008}. The approach makes use of the compatible critic by noting its solution is equal to the natural policy gradient. Theorem~\ref{th: natural actor-critic query complexity} shows a query complexity that depends on the log-policy gradient norm and the deviation between the critic prediction and the observed return. The formulation does not require separate calls for the policy and the critic, does not require explicit computation of the Fisher information matrix, and inherits the invariance property of natural policy gradient.
\end{itemize}

\begin{figure}[htbp!]
    \centering
    \includegraphics[width=0.5\linewidth]{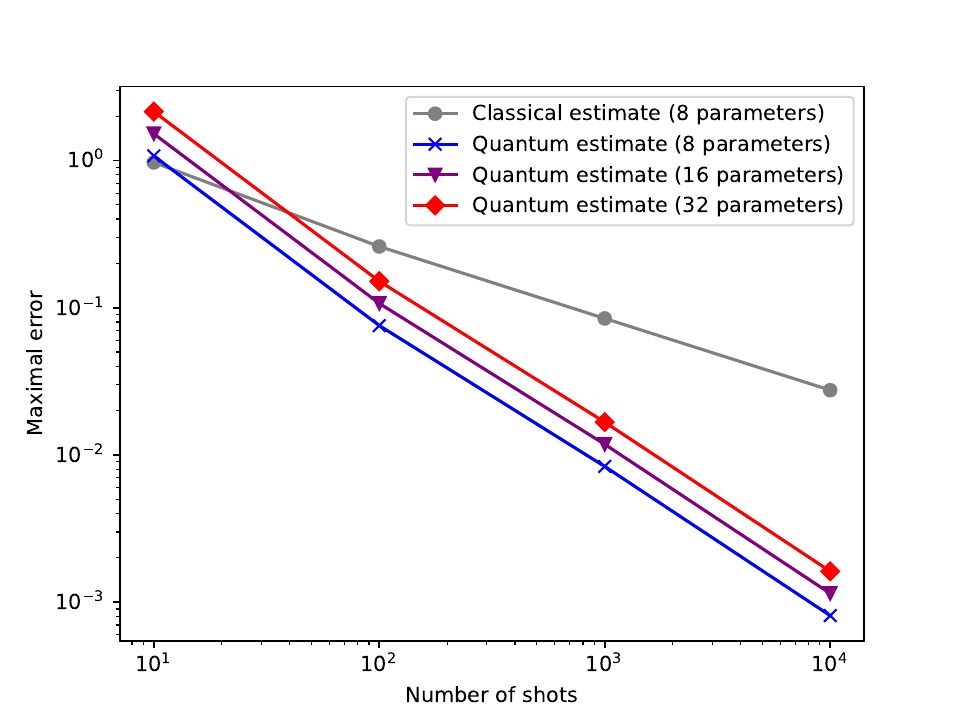}
    \caption{Illustration of the improved $\ell_{\infty}$ upper bound on the error from quantum estimation algorithms. To generate the classical estimate, the return of a randomly initialised Kronecker delta policy (see Section~\ref{sec: representer PQC}) was evaluated on a state control application (see Example~\ref{ex: quantum-accessible MDP}). The red line summarises the error upper bound of the value estimate compared to the true value as the maximal error across 100 estimates, averaged across the 5 settings. The quantum estimates (blue line) are illustrations based on the quadratic improvements and the effect of the parameter dimensionality $\epsilon  = \BigO(\sqrt{d})$ as it appears in the query complexity of numerical policy gradient.}
    \label{fig: error-classical-vs-quantum}
\end{figure}

Our query complexity results compare favourably to other methods to compute the policy gradients of PQCs, as shown in Table~\ref{tab: query complexity}. To illustrate the query complexity improvement visually, we also provide a numerical demonstration of error bounds on the return (see Fig.~\ref{fig: error-classical-vs-quantum}), which are directly related to numerical policy gradient algorithms.
\begin{table}
\caption{Query complexity of policy gradient estimation with PQCs. Our primary contributions include i) kernel-based policies which are parametrised by policy weights $\mathbb{R}^{N \times A}$, based on $N$ representers with $A$ action dimensions each, leading to $d=NA$ for classical parametrisations and  $d = \BigO(NAk)$ parameters for quantum parametrisations for per-dimension precision $k$; and ii) actor-critic algorithms which reduce the variance and provide alternative constants in analytical quantum policy gradient estimation. Notations include the following constants: $r_{\text{max}}$ denotes the maximal absolute reward; $T$ is the horizon; $\gamma$ is the discount factor; and $\epsilon$ is the tolerance for error in the gradient estimate. Further notations for constants appearing in the query complexity: $\mathcal{T}$ is the temperature of the softmax; $D$ is an upper bound on higher-order derivatives of the policy; $\Delta_Q$ is the maximal absolute deviation of the critic's prediction to a baseline estimate; $\sigma_{Q}$ is an upper bound on the standard deviation of the critic's prediction to the baseline estimate; upper bounds on $p$-norms are denoted as $B_p$ for the gradient of the log-policy, as $\sigma_{\nabla_p}$ for the standard deviation of the partial derivative of the log-policy, as $\kappa_{p}^{\text{max}}$ for the kernel computations across policy centres, and as $C_p$ for the gradient of the critic w.r.t. actions; $E$ is the maximal deviation of the critic's prediction to the observed return; and $\xi(p) = \max\{0,1/2-1/p\}$ is used for converting across $p$-norms.} \label{tab: query complexity}
\begin{tabular}{p{5cm} | p{5cm} l}
\toprule
\textbf{Algorithm} & \textbf{Oracle and estimation} & \textbf{Query complexity} \\ 
\midrule
 1. Policy gradient with Softmax-PQC \newline \citep{Sequeira2023a} & Return oracle, single-qubit parameter shift rule \citep{Schuld2019a}, and classical Monte Carlo   & $\tildeO\left( \frac{\mathcal{T}^2 r_{\text{max}}^2 T^2}{\epsilon^2 (1 - \gamma)^2})\right)$ \\
 2. Numerical QPG and Raw-PQC \newline \citep{Jerbi2022} & Return oracle, quantum gradient estimation via central differencing \citep{Cornelissen2019} &  $\tildeO\left(\sqrt{d} \frac{DT r_{\text{max}}}{\epsilon (1-\gamma)}\right)$ \\
 3. Analytical QPG and Softmax-PQC \newline \citep{Jerbi2022} & Analytical gradient oracle, bounded quantum multivariate Monte Carlo (Theorem 3.3 in \cite{Cornelissen2022}) & $\tilde{O}\left(d^{\xi(p)} \dfrac{B_p T r_{\text{max}}}{\epsilon(1-\gamma)} \right)$ \\ \hline
                                                  &                              &                                  \\
Proposed: Numerical QPG in RKHS & cf.2 \newline \phantom{cf.2} & cf.2 but $d = \BigO(NAk)$ \\
Proposed: Analytical QPG in RKHS  & cf.3 \newline \phantom{cf.3}  & cf.3  but $d = NA$\\
Proposed: CQRAC & Analytical gradient oracle, near-optimal quantum multivariate Monte Carlo (Theorem 3.4 in \cite{Cornelissen2022}) & $\tildeO\left(d^{\xi(p)} \frac{\Delta_Q B_p}{(1-\gamma)\epsilon}\right)$ for $d=NA$\\
                  &           &   $\tildeO\left(\frac{d^{\xi(p)} \sigma_Q \sigma_{\nabla_p}}{(1-\gamma)\epsilon}\right)$  for $d=NA$\\
Proposed: DCQRAC & cf.3 & $\tildeO\left(d^{\xi(p)} \frac{\kappa_{p}^{\text{max}} C_p}{(1-\gamma)\epsilon}\right)$  for $d=NA$\\
Proposed: CQRNAC & cf.3 & $\tildeO\left(\frac{d^{\xi(p)} E B_p}{(1-\gamma)\epsilon}\right)$  for $d=NA$\\
\bottomrule
\end{tabular}
\end{table}

\section{Preliminaries}
\subsection{Markov Decision Processes and classical policy gradient algorithms}
\label{sec: MDPs}
The Markov Decision Process (MDP) is the standard task-modelling framework for RL. The framework is defined by a tuple $(\mathcal{S},\mathcal{A},r,\gamma, P, T)$, where $\mathcal{S}$ is the state space, $\mathcal{A}$ is the action space, $r: \mathcal{S} \times \mathcal{A} \to [-r_{\text{max}},r_{\text{max}}]$ is the reward function, $\gamma \in (0,1)$ is the discount factor, and $P: \mathcal{S} \times \mathcal{A} \to \Delta(\mathcal{S})$ is the unknown true transition dynamics model outputting a distribution of states within the probability simplex $\Delta(\mathcal{S}) = \{ P \in \mathbb{R}^{\vert \mathcal{S} \vert} : P^{\intercal} \mathbf{1} = 1 \}$. Last, the horizon $T$ indicates the number of time steps.

The MDP proceeds in $T$-step episodes of the following nature. First, the agent is initialised to a particular state $s_0 \sim d_0$, where $d_0$ is the starting distribution, and the agent takes an action according to its policy $a_0 \sim \pi(\cdot \vert s)$. Then for $t=1,\dots,T-1$, the transition dynamics model reacts with $s_{t} \sim P(\cdot \vert s_{t-1}, a_{t-1})$ and the agent takes action $a_t \sim \pi(\cdot \vert s_t)$. The agent also receives a reward $r(s_t,a_t)$ for each $t=0,\dots,T-1$. 

The policy is learned by optimising the value, an objective which is based on the rewards the agent obtains in the episodes. In particular, the state-value is defined by the expected discounted cumulative reward when executing a policy from a given state $s \in \mathcal{S}$ for $T$ time steps,
\begin{equation}
V(s) = \mathbb{E}\left[\sum_{t=0}^{T-1} \gamma^t r(s_t,a_t) \vert s_0=s, a_t \sim \pi(\cdot \vert s_t), s_{t+1}\sim P(\cdot \vert s_{t},a_{t}) \right] \,,
\end{equation}
and the quantity $V(d_0) = \mathbb{E}_{s_0\sim d_0}[V(s_0)]$ is then often used as the agent's objective. Another useful notation is the state-action value (or Q-value), which indicates the value of executing a policy from a given state-action pair $(s,a) \in \mathcal{S} \times \mathcal{A}$; it is formulated as 
\begin{equation}
Q(s,a) = \mathbb{E}\left[\sum_{t=0}^{T-1} \gamma^t r(s_t,a_t) \vert s_0=s, a_0=a, a_t \sim \pi(\cdot \vert s_t), s_{t+1}\sim P(\cdot \vert s_{t},a_{t}) \right] \,.
\end{equation}
A few further notations are useful to work with MDPs. First, a notation that will often be used for state action pairs is $z = (s,a)$. Second, we use 
\begin{equation}
P(\tau) = d_0(s_0) \pi(a_0 \vert s_0)\prod_{t=1}^{T-1} P(s_t \vert s_{t-1}, a_{t-1}) \pi(a_t \vert s_t)
\end{equation}
to denote the probability of the $T$-step trajectory $\tau = s_0,a_0,\dots,s_{T-1},a_{T-1}$. Third, the notation
\begin{equation}
\label{eq: p_t-state}
\mathbb{P}_t(s \vert \pi) = \mathbb{E}_{\pi}[I(s_t=s)]  
\end{equation}
refers to the probability under policy $\pi$ that $s_t=s$ at time $t$, and analogously the notation
\begin{equation}
\label{eq: p_t-stateaction}
\mathbb{P}_t(s,a \vert \pi) = \mathbb{E}_{\pi}[I((s_t,a_t)=(s,a))]  
\end{equation}
refers to the probability under policy $\pi$ that $(s_t,a_t) = (s,a)$ at time $t$.

As the policy $\pi$ is parametrised by $\theta$, policy gradient algorithms aim to maximise the value of that policy by updating the policy parameters according to gradient ascent,
\begin{align*}
\theta \gets \theta + \eta \nabla_{\theta} V(d_0) \,.
\end{align*}

For MDPs, an optimal deterministic policy $\mu^*: \mathcal{S} \to \mathcal{A}$ is guaranteed to exist (see Theorem~6.2.7 in \cite{Puterman1994}) and we devise stochastic policies to explore the state-action space before converging to a (near-)optimal deterministic policy.

In practice, the policy gradient $\nabla_{\theta} V(d_0)$ is not known exactly but should be estimated from samples. Traditional policy gradient algorithms estimate the value based on classical Monte Carlo. For instance, in the limited rollout implementation of the REINFORCE algorithm \citep{Peters2008a}, the policy gradient is given by
\begin{align}
\label{eq: policy gradient}
\nabla_{\theta} V(d_0) &= \mathbb{E}\left[\sum_{t=0}^{T-1} \nabla_{\theta} \log(\pi(a_t \vert s_t)) \sum_{k=0}^{T-1} \gamma^k r_k \right] \,,
\end{align}
which has to be estimated from sampled trajectories. Due to the use of the Monte Carlo discounted sum of rewards, this formulation leads to high variance estimates and thereby large estimation errors. \textit{Actor-critic algorithms} reduce the variance by considering a critic $\hat{Q}(s,a)$ in the policy gradient definition,
\begin{align*}
 \nabla_{\theta} V(d_0) &:= \mathbb{E}\left[\hat{Q}(s,a) \nabla_{\theta} \log(\pi(a \vert s)) \right]\,,
\end{align*}
such that the discounted sum of rewards of Eq.~\ref{eq: policy gradient} is replaced by a function approximator that represents the state-action value.

\subsection{Reproducing Kernel Hilbert Space}
\label{sec: RKHS}
A kernel $K: \mathcal{X} \times \mathcal{X} \to \mathcal{Y}$ is a function that implicitly defines a similarity metric in a feature Hilbert space $\mathcal{H}_K$ through feature-maps of the form $\phi(x) = K(\cdot,x)$. Kernels have the defining property that they are positive definite and symmetric, such that $K(x,y) \geq 0$ and $K(x,y) = K(y,x)$ for all $x,y \in \mathcal{X}$. Reproducing kernels have the additional reproducing property, namely that if $f \in \mathcal{H}_K$, then 
\begin{equation}
\label{eq: reproducing}
f(x) = \langle f(\cdot), K(\cdot, x) \rangle \,.
\end{equation}
If a reproducing kernel $K$ spans the Hilbert space $\mathcal{H}_K$, in the sense that $\text{span}\{K(\cdot,x) : x \in \mathcal{X}\} = \mathcal{H}_K$, then $\mathcal{H}_K$ is called a \textit{reproducing kernel Hilbert space (RKHS)}.

\textbf{Operator-valued RKHS:} Traditionally, kernel functions are scalar-valued, i.e. $\mathcal{Y} = \mathbb{R}$ or $\mathcal{Y} = \mathbb{C}$. However, the RKHS can also be formulated to be operator-valued by formulating a kernel function such that $K(x,y)$ outputs a matrix in $\mathbb{C}^{A \times A}$, where $A$ is the output dimensionality. The paper will include two settings, namely the trivial case 
\begin{equation}
K(x,y) = \kappa(x,y) \mathbb{I}_A \,,
\end{equation} 
where $\kappa$ is a real- or complex-valued kernel, and the more general case 
\begin{equation}
K(x,y) = \kappa(x,y) \mathbf{M} \,,
\end{equation}
where a matrix $\mathbf{M}$ additionally captures scaling factors for each output dimension on the diagonal elements and the correlations between the output dimensions on non-diagonal elements.

\textbf{Quantum kernels.} 
In the context of quantum kernels, a variety of different Hilbert spaces need to be distinguished. In general, a quantum system with discrete basis can be described in terms of a $2^n$-dimensional Hilbert space, a vector space $\mathbb{C}^{2^n}$ with inner product $\langle v \vert w \rangle = \sum_{i=1}^{n} v_i^* w_i$. In the context of quantum kernels, two Hilbert spaces are often mentioned (see e.g. \citep{Schuld2021}). First, corresponding to each quantum kernel is a quantum encoding (or feature-map), which is often named this way since it encodes classical data $s$ into a quantum state $\ket{\phi(s)} \in \mathbb{C}^{2^n}$, corresponding to the density matrix $\rho = \ket{\phi(s)} \langle\phi(s) \vert  \in \mathbb{C}^{2^n \times 2^n}$. Table~\ref{tab: quantum-kernels} provides an overview of a few selected quantum kernels based on \citep{Schuld2021}. A second Hilbert space of interest is the space of quantum models, a space of functions $\mathcal{H}_{\kappa}$ defined for a particular kernel $\kappa$ such that each $f \in \mathcal{H}_{\kappa}$ takes the form of Eq.~\ref{eq: representer}. This space is directly related to the previous in the sense that the kernel can be expressed in terms of an inner product over feature-maps. They can also be expressed equivalently in terms of the density matrix as $f(s) =   \Tr(\rho H)$ where $H$ is an Hermitian operator representing the measurement. Due to spanning the space and having the reproducing property, the space of quantum models, i.e. $\mathcal{H}_K$, is an RKHS. We note that in our approach, we are interested in computing kernel functions over vector-valued eigenstates and eigenactions. With classical parametrisation of quantum circuit this is not a challenge. In quantum circuits, it becomes challenging to perform such computations coherently. Given the given the matching input types in Table~\ref{tab: quantum-kernels}, the basis encoding, i.e. the Kronecker delta, is a natural choice. We explore the Kronecker delta as well as more general inner product circuits  parametrised by the eigenstate.

\begin{table}[htbp!]
\caption{Overview of selected quantum kernels and their basic properties, including encoding, space complexity, and time complexity. The number of qubits in the computational basis is given by $n$. The notation $\text{round}(x)$ refers to rounding the number $x$ to the closest element in the computational basis. For vector-valued states, $n = kS$, where $S$ is the dimensionality of the state-space and $k$ is the per-dimension precision. $\ket{v_i}$ refers to the $i$'th computational basis state.}
\label{tab: quantum-kernels}
\begin{tabular}{p{6.0cm} p{6.0cm} p{1.2cm} p{1.2cm}}
\toprule
Encoding &  Kernel  &  Space \newline (qubits) & Time \newline (depth) \\
\midrule
Basis encoding \newline $\phi: \mathbb{R}^{S} \to \mathbb{C}^{2^n \times 2^n} \newline
\phantom{\phi:} x \mapsto \vert \text{round}(x) \rangle \langle \text{round}(x) \vert$ & Kronecker delta \newline $\kappa(x,x') = \vert \langle \text{round}(x) \vert \text{round}(x') \rangle \vert^2 = \delta_{x,x'}$ & $\BigO(n)$ & $\BigO(1)$          \\  
Amplitude encoding \newline $\phi: \mathbb{C}^{2^n} \to \mathbb{C}^{2^n \times 2^n} \newline
\phantom{\phi:}x  \mapsto \vert x \rangle \langle x \vert = \sum_{i,j=0}^{2^{n}-1} x_i x_j^* \ket{v_i} \langle v_j \vert$ & Fidelity kernel/Quantum kernel of pure states \newline $\kappa(x,x') = \vert \langle x \vert x' \rangle \vert^2$ & $\BigO(n)$ &   $\BigO(2^{n})$         \\
Repeated amplitude encoding \newline $\phi: \mathbb{C}^{2^n} \to \mathbb{C}^{2^{rn} \times 2^{rn}} \newline
\phantom{\phi:} x \mapsto \left(\vert x \rangle \langle x \vert\right)^{\otimes r}$ & $r$-power quantum kernel of pure states \newline $\kappa(x,x') = \left(\vert \langle x \vert x' \rangle \vert^2\right)^r$ & $\BigO(rn)$ &   $\BigO(2^n)$    \\  
Rotation encoding \newline $\phi:  \mathbb{R}^{n} \to \mathbb{C}^{2^n \times 2^n} \newline
\phantom{\phi:} x \mapsto \vert \varphi(x) \rangle \langle \varphi(x) \vert$, 
\newline $\ket{\varphi(x)}=\sum_{q_1,\cdots,q_n = 0}^{1} \prod_{j=1}^{n} \cos(x_j)^{q_j} \sin(x_j)^{1-q_j} \vert q_1,\dots , q_n \rangle$ & Variant of Squared cosine kernel \newline $\kappa(x,x') = \prod_{j=1}^{n} \vert \cos(x_j - x_{j}') \vert^2  $ & $\BigO(n)$ &   $\BigO(1)$        \\
\bottomrule
\end{tabular}
\end{table}

\subsection{Gradient estimation and approximations}
The policy of the RL algorithm will be parametrised by $\theta$, which is a $d$-dimensional set of variables. Sets of the form $\{1,2,\dots,n\}$ are written as $[n]$ for short. We define the multi-index notation $\alpha = (\alpha_1,\dots,\alpha_n) \in [d]^n$ for $\alpha_i \in \mathbb{N}^+$. The notation is useful for higher-order partial derivatives of the form $\partial_{\alpha} f(x) = \frac{\partial^{n} }{\partial_{\alpha_1 \alpha_2 \dots \alpha_n}} f(x)$. We also use the following notation for truncation with respect to the $\ell_2$ norm, namely
\begin{align*}
\trunc{x}{a}{b} = \begin{cases}
x & \text{for } \norm{x}{2} \in [a,b] \\
0 & \text{otherwise.}
\end{cases}
\end{align*}

In addition to standard big O notations, we also use $\epsilon = \BigO_P(n^{-x})$ to denote the rate of convergence in probability, i.e. that there exists a $\delta > 0$ such that for all $n$ and with probability at least $1 - \delta$, the error satisfies $\epsilon \leq C n^{-x}$ for some constant $C > 0$. Moreover, for two positive sequences $x_n$ and $y_n$, the notation $x_n \asymp y_n$ is used to indicated that $C \leq x_n / y_n \leq C'$ for some constants $C,C' > 0$.

\subsection{The quantum-classical setup}
\label{sec: quantum-classical setup}
The above learning representation is implemented in a quantum-classical setup, in which the environment interactions occur on a quantum device whereas learning parameters are stored and updated on a classical device. The interaction is assumed to follow the same conventions as in the quantum policy gradient setting, where the agent obtains $T$-step trajectories from queries to a set of quantum oracles \citep{Jerbi2022}. 
\begin{definition} 
\label{def: oracles}
\textbf{Oracles for Quantum-Accessible MDPs.} The following types of quantum oracles are used, with the first four being essential and the last two building on the previous:
\begin{itemize}
\item \textbf{Transition oracle} $O_P: \ket{s, a} \ket{0} \to \ket{s, a} \sum_{s' \in \mathcal{S}} \sqrt{P(s' \vert s,a )} \ket{s'}$.
\item \textbf{Reward oracle} $O_R : \ket{s, a} \ket{0} \to \ket{s, a} \ket{r(s,a)}$.
\item \textbf{Initial state oracle} $O_{d_0}: \ket{0} \to \sum_{s \in \mathcal{S}} \sqrt{d_0(s)} \ket{s}$. 
\item \textbf{Policy evaluation oracle} (see e.g. Fig.~\ref{fig: Pi})  $\Pi: \ket{\theta} \ket{s} \ket{0} \to \ket{\theta} \ket{s} \sum_{a \in \mathcal{A}} \sqrt{\pi(a \vert s)}  \vert a \rangle$. The oracle applies the policy with parameters $\theta$ coherently to the superposition over states.
\item \textbf{Trajectory oracle} $U_P: \ket{\theta} \ket{0} \to \ket{\theta} \ket{s_0} \sum_{\tau} \sqrt{P(\tau)} \ket{s_0,a_0, s_1, a_1, \dots, s_{T-1}, a_{T-1}}$, \\
where $P(\tau) = d_0(s_0) \pi(a_0 \vert s_0) \prod_{t=1}^{T-1}  P(s_{t} \vert s_{t-1}, a_{t-1}) \pi(a_{t} \vert s_{t})$. The oracle uses 1 call to $O_{d_0}$, $T$ calls to $\Pi$, and $T-1$ calls to $O_P$ to define a superposition over trajectories.
\item \textbf{Return oracle} $U_R: \ket{\tau} \ket{0} \to \ket{\tau} \ket{R(\tau)}$. This oracle computes the discounted return of the trajectory superpositions, and is based on $T$ calls to $O_R$. 
\end{itemize}
\end{definition}

The oracles are then combined as subroutines of a quantum gradient estimation algorithm, which returns an $\epsilon$-close approximation to the policy gradient $\bar{X} \approx \nabla_{\theta} V(d_0)$. The parameter vector $\theta$ can then be updated classically according to the policy gradient update $\theta \gets \theta + \eta \bar{X}$. The quantum circuit is then updated with the new $\theta$ for the subsequent episode(s).

\subsection{Vector-valued state and action spaces}
\label{sec: state-action space}
To represent a rich class of state and action spaces, we represent classical states $\mathcal{S}$ as $S k$-bit representations of vectors in $\mathbb{C}^S$, classical actions $\mathcal{A}$ as $A k$-bit representations of vectors in $\mathbb{C}^A$, and classical rewards $\mathcal{R}$ as $k$-bit representations of scalars in $\mathbb{R}$. Using similar terminology as in \cite{Dong2008}, we refer to these as eigenstates, eigenactions, and eigenrewards, and these form the computational basis. The agent will be in a superposition over eigenstates, i.e. a quantum state of the form $\ket{s'} = \sum_{s \in \mathcal{S}} c(s) \ket{s} \in \mathbb{C}^{2^{Sk}}$, where  $\sum_{s \in \mathcal{S}} \vert c(s) \vert^2 = 1$ and any quantum state in this context $\vert s' \rangle = \ket{s'[0][0],\dots,s'[0][k-1],s'[1][0],\dots,s'[1][k-1],\dots,s'[S-1][k-1]}$. Similarly, a quantum action can be written as a superposition of eigenactions $\ket{a'} = \sum_{a \in \mathcal{A}} c(a) \ket{a} \in \mathbb{C}^{2^{Ak}}$ where  $\sum_{a \in \mathcal{A}} \vert c(a) \vert^2 = 1$ and $\ket{a'} = \ket{a'[0][0],\dots,a'[0][k-1],a'[1][0],\dots,a'[1][k-1],\dots,a'[A-1][k-1]}$. Rewards are superpositions of the form  $\vert r' \rangle =  \sum_{r \in  \mathcal{R}} c(r) \ket{r}$ where $\sum_{r \in \mathcal{R}} \vert c(r) \vert^2 = 1$ and $\ket{r'} = \ket{r'[0],\dots, r'[k-1]}$.\footnote{While we assume the reward function is deterministic, the rewards are in superposition due to the dependency on the trajectory superposition.}  As in the above, the remainder of the document will use the double square bracket notation to indicate the dimension and qubit index in the first and second bracket, respectively. When using a single bracket, e.g. $s[j]$, it refers to all qubits in the $j$'th dimension. A related notation that will be used is $\ket{0}$ instead of $\ket{0}^{\otimes n}$ when this is clear from the context.

As an illustrative example of quantum-accessible MDPs with vector-valued state and action spaces, consider the following state control environment, which is used in simulation experiments (see Section~\ref{sec: learnability RPQC}). The environment is based on a trajectory oracle $U_P$ and a return oracle $U_R$, which is implemented based on at most $T$ applications of the more basic oracles, resulting in a circuit comparable to the circuits in Fig.~\ref{fig: diagram}.
\begin{example}
\label{ex: quantum-accessible MDP}
Consider a $T$-step quantum-accessible MDP with $T=3$. For each time step $t \in \{0,\dots,T-1\}$ there are $(S+A+1)k$ qubits, where $S = A = 2$ and $k=1$ is the number of bits per dimension. $O_{d_0}$ is based on the $2$-qubit Hadamard-Welsch gate, preparing an equal superposition over states. Then two consecutive applications of $\Pi$, $O_R$, and $O_P$ follow, and at the last time step $\Pi$ and $O_R$ are performed. The task of the agent is to control the environment state to be $\ket{0,0}$. The environment gives rewards to the agent using the reward oracle $O_R$, which is an $R_Y(\pi)$-gate controlled on the state register reading $\ket{0,0}$. The transition oracle $O_P$ first applies a set of CX-gates, so that the $t$'th state register reads the same as $\ket{s_{t-1}}$, and then applies for each $i=0,1$ an $R_Y(\pi/2)$ gate on the $i$'th state dimension $\ket{s_t[i]}$, using the $i$'th action dimension $\ket{a_{t-1}[i]}$ as a control with control state $\ket{1}$.
\end{example}
It is clear that at least $(S+A+1) kT$ qubits are required for such circuits. In quantum policy gradient algorithms, additional qubits are required for quantum gradient estimation, more advanced quantum oracles (see e.g. Fig.~\ref{fig: U_X}), and in some cases also the policy evaluation oracle (see Section~\ref{sec: QKPs}).

\section{Background}
\label{sec: background}
\begin{figure}[htbp!]
\subfigure[Quantum Kernel Policies]{
\includegraphics[width=0.99\linewidth]{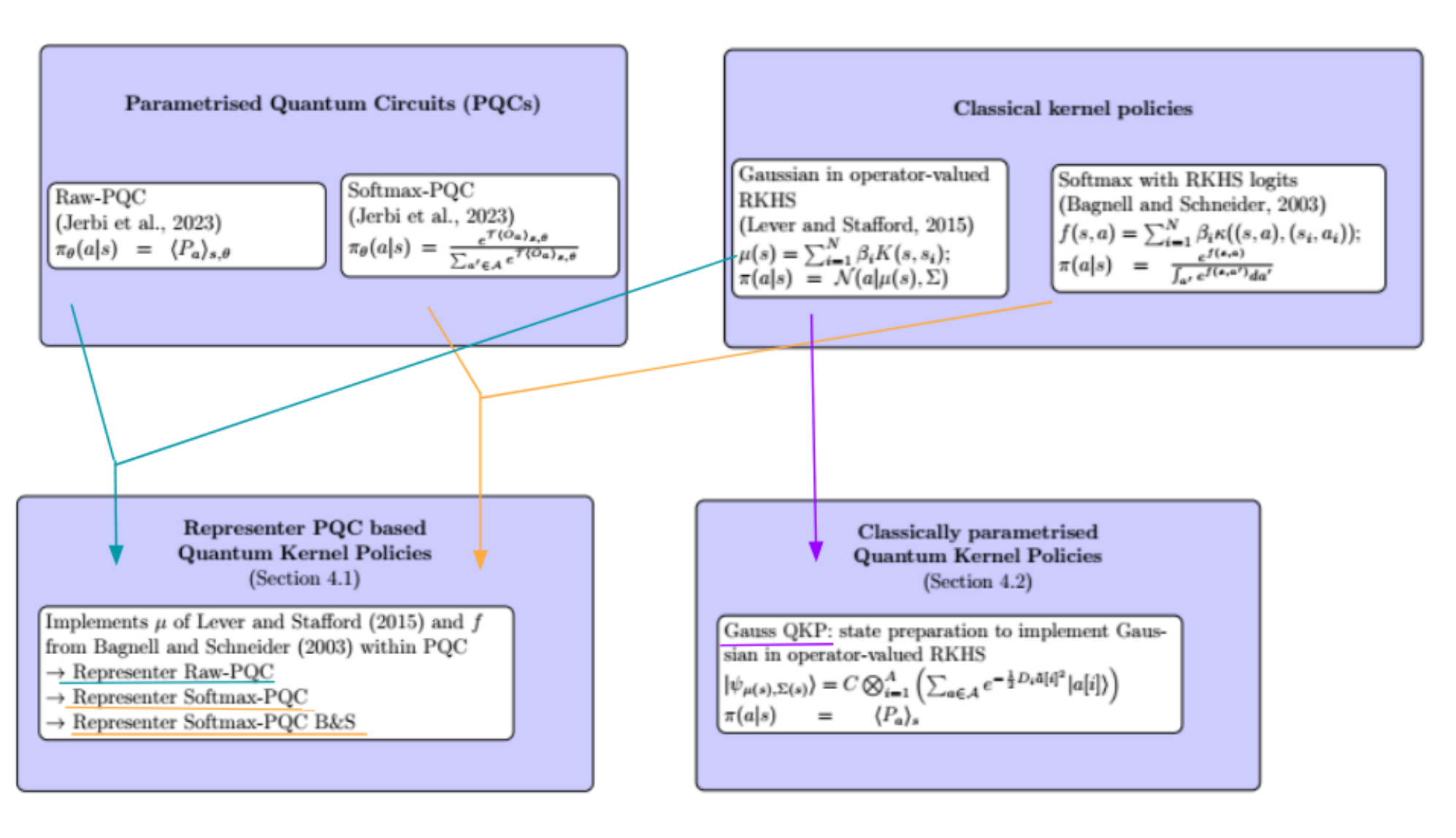} 
}
\subfigure[Quantum policy gradient in RKHS]{
\includegraphics[width=0.99\linewidth]{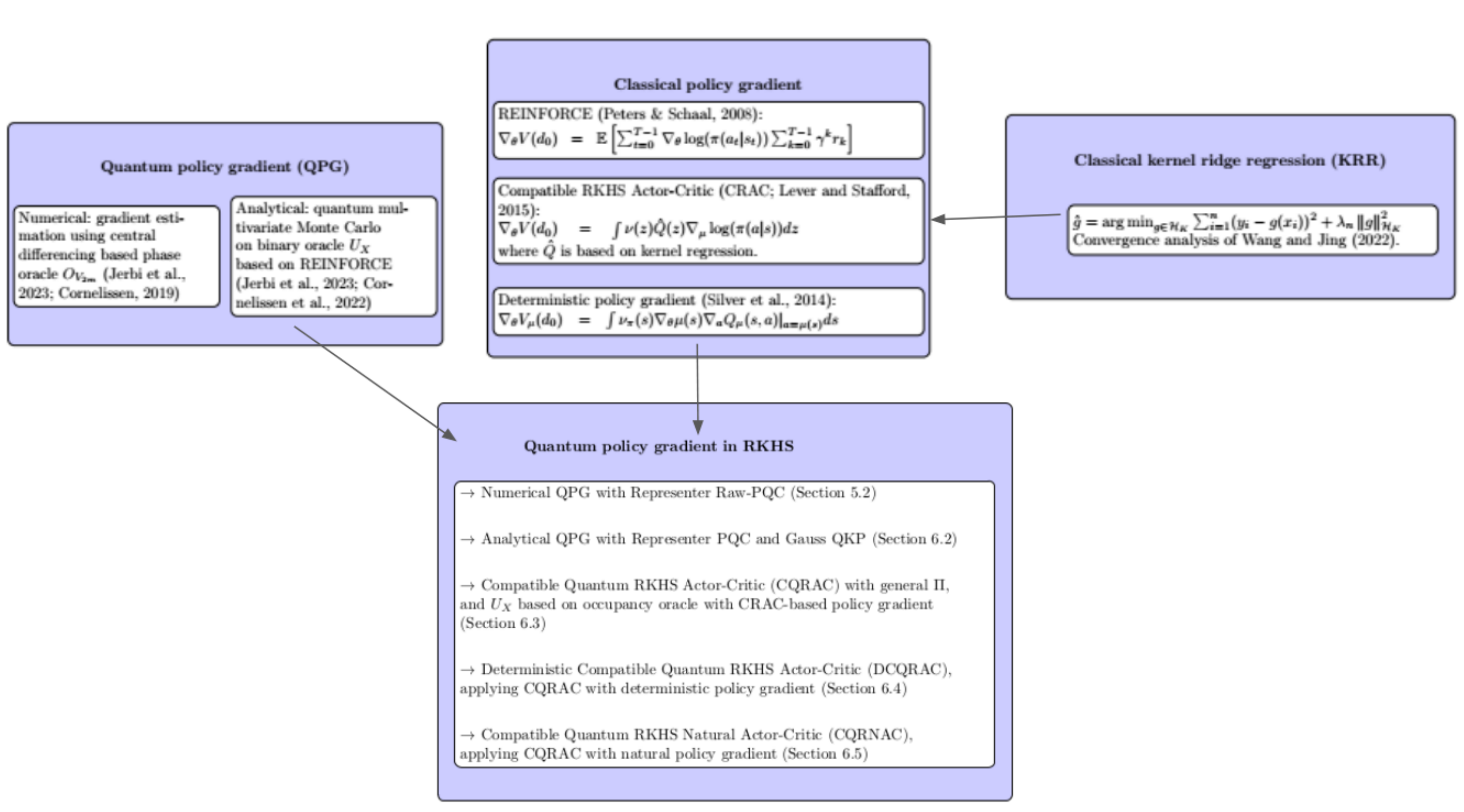}
}
\caption{Diagram illustrating the background as it relates to this work.}
\label{fig: diagram-background}
\end{figure}

Our work will make use of formalisms introduced by four classes of prior works, as illustrated in Fig.~\ref{fig: diagram-background}. The first set of formalisms is related to the quantum policy gradient algorithms due to \cite{Jerbi2022}, which allows us to use the above-mentioned quantum oracles to efficiently compute the policy gradient using both numerical and analytical techniques. The second set of formalisms pertains to the work by \cite{Lever2015}, who formulate classical Gaussian kernel-based policies within an operator-valued RKHS framework, the extension of which leads to our Compatible Quantum RKHS Actor-Critic algorithm. The third set of formalisms is based on the work of \cite{Bagnell2003}, who formulate softmax policies within RKHS for REINFORCE, an approach we also cover in our query complexity analysis. Finally, to bound the error of the classical critic in our actor-critic algorithms, we also make use of results on the convergence rate of kernel ridge regression by \citep{Wang2022}.

\subsection{Quantum policy gradient}
\cite{Jerbi2021} propose quantum policy gradient algorithms for numerical and analytical gradient estimation for quantum-accessible MDPs as mentioned in Definition~\ref{def: oracles}. 

\subsubsection{Quantum policies}
When trajectories are sampled within a quantum-accessible MDP, the policy $\pi$ is evaluated according to the policy evaluation oracle $\Pi$ and can be expressed classically in terms of the expectation of a PQC. Previous work  \citep{Jerbi2021,Jerbi2022} formulates three variants of PQCs to support their derivations. The Representer PQCs that we propose in Section~\ref{sec: QKPs} can be cast to such PQCs, exploiting their properties.

The Raw-PQC as defined below provides a circuit for coherent policy execution (directly applicable to $\Pi$) and allows us to exploit bounds on higher-order derivatives for query-efficient numerical policy gradient algorithms. 
\begin{definition}
\label{def: raw-pqc}
\textbf{Raw-PQC.} The Raw-PQC \citep{Jerbi2021,Jerbi2022} defines 
\begin{equation}
\label{eq: Raw-PQC}
\pi_{\theta}(a \vert s) = \langle P_a \rangle_{s,\theta} \,,
\end{equation}
where $P_a$ is the projection associated to action $a$ such that $\sum_a P_{a} = \mathbb{I}$, $P_{a}P_{a'} = \delta_{a,a'} P_{a}$, and the expectation $\langle P_{a} \rangle_{s,\theta} = \langle \psi_{s,\theta} \vert P_a  \vert \psi_{s,\theta} \rangle$ is the probability of being projected onto the basis state $\vert a \rangle$.
\end{definition}

A second class of policies derived from PQCs is the Softmax-PQC which implements softmax policies from a PQC.
\begin{definition}
\label{def: softmax-pqc}
\textbf{Softmax-PQC.} The Softmax-PQC \citep{Jerbi2021} defines the policy as 
\begin{equation}
\label{eq: Softmax-PQC}
\pi_{\theta}(a \vert s) = \frac{e^{\mathcal{T} \langle O_a \rangle_{s,\theta} }}{\sum_{a' \in \mathcal{A}} e^{\mathcal{T} \langle O_a \rangle_{s,\theta}}} \,,
\end{equation}
where $\mathcal{T}>0$ is a temperature parameter. Defining $\theta=(w,\phi)$, the observables in Eq.~\ref{eq: Softmax-PQC} are given by  
\begin{align*}
\langle O_a \rangle_{s,\phi} = \langle \psi_{s,\phi}  \vert \sum_{i=1}^{N_w} w_i H_{a,i} \vert  \psi_{s,\phi} \rangle \,,
\end{align*}
where $w_{a,i} \in \mathbb{R}$ and $H_{a,i}$ is a Hermitian operator. Both $w$ and $\phi$ are trainable parameters, where $\phi$ refers to parameters within the circuit (e.g. rotation angles).
\end{definition}
A particularly useful special case is the Softmax-1-PQC as defined below, which allows us to exploit an upper bound on its analytical gradient for query-efficient analytical policy gradient algorithms.
\begin{definition}
\label{def: softmax-1-pqc}
\textbf{Softmax-1-PQC.} The Softmax-1-PQC \citep{Jerbi2022} is an instance of Softmax-PQC for which $\phi = \emptyset$ and for all $a \in \mathcal{A}$, $H_{a,i} = P_{a,i}$ is a projection on a subspace indexed by $i$ such that $\sum_{i=1}^{N_w} P_{a,i} = \mathbb{I}$ and $P_{a,i}P_{a,j} = \delta_{i,j} P_{a,i}$ for all $i=1,\dots,N_w$.
\end{definition}

\subsubsection{Numerical policy gradient}
To estimate the policy gradient with minimal query complexity, we make use of numerical gradient estimation based on quantum gradient estimation of Gevrey functions \citep{Cornelissen2019}. To implement the policy in this case, we design circuits which perform a set of controlled rotations based on real-valued parameters $\theta$ such that their expectation is a representer formula for some particular kernel.

This numerical approach is based on central differencing,  which implements a quantum circuit that obtains the value for different settings of the policy parameters to estimate the gradient of the value. Doing so requires a phase oracle for the value as defined below.
\begin{definition}
\label{def: phase oracle}
A \textbf{phase oracle} of the value function (Lemma~2.3 and Theorem~3.1 in \cite{Jerbi2022}; Corollary~4.1 in \cite{Gilyen2019}) encodes the phase of the value function $V(d_0; \theta) = \sum_{\tau} P(\tau) R(\tau)$ into the input register, according to
\begin{align*}
O_V: \vert \theta \rangle \to \vert \theta \rangle e^{i \tilde{V}(d_0; \theta)} \,,
\end{align*}
where $\tilde{V}(d_0; \theta) = \frac{V(d_0; \theta) (1-\gamma)}{r_{\text{max}}} \in [-1,1]$ is the normalised value, and $\theta$ parametrises the policy (and thereby $P(\tau)$ and $V(d_0; \theta)$). The phase oracle can be obtained up to $\epsilon$-precision within $\BigO(\log(1/\epsilon))$ queries to a \textbf{probability oracle} of the form:
\begin{align*}
O_{PV}: \vert \theta \rangle \vert 0 \rangle \to  \vert \theta \rangle  \left( \sqrt{\tilde{V}(d_0; \theta)} \vert \psi_0(\theta) \rangle \vert 0 \rangle  + \sqrt{1 - \tilde{V}(d_0; \theta)} \vert \psi_1(\theta) \rangle \vert 1 \rangle    \right) \,,
\end{align*}
where $\vert \psi_0(\theta) \rangle$ and $\vert \psi_1(\theta) \rangle$ are (often entangled) states in an additional register.
\end{definition}

The central differencing technique can be used on functions which satisfy the Gevrey condition.
\begin{definition}
For any function $f: \mathbb{R}^d \to \mathbb{R}$, the \textbf{Gevrey condition} is a smoothness condition according to which, for some parameters $M > 0$, $c > 0$, and $\sigma \in [0,1]$, the higher order derivative with respect to multi-indices $\alpha \in [d]^p$ satisfies
\begin{equation}
\label{eq: gevrey}
\vert \partial_{\alpha} f(x) \vert \leq \frac{M}{2} c^p (p!)^{\sigma} \,
\end{equation}
for all $p \in \mathbb{N}_0$ and all $x \in \mathcal{X} \subset \mathbb{R}^d$.
\end{definition}

The value function is one such function, as shown in the lemma below, and its Gevrey smoothness depends critically on the higher-order derivatives of the policy as well as parameters of the MDP.
\begin{lemma}
\label{lem: Gevrey value}
\textbf{Gevrey value function (Lemma F.1 in \cite{Jerbi2022}).} The value $V(\theta)  := V(d_0; \theta)$ as a function of the policy parameters \textbf{satisfies the Gevrey condition with $\sigma = 0$, $M = \frac{4 r_{\text{max}}}{1 - \gamma}$, and $c = DT^2$} where $D$ is an upper bound on the higher-order derivative of the policy:
\begin{align}
\label{eq: D higher-order bound}
D &= \max_{p \in \mathbb{N}_0} D_p  \nonumber \\
D_p &= \max_{\theta \in \mathcal{S}, \alpha \in [d]^p} \sum_{a \in \mathcal{A}} \vert \partial_{\alpha} \pi_{\theta}(a \vert s) \vert \,,
\end{align}
where $\alpha \in [d]^p$.
\end{lemma}

Following the Gevrey smoothness with the above parameters for $\sigma$, $M$, and $c$, and phase oracle access to $V(\theta)$, quantum gradient estimation of Gevrey functions \citep{Cornelissen2019} provides precise estimates with limited query complexity.
\begin{lemma}
\label{lem: QPG Gevrey}
\textbf{Quantum policy gradient estimation of Gevrey value functions (Theorem 3.1 in \cite{Jerbi2022})}. Quantum gradient estimation computes an $\epsilon$-precise estimate of $\nabla_{\theta} V(d_0)$  such that $\vert \vert \bar{X} - \nabla_{\theta} V(d_0) \vert \vert_{\infty} \leq \epsilon$ with failure probability at most $\delta$ within 
\begin{equation}\label{eq: query-complexity numerical}
\tildeO\left(\sqrt{d} \frac{DT^2r_{\text{max}}}{\epsilon (1-\gamma)}\right) \,,
\end{equation}
 yielding a quadratic improvement over the query complexity of the classical numerical gradient estimator (see Lemma G.1 in \cite{Jerbi2022} and \refSI{1}):
\begin{equation}
\tildeO\left(d \left( \frac{DT^2r_{\text{max}}}{\epsilon (1-\gamma)}\right)^2\right) \,.
\end{equation}
queries to $U_P$ and $U_R$ (i.e. $\BigO(T)$ time steps of interaction with the quantum environment).
\end{lemma}

To demonstrate Lemma~\ref{lem: QPG Gevrey}, \cite{Jerbi2022} apply quantum gradient estimation of Gevrey functions (Algorithm~3.7 and Theorem~3.8 of \citep{Cornelissen2019}) on the phase oracle of the value function ($O_V$ in Definition~\ref{def: phase oracle}), which yields an $\epsilon$-precise estimate of $\vert \vert \bar{X} - \nabla_{\theta} V(d_0) \vert \vert_{\infty} \leq \epsilon$ with failure probability at most $\delta$ within 
\begin{equation}
\label{eq: gevrey estimation}
 \tildeO\left(Mcd^{\text{max}\{\sigma,1/2\}}\right) 
\end{equation}
queries. Filling in $\sigma$, $M$, and $c$ into Eq.~\ref{eq: gevrey estimation} based on their values in Lemma~\ref{lem: Gevrey value} yields the desired result.

\subsubsection{Analytical policy gradient}
\label{sec: analytical policy gradient background}
We further make use of an alternative policy gradient estimation based on quantum multivariate Monte Carlo \citep{Cornelissen2022}.

The technique as implemented by \cite{Jerbi2022} uses an analytical expression based on the policy gradient theorem \citep{Sutton2018b}, following the limited rollout implementation of the REINFORCE algorithm \citep{Peters2008a},
\begin{equation}
\label{eq: REINFORCE}
\nabla_{\theta} V(d_0) = \mathbb{E} \left[\sum_{t=0}^{T-1} \nabla_{\theta} \log(\pi(a_t \vert s_t)) \sum_{k=0}^{T-1} \gamma^k r_k \right]  \,.
\end{equation}

The quantity within the expectation of Eq.~\ref{eq: REINFORCE} is highly stochastic with high variance. To estimate the analytical policy gradient from a quantum oracle, we conduct quantum experiments with binary oracle access. 
\begin{definition}
\label{def: binary oracle}
A \textbf{binary oracle} of the random variable $X: \Omega \to \mathbb{R}^d$ obtained from a quantum experiment (see Definition 2.14.1 and 2.14.2 in \cite{Jerbi2022}) is given by 
\begin{align*}
    U_{X,\Omega}: \ket{0} \to \sum_{\omega \in \Omega} \sqrt{P(\omega)}  \ket{\omega} \ket{X(\omega)} \,,
\end{align*}
where $\omega \in \Omega$ is the outcome of the experiment, and $\vert X(\omega) \rangle$ encodes $X(\omega)$ into a binary representation. 
\end{definition}

The trajectory oracle $U_P$ and $U_R$ can be used to form a binary oracle for the analytical expression in Eq.~\ref{eq: REINFORCE}, in which case each outcome $\omega \in \Omega$ is a trajectory $\tau=s_0,a_0,s_1,a_1,\dots,s_{T-1},a_{T-1}$. Similarly, we also formulate alternative oracles for actor-critic formulations, which are based on state(-action) occupancies, in which case $\omega \in \mathcal{S}$ or $\omega \in \mathcal{S} \times \mathcal{A}$, reflecting the discounted frequency of different states or state-action pairs. Each such state(-action) can then be coupled to an analytical expression depending on the critic's prediction, which is implemented in the final computation of $U_{X,\Omega}$ by calling another binary oracle $O_{X,\Omega}$ which computes the expression controlled on the state(-action). The actor-critic algorithm defined in this way helps to reduce the variance of the policy gradient compared to the expression depending on the high-variance cumulative reward. 

As we will design kernel policies, the analytical policy gradient will include an expression in $\mathbb{C}^{d}$ or $\mathbb{R}^d$ depending on the kernel and the action space. While our exposition will focus on the real-valued case for simplicity, for analytical gradient estimation, we treat the complex-valued case as $\mathbb{R}^{2d}$, which is straightforward since outcome-dependent formulas of the policy gradient can be given by a binary oracle and the expectation of a complex random variable can be decomposed into the expectations of real and imaginary parts. 

Work in quantum Monte Carlo \citep{Montanaro2017,VanApeldoorn2021,Cornelissen2022} uses quantum algorithms such as phase estimation and amplitude estimation to efficiently estimate the mean of a random variable using quantum oracles (i.e. binary, probability, and/or phase oracles). Results typically provide quadratic query complexity speedups over classical estimators. Recently, multivariate quantum Monte Carlo has been proposed as a technique to estimate the mean of multi-dimensional vectors. Using techniques proposed by Cornelissen \citep{Cornelissen2022}, this approach has been used to compute high-precision estimates of the policy gradient \citep{Jerbi2022}.
\begin{lemma}
\label{lem: analytical policy gradient}
\textbf{Quantum multivariate Monte Carlo for REINFORCE  (Theorem 4.1 of \cite{Jerbi2022})} Let $\epsilon > 0$ and $p>0$. QBounded (Theorem 3.3 of \cite{Cornelissen2022}) yields an $\epsilon$-precise estimate of $\nabla_{\theta} V(d_0)$ w.r.t $\ell_{\infty}$-norm within
\begin{equation}
\label{eq: query complexity analytic}
\BigO\left(d^{\xi(p)} \dfrac{B_p T r_{\text{max}}\log(d/\delta)}{\epsilon(1-\gamma)} \right) 
\end{equation}
queries to $U_P$ and $U_R$ (i.e. $\BigO(T)$ time steps of interaction with the quantum environment), where $\xi(p) = \max(\{0,1/2 - 1/p\})$ and $B_p \geq \norm{\nabla_{\theta} \log(\pi(a_t \vert s_t))}{p}$. Conversely, the classical policy gradient has query complexity following from \refSI{1},
\begin{equation}
\label{eq: query complexity analytic classic}
\BigO\left( \left(\dfrac{B_p T r_{\text{max}} \log(d/\delta)}{\epsilon(1-\gamma)} \right)^2 \right) \,.
\end{equation}
A full quadratic query complexity speedup is obtained for $p \in [1,2]$.
\end{lemma}
We therefore use similar techniques to prove quadratic improvements for kernel policies and actor-critic algorithms, exploiting variance reduction and smoothness properties.

\subsection{Gaussian RKHS policies and Compatible RKHS Actor-Critic}
\label{sec: background-RKHS-AC}
To design rich Gaussian RKHS policies and formulate an effective quantum actor-critic algorithm, we extend the Compatible RKHS Actor-Critic approach by Lever and Stafford \citep{Lever2015}. The approach forms Gaussian policies with mean based on a representer formula, which is updated by the policy gradient and the parameters of which are regularly sparsified. The algorithm uses a kernel-regression based critic instead of the empirical returns in the computation of the policy gradient. The kernel regression approximator is compatible, i.e. the policy gradient computed using the optimal function in the kernel regression RKHS is equivalent to the true policy gradient, and using the critic in this way helps to lower the variance of the policy gradient estimate.

Being based on a representer formula, the policy parametrisation in this approach is data-driven, i.e. based on the state-action pairs as parameters, which is also referred to as a non-parametric approach. Since the approach assumes actions and states, policies are parametrised by $N$ policy weights $\beta_1, \dots, \beta_{N} \in \mathcal{A}$ and  $N$ policy centres $c_1, \dots, c_{N} \in \mathcal{S}$ for $i=1,\dots,N$. The mean $\mu(s)$ for given state $s \in \mathcal{S}$ is defined based on an operator-valued kernel $K$,
\begin{equation}
\label{eq: GQK mean}
\mu(s) = \sum_{i=1}^{N} K(c_i,s)  \beta_{i} \,,
\end{equation}
which is an action in $\mathcal{A}$. The Gaussian RKHS policy is then defined by the Gaussian distribution with mean $\mu(s)$ and covariance matrix $\Sigma$:
\begin{align}
\label{eq: GQK policy}
\pi(a \vert s) &= \mathcal{N}(\mu(s),\Sigma)  \nonumber \\
               &= \frac{1}{Z} \exp\left(-\frac{1}{2} (\mu(s) - a)^{\intercal} \Sigma^{-1} (\mu(s) - a)\right) \,,
\end{align}
where $Z$ is the normalisation constant. We formulate such policies in a quantum circuit by computing the mean and covariance classically, and then applying state preparation techniques. 

This interpretation allows defining a functional gradient based on the Fr\'{e}chet derivative, a bounded linear map $Dg \vert_{\mu}: \mathcal{H}_K \to \mathbb{R}$ with $\lim_{\norm{h}{} \to 0} \dfrac{\norm{g(\mu + h) - g(\mu)}{\mathbb{R}} - Dg\vert_{\mu}(h)}{\norm{h}{\mathcal{H}_K}} = 0$. Specifically, their result provides (see \refSI{2})
\begin{align*} 
Dg\vert_{\mu}: & h \to (a - \mu(s) ) \Sigma^{-1} h(s)   \\
             &= \langle K(s, \cdot)\Sigma^{-1} (a - \mu(s)) , h(\cdot) \rangle 
\end{align*}
for any operator-valued kernel $K$, such that the gradient is given by 
\begin{equation} 
\label{eq:log-policy-grad}
\nabla_{\mu} \log(\pi(a \vert s)) = K(s, \cdot)\Sigma^{-1} (a - \mu(s)) \,.
\end{equation}
This being a functional gradient with respect to $\mu$, the $\nabla_{\mu} \log(\pi(a \vert s)) \in \mathcal{H}_K$ and is of the same form as the function $\mu(\cdot)$ in Eq.~\ref{eq: GQK mean}. In practice, the gradient can be formulated in terms of an $N \times A$ parameter matrix, and for our purposes we make use of a vectorised form of the analytical gradient w.r.t the policy weights $\beta$ (see \refSI{3}).

To maintain a sparse set of centres and weights, Lever and Stafford propose to periodically apply vector-valued kernel matching pursuit \citep{Mallat1993} in which feature vectors $\{K(c_i,\cdot)\}_{i=1}^{N}$ and weights $\{\beta_i\}_{i=1}^{N}$ are stored based on the error of its corresponding function approximator $\hat{\mu}$. Using the technique, one greedily and incrementally adds the next centre $c_i$ and weight $\beta_i$, when added, yields the lowest mean squared error (MSE): 
\begin{equation}
\label{eq: matching-pursuit}
\min_{c,\beta} \sum_{s_i \in \mathcal{I} \subset \mathcal{S}} \vert \vert \mu(s_i) - (\hat{\mu} + \beta K(c,\colon))(s_i) \vert\vert_2^2 \,,
\end{equation}
where basis functions $K(c,\colon)$ are stored from observed states $c \in \mathcal{S}$ and policy weights $\beta \in \mathcal{A}$ are stored based on observed actions. The resulting estimator approximates the original policy $\mu$ to reduce the number of representers, $N$. A lower $N$ is obtained when meeting a stopping criterion, e.g. based on an MSE improvement below a threshold $\epsilon_{\mu}$. Adaptively restricting the number of basis functions using such a threshold allows tailoring the complexity of the function approximator $\hat{\mu}$ to the complexity of $\mu$. Since our mean policy function $\mu$ is computed classically, we straightforwardly apply this technique to our setting. Note that due to updating $\mu \gets \hat{\mu}$ after sparsification, we only use the notation $\hat{\mu}$ in the context of this approximation process.

With the above policies in mind, we seek to compute the policy gradient within a quantum circuit. The algorithm defines the policy gradient for a Gaussian policy based on the critic $\hat{Q}$ as
\begin{align}
\label{eq: integral background}
\nabla_{\mu} V(d_0) &= \int \nu(z) Q(z) \nabla_{\mu} \log(\pi(s,a)) dz \nonumber \\
                        &= \int \nu(z) Q(z)  K(s, \cdot)\Sigma^{-1} (a - \mu(s))  dz \nonumber \\
                        &\approx \int \nu(z) \hat{Q}(z)  K(s, \cdot)\Sigma^{-1} (a - \mu(s))  dz
\end{align}
where $z \in \mathcal{S}\times \mathcal{A}$, $\hat{Q}:\mathcal{S}\times\mathcal{A} \to \mathbb{R}$ is the critic, and the occupancy measure $\nu$ is defined based on $\mathbb{P}_t$  (see Eq.~\ref{eq: p_t-stateaction}) according to
\begin{equation}
\label{eq: occupancy measure}
\nu(z) := \sum_{t=0}^{T-1} \gamma^{t} \mathbb{P}_t(z \vert \pi) \,,
\end{equation}
which sums the discounted probability at each time based on the policy $\pi$ parameterised by $\mu$ and $\Sigma$. The integral is then approximated based on samples from a related occupancy distribution $(1-\gamma) \nu(s,a)$. To approximate this integral, we perform analytical gradient estimation (as explained in Section~\ref{sec: analytical policy gradient background}). Specifically, we will construct a binary oracle $U_{X,\mathcal{S}\times\mathcal{A}}$ (an instance of Definition~\ref{def: binary oracle}), which computes the state-action occupancy measure as a superposition within a quantum circuit and which calls $O_{X,\mathcal{S}\times\mathcal{A}}$ based on $\hat{Q}(z)  K(s, \cdot)\Sigma^{-1} (a - \mu(s))$. This yields a random variable $X(s,a)$, the expectation of which is estimated with Multivariate Monte Carlo, finally yielding the policy gradient estimate $\bar{X} \approx \nabla_{\mu} V(d_0)$. 

In this context, the critic $\hat{Q}(z)$ is parametrised classicaly by $\mu$ and $\Sigma$ to form a compatible function approximator of $Q(s,a)$ using a kernel regression technique (e.g. kernel matching pursuit \citep{Vincent2002}) with the compatible kernel
\begin{align*}
K_{\mu}((s,a),(s',a')) := K(s, s')(a - \mu(s))^{\intercal} \Sigma^{-1} (a' - \mu(s')) \,.
\end{align*}
This leads to a critic of the form
\begin{equation}
\label{eq: compatible-critic}
\hat{Q}(s,a) = \langle w, \nabla_{\mu} \log(\pi(a \vert s)) \rangle \,,
\end{equation}
where $w \in \mathcal{H}_K$ and  $\nabla_{\mu} \log(\pi(a \vert s)) =  K(s, \cdot)\Sigma^{-1} (a - \mu(s)) \in \mathcal{H}_K$. The objective of the critic is to minimise the MSE:
\begin{align}
\label{eq: critic MSE}
\hat{Q}(s,a) = \argmin_{\hat{Q} \in \mathcal{H}_{K_{\mu}}} \int \tilde{\nu}(z) \frac{1}{1-\gamma} \left(Q(z) - \hat{Q}(z)) \right)^2 dz \,,
\end{align}
where $Q(z) = \mathbb{E}[R(\tau \vert s,a)]$.  
Similar to the proof of Lever and Stafford \citep{Lever2015}, \refSI{4.1} demonstrates that indeed the critic $\hat{Q}$ as defined in Eq.~\ref{eq: compatible-critic} is \textit{compatible}, in the sense that the optimal approximation in Eq.~\ref{eq: integral background} yields an exact equality to $\nabla_{\mu} V(d_0)$:
\begin{equation}
\label{eq: compatible def}
\int \nu(z) Q(z) \nabla_{\mu} \log(\pi(a \vert s)) dz =  \int \nu(z) \hat{Q}(z) \nabla_{\mu} \log(\pi(a \vert s)) dz     \,.   
\end{equation} 
The optimal solution for the critic within the Compatible RKHS Actor-Critic implementation is consistent with the natural policy gradient, which is robust to the choice of the coordinates by taking into account the curvature of the probability manifold that they parametrise. We give a proof of the natural policy gradient interpretation in \refSI{4.2} with reasoning based on a related proof by \cite{Kakade2002}. While Compatible RKHS Actor-Critic does not implement the natural policy gradient, the approach with functional gradient ascent and sparsification in RKHS benefits from the design of smooth policies, efficient parametrisation and feature design, and domain-specific kernels. However, one may leverage both properties by designing a natural actor-critic as we show in Section~\ref{sec: natural AC}.

\subsection{Softmax RKHS policy} 
A second kernel-based policy of interest is the softmax formulation of \cite{Bagnell2003}, which was proposed for REINFORCE. It is formulated as
\begin{equation}
\label{eq: softmax RKHS policy}
\pi(a \vert s) = \frac{1}{Z} e^{\mathcal{T} f(s,a)}
\end{equation}
where $Z = \sum_{a \in \mathcal{A}} e^{\mathcal{T} f(s,a)}$, $\mathcal{T} > 0$ is the temperature parameter, $f: \mathcal{S} \times \mathcal{A} \to \mathbb{R}$ is a state-action dependent function in RKHS according to
\begin{equation}
\label{eq: f}
f(s,a) = \sum_{i=1}^{N} \beta_i \kappa((s_i,a_i),(s,a))\,,
\end{equation}
for $\beta_i \in \mathbb{R}$, $Z = \sum_{a} e^{\mathcal{T} f(s,a)}$, and real-valued kernel $\kappa: \mathcal{S} \times \mathcal{A} \to \mathbb{R}$. That is, now the policy centres are state-action pairs and the policy weights are scalars. We will use this form as a warm-up example, where we straightforwardly apply the analytical gradient estimation technique from Jerbi at al. \citep{Jerbi2022}.

\subsection{Convergence rate of kernel ridge regression}
As already seen in Section~\ref{sec: background-RKHS-AC}, function approximation using kernel regression is a key component of Compatible RKHS Actor-Critic. For some classes of kernels, optimal convergence rates can be demonstrated for kernel regression methods, and for kernel ridge regression in particular. Kernel ridge regression seeks to estimate a function $f: \mathcal{X} \to \mathcal{Y}$ by optimising the objective
\begin{align*}
\hat{f} = \argmin_{g \in \mathcal{H}_K} \frac{1}{n} \sum_{i=1}^{n} (y_i - g(x_i))^2 + \lambda_n \norm{g}{\mathcal{H}_K}^2 \,,
\end{align*}
where $y_i \in \mathcal{Y}$ and $g(x_i) = \sum_{j=1}^{N} \beta_j \kappa(x_i,x_j) \in \mathcal{Y}$ are the target output and the predicted output, respectively, and $\{x_j\}_{j=1}^N \subset \mathcal{X}$. Its optimal coefficients are given by $\beta = (\mathbf{K} + n \lambda_n \mathbb{I}_n)^{-1} Y$, where $\mathbf{K}$ is the Gram matrix and $Y=(y_1,\dots,y_n)^{\intercal}$.

Now we turn to reviewing useful results about kernel regression that can be used to assess the convergence rate of the critic. We will denote $\mathcal{X}$ as the input space and $f: \mathcal{X} \to \mathcal{Y}$ as a function in the RKHS, where for our purposes $\mathcal{X} = \mathcal{S}$ or $\mathcal{S} \times \mathcal{A}$ and $\mathcal{Y}=\mathbb{R}$.

First we provide the definition of a Sobolev space and quasi-uniform sequences, which are the two assumptions required for the convergence rate proof.
\begin{definition}
\textbf{Sobolev space.} A Sobolev space $H^l(\mathcal{X})$ with smoothness degree $l$ is a Hilbert space defined by 
\begin{align*}
H^l(\mathcal{X}) = \{ f \in L^2(\mathcal{X}): \partial_{\alpha}f \in L^2(\mathcal{X}) \text{ for } \vert \alpha \vert \leq l \} \,,
\end{align*}
where $\alpha$ is a multi-index and $\partial_{\alpha} f = \frac{\partial^{n} }{\partial_{\alpha_1 \alpha_2 \dots \alpha_n}} f$. The RKHS spanned by the Mat\'ern kernel in Eq.~2.9 of \citep{Tuo2020} is an example Sobolev space.
\end{definition}

\begin{definition}
\textbf{Quasi-uniform sequence (Definition 2.5 in \cite{Tuo2020} and Example 3.2 in \cite{Wang2022}).} A sequence $x_1,\dots,x_n$ is quasi-uniform if there exists a universal constant $U>0$ such that for all $n > 0$
\begin{align*}
h_n / q_n \leq U \,,
\end{align*}
where $h_n = \max_{x \in \mathcal{X}} \min_{i \in [n]} \norm{x - x_i}{2}$ is the fill distance and $q_n = \min_{i,j \in [n]} \norm{x_i - x_j}{2}$ is the separation distance.
\end{definition}

With these definitions in place, we now turn to reviewing an existing result on $L_2$ norm convergence rates, which we will use to assess the number of samples needed for obtaining $\epsilon$-precise critic functions.
\begin{lemma}
\label{lem: KRR bounds}
\textbf{Convergence rates for kernel ridge regression (Theorem 5.3 and 5.4 in \cite{Wang2022}).} Let $f \in H^l(\mathcal{X})$ be a function in a Sobolev space over $\mathcal{X}$, a convex and compact subset of $\mathbb{R}^d$, and let $l > d/2$. Moreover, let the input samples $x_1,\dots,x_n$ be quasi-uniform in $\mathcal{X}$ and let $y_i = f(x_i) + e_i$ be noisy output samples, where the random errors ($e_i$) are sub-Gaussian. Define the kernel ridge regression estimator
\begin{align*}
\hat{f} = \argmin_{g} \frac{1}{n} \sum_{i=1}^{n} (y_i - g(x_i))^2 + \lambda_n \norm{g}{\mathcal{H}_K}^2 \,,
\end{align*}
where $\lambda_n \asymp n^{-\frac{2\hat{l}}{2l +d}}$. Moreover, let $\hat{l} \geq l/2$ be the smoothing factor in the RKHS of the estimator, $\mathcal{H}_{\kappa}$, where $\kappa: \mathcal{X} \times \mathcal{X} \to \mathbb{R}$ is a kernel that is subject to algebraic decay conditions (see C2 and C3 in \cite{Wang2022}; e.g. a Mat\'ern kernel). Then the $L_2$ error converges in probability according to
\begin{align}
\label{eq: KRR bound}
\norm{\hat{f} - f}{L_2} = \BigO_{P}\left( n^{-\frac{l}{2l+d}} \right) \,.
\end{align}
\end{lemma}

\section{Quantum kernel policies}
\label{sec: QKPs}
To design kernel policies for quantum-accessible MDPs (see Definition~\ref{def: oracles}), a suitable policy evaluation oracle must be constructed. To this end, we design two types of \textit{quantum kernel policies (QKPs)}, which implement representer theorem based policies (see Eq.~\ref{eq: representer}) within a quantum circuit.  A first class of circuits, further called Representer PQCs, implements the representer formula coherently within the circuit, in the sense that the expectation can be written as a representer formula for a particular scalar-valued kernel $\kappa$. Circuits in this class are PQCs which are parametrised in a quantum sense (i.e. directly though rotation angles in the circuit), and a subset of these are suitable for numerical optimisation without any policy estimation while others are proposed for analytical gradient based optimisation. A second class of circuits, called Gaussian quantum kernel policies, prepares a Gaussian wave function based on  classically parametrised mean and covariance functions. Circuits in this class are proposed for analytical gradient based optimisation. In this implementation, classical kernels, such as Mat\'ern kernels and radial basis function kernels, are also supported since their computation can be stored in binary memory. An overview of selected policies and their properties can be found in Table~\ref{tab: QKPs}.

\begin{table}[htbp!]
    \centering
        \caption{Quantum kernel policies and their properties. Notations: \textbf{Param} refers to the parametrisation of the policy; \textbf{Feature} refers to the feature-map computed within the circuit; \textbf{Measure} refers to the measurement operator of the policy; \textbf{Prep} refers to the preparation procedure, which can be direct or require estimation of the policy; \textbf{Complexity} defines the complexity of estimation procedures and the state preperation. For the estimation, we summarise the policy/gradient estimation complexity. For state preparation, we summarise the auxiliary qubit/gate count, where none indicates that the state-action register is sufficient and the gate count is computed per eigenstate.}
    \label{tab: QKPs}
    \begin{tabular}{p{2.0cm} p{1.0cm} p{1.9cm} p{2.0cm} p{2.5cm} p{4.2cm}}
    \toprule 
    \textbf{Policy}        & \textbf{Param} & \textbf{Feature}  & \textbf{Measure} &  \textbf{Prep} & \textbf{Complexity}   \\ 
    \midrule
    1. Representer Raw-PQC   & quantum & state representer formula  & standard action basis & no estimation \newline   GIP \newline Kronecker &  / \newline $\BigO(N n_f)$/$\BigO(N(n_{f}2^{n_f} + Ak))$ \newline none / $\BigO(Ak)$   \\
     2. Representer Softmax-PQC &  both  & state kernel or representer formula &  weighted action basis  & estimation \newline state preparation & $\vert \mathcal{S} \times \mathcal{A} \vert$ observables  \newline cf. Table~\ref{tab: quantum-kernels} or  cf.~1 \ \\
     3. Representer Softmax-PQC B\&S & both & state-action kernel or representer formula & weighted scalar basis &  estimation \newline state preparation &  $\vert \mathcal{S} \times \mathcal{A} \vert$ observables \newline cf. Table~\ref{tab: quantum-kernels} with $n=(S+A)k$ or cf.~1 \\
     4. Gauss-QKP  & classical & state representer formula & standard action basis &  no estimation \newline
     Kitaev-Webb preparation & / \newline none/$\BigO(2^{Ak})$\\
     \bottomrule
    \end{tabular}
\end{table}

\subsection{Representer PQCs}
\label{sec: representer PQC}
We first define the Representer PQC condition, which is a condition on the quantum circuit which states that its expectation implements a representer formula for a particular kernel. 
\begin{definition}
\label{def: RPQC condition}
Let $\kappa: \mathcal{X} \times \mathcal{X} \to \mathbb{R}$ be a kernel. A quantum circuit satisfies the \textbf{Representer PQC condition} for kernel $\kappa$ with respect to the output space $\mathcal{Y}$ if for any $\theta \in \Theta$, there exists a set of centres $\{c_i\}_{i=1}^{N} \subset \mathcal{X}$ and some set of weights $\{\beta_i\}_{i=1}^{N} \subset \mathcal{Y}$ such that 
\begin{align}
\label{eq: RPQC condition}
\langle P \rangle_{x,\theta} = \sum_{i=1}^{N} \beta_i \kappa(x,c_i) 
\end{align}   
for any $x \in \mathcal{X}$. The weight $\beta_i$ is also called the \textbf{associated classical policy weight} for policy centre $c_i$. 
\end{definition}

To explain Representer PQCs, we first note that in supervised learning, implementing representer formulas for quantum circuits has traditionally been done by defining a measurement operator $H=\sum_{i=1}^{N} w_i \rho(c_i)$ in terms of kernel expansions in the data \citep{Schuld2021}:
\begin{equation}
\label{eq: representer traditional}
\langle P \rangle_{s,w} = \langle \phi(s) \vert H \vert \phi(s)  \rangle = \Tr\left(\left(\sum_{i=1}^{N} w_i \rho(c_i)\right) \rho(s) \right) \,,
\end{equation}
where $w_i \in \mathbb{R}$ and $c_i,s \in \mathcal{X}$ for some input space $\mathcal{X}$, and outputs are real values. 

To represent vector-valued actions, one option is to define separate measurements for each centre, i.e.
\begin{equation}
\label{eq: representer traditional multi}
\langle P \rangle_{s,\beta}  = \sum_{i=1}^{N} \beta_i \langle \phi(s) \vert \rho(c_i) \vert \phi(s) \rangle \,,
\end{equation}
where $\beta_i \in \mathcal{A}$ for all $i=1,\dots,N$. 

Using Eq.~\ref{eq: representer traditional} or Eq.~\ref{eq: representer traditional multi} has several limitations for our purposes. First, to make use of the expressiveness afforded by the weight parameter $w \in \mathbb{R}^{N}$, a state preparation algorithm must be applied before its use in the policy evaluation oracle $\Pi$. Eq.~\ref{eq: representer traditional multi} is particularly problematic as it requires $N$ distinct inner product estimations before the state preparation. Second, the stochasticity of such circuits is more challenging to control without the use of any rotation angles inside the circuit. Third, these circuits are limited to quantum kernels, and we seek to form an approach which also allows classical kernels. We will specifically refer to Eq.~\ref{eq: representer traditional}--\ref{eq: representer traditional multi} as \textit{non-coherent Representer PQCs} as opposed to \textit{coherent Representer PQCs} that we will introduce.

\begin{figure}[htbp!]
    \centering
    \subfigure[Kronecker delta]{
    \includegraphics[width=0.90\linewidth]{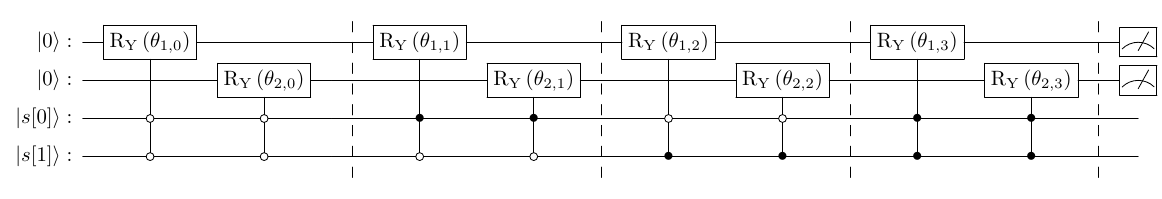} 
    }
    \subfigure[GIP subcircuit]{
    \includegraphics[width=1.0\linewidth]{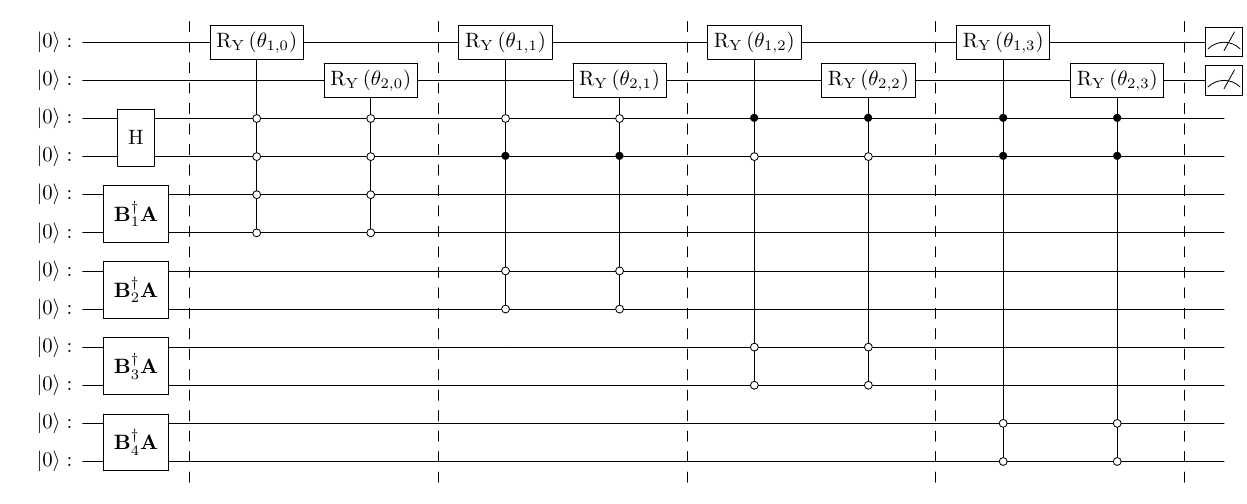} 
    }
    \subfigure[GIP subcircuit with bias]{
    \includegraphics[width=1.0\linewidth]{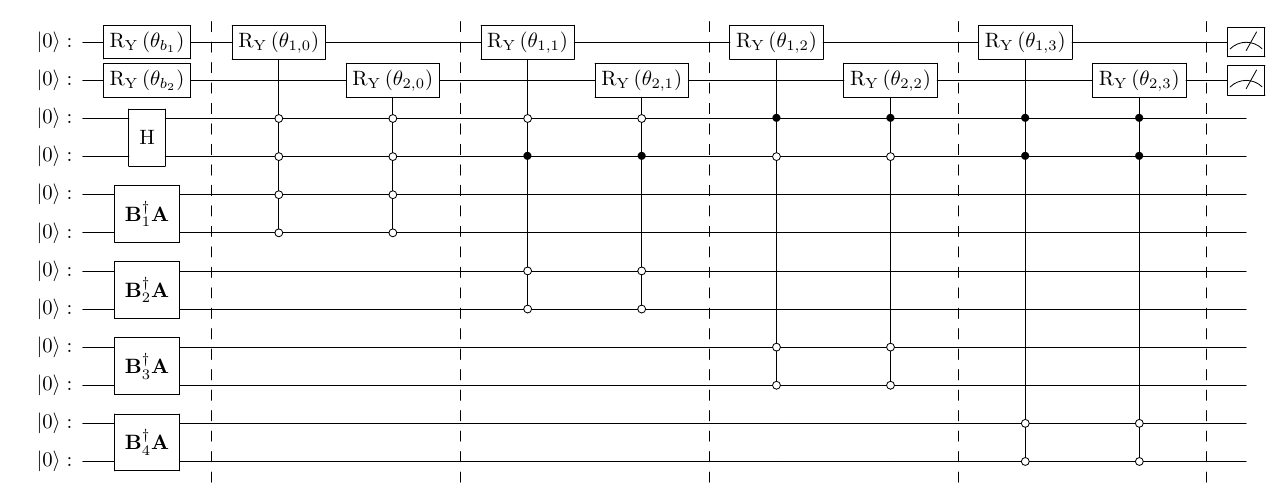} 
    }
    \caption{Implementation of a Representer PQC with two-qubit states and two-qubit actions. \textbf{a)} Kronecker delta kernel is implemented such that for each possible eigenstate, a separate set of rotation angles is applied to the action qubits. \textbf{b)} Subcircuit applied to a particular eigenstate $s \in \mathcal{S}$ to generalise the Representer PQC to general kernels based on an inner product operator. \textbf{c)} If applying the policy directly for coherent policy evaluation, an initial set of rotation gates can determine a learnable action bias. Note 1: to allow state-dependent policies in b--c, different such subcircuits can be joined into a common circuit for superposition states using multi-control. Note 2: to build the policy evaluation oracle $\Pi$, one should omit the measurements, and a quantumly parametrised $\Pi$ requires further controlled gates to rotate the angles controlled on $\ket{\theta}$.}
    \label{fig: representer-pqc}
\end{figure}

\textbf{Kronecker Representer PQC.} To demonstrate a coherent Representer PQC, we formulate a simple proof-of-concept based on the Kronecker delta kernel $\kappa(s,s')=\delta_{s,s'}$. Due to simply computing equality in the computational basis, the kernel can be implemented as multi-controlled gates $R_{Y}$ gates as shown in  Fig.~\ref{fig: representer-pqc}a.  Since there is no overlap between states in the Kronecker delta, one ideally sets $N = \vert \mathcal{S} \vert = 2^{Sk}$ as in Fig.~\ref{fig: representer-pqc}a; this allows each eigenstate its own action distribution. The circuit does not require any auxiliary register in addition to the state and action register. The parameter and gate counts are given by $\BigO(NAk)$.

Since the circuit controls directly on unique eigenstates, the Kronecker Representer PQC can be written as the policy evaluation oracle
\begin{align}
\label{eq: kronecker RPQC}
 \Pi_{\theta}^{\text{Kron}} \ket{s} \ket{0} &=  \ket{s} \sum_{i=1}^{N} \delta_{s,c_i} \ket{\beta_i'} \,,
\end{align}
where $\ket{s}$ is an eigenstate, $\theta$ represents the rotation angles, $N$ is the number of representers. The \textit{quantum policy weight} of centre $i$,
\begin{equation}
\label{eq: quantum policy weight}
\ket{\beta_i'} = \sum_{a \in \mathcal{A}} \psi_i(a) \ket{a} \,,
\end{equation}
has amplitudes defined by
\begin{align}
\label{eq: amplitude beta}
\psi_i(a) := \prod_{j=1}^{Ak} (-1)^{q_{j}(a)} \cos(\theta_{j,i}/2)^{1-q_j(a)} \sin(\theta_{j,i}/2)^{q_{j}(a)} \,, 		  
\end{align}
where $q_j(a)$ indicates the $j$'th qubit of eigenaction $a$, and $\theta_{j,i}$ is the rotation angle for the $j$'th action qubit and the $i$'th centre. The quantum policy weight can be interpreted as a superposition over classical policy weights (i.e. over eigenactions). As shown in Lemma~\ref{lemma: kronecker RPQC}, evaluating the expectation reveals the associated classical policy weight for the Kronecker delta, which is given by 
\begin{equation}
\label{eq: associated classical policy weight}
\beta_i = \sum_{a \in \mathcal{A}} \psi_i(a)^2 a \,,
\end{equation}
and the circuit provably satisfies the Representer PQC condition. Note in Eq.~\ref{eq: kronecker RPQC} that if the rotation angles are $\pi$ or $0$, the policy becomes deterministic for state $s$, yielding a particular eigenaction while angles in between yield stochastic policies, with $\pi/2$ yielding a uniform superposition. 

\begin{lemma}
\label{lemma: kronecker RPQC}
Let $\kappa: \mathcal{S} \times \mathcal{S} \to \mathbb{R}$ be the Kronecker delta kernel and let $\mathcal{A} = \{-2^{k-1}x,\dots,0,\dots,(2^{k-1} -1) x\}^{A}$ be an action space for some scalar $x \in [0,1]$ (e.g. a negative power of 2). Then any policy evaluation oracle $\Pi_{\theta}^{\text{Kron}}$ constructed according to Eq.~\ref{eq: kronecker RPQC}\\
a) satisfies the Representer PQC condition for $\kappa$ w.r.t $\mathbb{R}^A \cap [-2^{k-1}x,(2^{k-1}-1)x]^A$; and \\
b) approximates the Representer PQC condition for $\kappa$ w.r.t $\mathcal{A}$ with $\ell_{\infty}$ error of at most $x/2$.
\end{lemma}
\begin{proof}
Following its definition in Eq.~\ref{eq: kronecker RPQC}, the expectation of the circuit on the action qubits, with measurement in the computational basis, is given for any state $s \in \mathcal{S}$ as
\begin{align*}
\langle P \rangle_{s,\theta} &= \sum_{a \in \mathcal{A}} a \, \langle P_a \rangle_{s,\theta} \\
                             &= \sum_{a \in \mathcal{A}} a \, \left(\sum_{i=1}^{N} \delta_{s,c_i} \psi_i(a)^2 \right)  \tag{(definition in Eq.~\ref{eq: kronecker RPQC} and Eq.~\ref{eq: quantum policy weight}} \\
                            &= \sum_{a \in \mathcal{A}} a \left(\sum_{i=1}^{N} \kappa(s,c_i) \psi_i(a)^2 \right)  \tag{(Kronecker delta kernel}\\
                             &= \sum_{i=1}^{N} \kappa(s,c_i) \left( \sum_{a \in \mathcal{A}} \psi_i(a)^2 a \right) \tag{rearranging} \\
                             &= \sum_{i=1}^{N} \kappa(s,c_i) \beta_i \quad \tag{associated classical policy weight in Eq.~\ref{eq: associated classical policy weight}} \,.
\end{align*}
Since $s$ is arbitrarily chosen and the coefficient $\psi_i(a)^2$ is independent for each state (due to only factoring in $\psi_i(a)^2$ when $ \delta_{s,c_i}=1$), this result matches Eq.~\ref{eq: RPQC condition} for any $s \in \mathcal{S}$ for the given kernel $\kappa$. As last part of the proof, we clarify the dependency on the output space. For part a), note that since $\psi_i(a)^2$ is a squared amplitude for any $a \in \mathcal{A}$, it is a real value in $[0,1]$, which makes it possible for $\beta_i$ to take any number in $\mathbb{R}^A \cap [-2^{k-1}x,(2^{k-1} -1)x]^A$. For part b), note that since actions are evenly spaced, the maximal distance to the closest action in $\mathcal{A}$ is given by the midpoint of two computational basis states; that is, $\norm{\beta_i - \text{round}(\beta_i)}{\infty} \leq x/2$ for any $i \in [N]$.
\end{proof}

\textbf{General Inner Product (GIP) Representer PQC.} Since the Kronecker delta kernel has no correlations between different states, the Kronecker delta Representer PQC would typically require all states to come with their own rotation angles. To make use of correlations between different states, the concept of the coherent Representer PQC can be generalised to other kernels by using the approach of  \cite{Markov2022} to prepare general inner products in the amplitude of $\ket{0}$ of an auxiliary register. With $n_f$ being the number of qubits of the feature space, the inner product between two feature maps can be implemented using two operators $\mathbf{A}$ and $\mathbf{B}$ as
\begin{align}
\label{eq: inner product}
\ket{\varphi_A} = \mathbf{A} \ket{0} = \sum_{i=0}^{2^{n_f} - 1} c_\mathbf{A}(i) \ket{i} \nonumber \\
\ket{\varphi_B} = \mathbf{B} \ket{0} = \sum_{i=0}^{2^{n_f} - 1} c_\mathbf{B}(i) \ket{i} \nonumber \\
\mathbf{B}^{\dagger} \mathbf{A} \ket{0} = \langle \varphi_\mathbf{B} \vert \varphi_\mathbf{A}  \rangle \ket{0} + \sum_{i=1}^{2^{n_f} - 1} c_{G}(i) \ket{i} \,,
\end{align}
where $c_\mathbf{A}(i),c_\mathbf{B}(i)\in \mathbb{C}$ are the amplitudes for all $\ket{i}$, and since only the $\ket{0}$ is associated with amplitude of interest, the amplitudes $c_{G}(:)$ are considered as garbage.

To form such a PQC, which we call a General Inner Product (GIP) Representer PQC, a subcircuit is formed for each $s \in \mathcal{S}$. In each such subcircuit, there are three distinct types of registers. First, there are $N$ inner product registers, which compute the inner products $\langle \phi(s) \vert \phi(c_i) \rangle$ for all $i\in [N]$ by applying the operators $\mathbf{A}$ and $\mathbf{B}_1,\dots,\mathbf{B}_N$. Second, there is one index register, which applies Hadamard gates to form an equal superposition over $\log(N)$ qubits. Each basis state in the index register is linked to a particular inner product register. Third, the action register performs $N$ distinct controlled $R_Y$ gates for each of its $Ak$ action qubits, resulting in $\BigO(NAk)$ total parameters. The control state of the $i$'th rotation of any of the action qubits is given by $\ket{i-1}$ and $\ket{0}$ on the index register and the $i$'th inner product register, respectively. One such sub-circuit with $N=4$ policy centres is shown in Fig.~\ref{fig: representer-pqc}b; the different sub-circuits are joined by multi-control (analogous to Fig.~\ref{fig: Pi}). Overall, the circuit requires $N \times n_f$ additional qubits, where $n_f$ is the number qubits per inner product. Preparing arbitrary quantum states requires depth $\BigO(2^{n_f})$ and space $\BigO(n_f)$ such that the overall gate complexity is $\BigO(N (n_{f} 2^{n_f} + Ak))$. 

\begin{lemma}
\label{lemma: innerproduct RPQC}
Let $\kappa: \mathcal{S} \times \mathcal{S} \to \mathbb{R}$ be a kernel defined by $\kappa(s,s') := \vert \langle  \phi(s) \vert \phi(s') \rangle \vert^2$ for some feature-map $\phi: \mathcal{S} \to \mathbb{C}^{n}$. Then the GIP Representer PQC $\Pi_{\theta}^{\text{GIP}}$ implementing Eq.\ref{eq: inner product} as $\langle  \phi(s) \vert \phi(c_j) \rangle$ for all $j \in [N]$ for each eigenstate $s \in \mathcal{S}$ according to Fig.~\ref{fig: representer-pqc}b) is a Representer PQC, satisfying the Representer PQC condition for $\kappa$ w.r.t $\mathcal{A}$. The associated classical policy weight of the GIP Representer PQC is given for each $j \in [N]$ by 
\begin{equation}
\label{eq: associated classical policy weight GIP}
\beta_j = \frac{1}{N} \sum_{a \in \mathcal{A}} \psi_j(a)^2 a \,.
\end{equation}
\end{lemma}
The proof of this lemma is given in \refSI{5}. It follows a similar argumentation to Lemma~\ref{lemma: kronecker RPQC} but has more involved computations.

Lemma~\ref{lemma: innerproduct RPQC} provides a proof that the expectation of GIP Representer PQCs can be expressed in terms of a representer formula. We now further discuss their distributional properties. First, note that in the circuit in Fig.~\ref{fig: representer-pqc}b, the action $\ket{0,0}$ will be overrepresented. That is, the action $\ket{0,0}$ will have an amplitude at least $1 - \frac{1}{\sqrt{N}} \sum_{j=1}^{N} \langle \phi(s) \vert \phi(c_j) \rangle$ due to the rotations not occurring in garbage states, and as such $\ket{0,0}$ can be seen as a default action that is taken with some probability. One way to mitigate this problem is by allowing a learnable set of parameters to determine the bias, e.g. by using initial rotation angles added independently of the policy centres (see Fig.~\ref{fig: representer-pqc}c), which is quite expressive since it is a superposition over actions. If actions are allowed to be negative floats (setting e.g. $\mathcal{A} = \{-2^{k-1}x,\dots,0,\dots,(2^{k-1} -1) x\}^{A}$), the expectation of the GIP Representer PQC will be a representer formula that allows to deviate around the expected action bias. Second, like Kronecker delta PQCs, another property of GIP Representer PQCs is that their variance depends on the amplitude; in particular, amplitudes of $1/\sqrt{2}$ correspond to high variance, particularly for the most significant bits. Note that when the policy evaluation oracle is to be constructed from estimates of the PQC, the above properties may not have any impact. For instance, in the Representer Softmax-PQCs (see e.g. Definition~\ref{def: representer softmax-pqc}), the bias is not needed as the expectation can be normalised and the variance of the policy evaluation depends only on action-specific expectations of the circuit.

Below we present a few policy formulations based on Representer PQCs, the properties of which are summarised in Table~\ref{tab: QKPs}. Based on the distinction between Raw-PQCs (Definition~\ref{def: raw-pqc}) and Softmax-PQCs (Definition~\ref{def: softmax-pqc}), they vary in their applicability, in terms of coherent computation within a numerical gradient optimisation versus the need for estimating of the policy and the log-policy gradient within an analytical gradient optimisation (see e.g. Appendix B of \cite{Jerbi2021}; \cite{Sequeira2023a}).

A coherent Representer PQC can be formulated as a special case of the Raw-PQC in Definition~\ref{def: raw-pqc}. This makes it suitable for optimisation with numerical gradient without estimating $\pi$ or $\nabla_{\theta} \log(\pi(a \vert s))$, as the circuit can be computed coherently when removing the measurements in Fig.~\ref{fig: representer-pqc}.
\begin{definition}
\label{def: RRPQC policy}
\textbf{Representer Raw-PQC.} A policy $\pi$ is a \textbf{Representer Raw-PQC} if its action probabilities are given by direct measurements in the computational basis, i.e.
\begin{equation}
\label{eq: RRPQC policy}
\pi(a \vert s) = \langle \psi_{\theta,s } \vert P_a \vert \psi_{\theta,s }  \rangle \,,
\end{equation}
where $P_a$ is the projection associated to action $a$ such that $\sum_a P_{a} = \mathbb{I}$, $P_{a}P_{a'} = \delta_{a,a'} P_{a}$, and $\ket{\psi_{\theta,s}}$ is the quantum state prepared for state $s$ and parameter $\theta$ by a coherent Representer PQC -- which is coherent and satisfies Eq.~\ref{eq: RPQC condition}). 
\end{definition}

We also formulate two PQCs suitable for softmax policies, which are closely related to the Softmax-PQC in Definition~\ref{def: softmax-pqc}. This is suitable for optimisation with analytical gradient as it requires estimation of $\pi$ and $\nabla_{\theta} \log(\pi(a \vert s))$ and state preparation.
\begin{definition}
\label{def: representer softmax-pqc}
\textbf{Representer Softmax-PQC and Representer Softmax-1-PQC.} A policy $\pi$ is a \textbf{Representer Softmax-PQC} if its policy probabilities are defined by Eq.~\ref{eq: Softmax-PQC}, with observables defined as 
\begin{align}
\label{eq: representer softmax-pqc}
\langle O_a \rangle_{s,\theta} = \langle \psi_{\phi,s}  \vert \sum_{i=1}^{N_w} w_{a,i} H_{a,i} \vert  \psi_{\phi,s} \rangle \,,
\end{align}
where $\theta = (w,\phi)$, $w_{a,i} \in \mathbb{R}$, $H_{a,i}$ is a Hermitian operator, $\phi$ is the set of rotation angles within the circuit, and $\ket{\psi_{\phi,s}}$ is the quantum state prepared by a Representer PQC (Eq.~\ref{eq: RPQC condition}) for state $s$ and parameter $\phi$. The \textbf{Representer Softmax-1-PQC} restricts the Representer Softmax-PQC in Eq.~\ref{eq: representer softmax-pqc} through the choice of parameters and projectors; it sets $H_{a,i} = P_{a,i}$ to be a projection on a subspace indexed by $i$ such that $\sum_{i=1}^{N_w} P_{a,i} = \mathbb{I}$, $P_{a,i}P_{a,j} = \delta_{i,j} P_{a,i}$ for all $i=1,\dots,N_w$, and $\phi$ to be the empty set  $\emptyset$. 
\end{definition}

Note that with additional controls on eigenactions $\ket{a}$ for all $a \in \mathcal{A}$ and an alternative interpretation of the outputs in terms of $f(s,a)$ rather than an action, a representer formula with state-action kernel of the form $K((s,a),(s',a'))$ can be incorporated within the circuit to optimise the function $f$ in the approach by Bagnell and Schneider \citep{Bagnell2003} (see Eq.~\ref{eq: f}). Since the function $f$ is one-dimensional, this computation lends itself well to the approach of Eq.~\ref{eq: representer traditional}, where we can define measurements in terms of quantum feature-maps over centres in the input space (now $\mathcal{X} = \mathcal{S} \times \mathcal{A}$) to design an appropriate circuit. This policy definition is also non-coherent, which implies it is necessary to estimate $\pi$ and prepare its corresponding wave function.
\begin{definition}
\label{def: representer softmax-pqc B&S}
\textbf{Representer Softmax-PQC B\&S.} A policy $\pi$ is a \textbf{Representer Softmax-PQC B\&S} if its policy probabilities are formed according to
\begin{equation}
\label{eq: Softmax-PQC B&S}
\pi_{\theta}(a \vert s) = \frac{e^{\mathcal{T} \langle O \rangle_{s,a,\theta} }}{\sum_{(s',a') \in \mathcal{S} \times \mathcal{A}} e^{\mathcal{T} \langle O \rangle_{s,a,\theta}}} \,,
\end{equation}
based on the observable 
\begin{align}
\label{eq: representer softmax-pqc B&S}
\langle O \rangle_{s,a,\theta} = \langle \psi_{\phi,s,a}  \vert \sum_{i=1}^{N_w} w_i H_{i} \vert  \psi_{\phi,s,a} \rangle \,,
\end{align}
where $\theta=(w,\phi)$, $\ket{\psi_{\phi,s,a}}$ is the quantum state prepared by a Representer PQC (Eq.~\ref{eq: RPQC condition}) for state $s$, action $a$, and parameter $\phi$, a set of rotation angles within the circuit, and $H= \sum_{i=1}^N w_i H_i$ is a measurement operator that defines for each $i \in [N]$ a weight $w_i \in \mathbb{R}$. The \textbf{Representer Softmax-1-PQC B\&S} restricts the Representer Softmax-PQC B\&S such that $\phi = \emptyset$ and $H_i = \rho(c_i)$, which corresponds to the density matrix of the centre $c_i$.
\end{definition}

Representer Softmax-1-PQC B\&S has a convenient analytical form for the gradient following \cite{Bagnell2003} (see \refSI{6} for a proof of the functional gradient; the vectorised gradient is analogous),
\begin{equation}
\label{eq: Bagnell analytic}
\nabla_{\theta} \log(\pi(a \vert s)) = \mathcal{T} \left( \kappa((s,a),\cdot) - \mathbb{E}_{a' \sim \pi(\cdot \vert s)} \kappa((s,a'),\cdot) \right) \,.
\end{equation}
Thereby this formulation also avoids the additional computations required to estimate $\nabla_{\theta} \log(\pi(a \vert s))$.

\subsection{Gaussian quantum kernel policies}
The Gaussian quantum kernel policy (\textbf{Gauss-QKP}) is a policy that extends the formulation of Lever and Stafford \citep{Lever2015} (see Eq.~\ref{eq: GQK policy}) by formulating it in terms of a quantum wave function. A benefit of this formulation is that gradient computations for $\nabla_{\beta} \log(\pi(a \vert s))$, and even Fisher information computations if needed, are analytically given without requiring additional estimations. 

\begin{figure}
    \centering
    \includegraphics[width=0.99\linewidth]{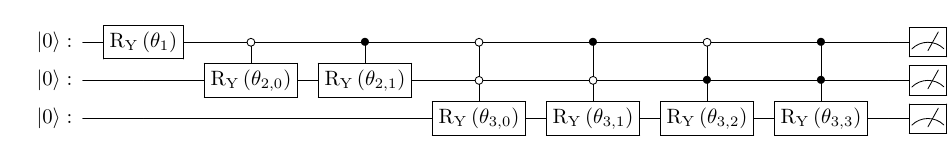}
    \caption{Circuit for the Gaussian quantum kernel policy at a given state $s$. The policy is parametrised by $m := \mu(s)$ and standard deviation $v:=\sqrt{\Sigma(s)}$, where $\theta_{i,j}$ represents the rotation angle for the $i$'th qubit and $j$ represents the control state. For instance, $\theta_{2,0} = 2 \cos^{-1}(\sqrt{F(m_2/2,v_2/2) / F(m_2,v_2)})$ corresponds to the angle when the first qubit is $\ket{0}$ while $\theta_{2,1} = 2 \cos^{-1}(\sqrt{F((m_2 - 1)/2,v_2/2) / F(m_2,v_2)})$ corresponds to the angle when the first qubit is in state $\ket{1}$. Note: the measurements are useful for defining the policy statistics but are removed when calling the circuit coherently within the policy evaluation oracle $\Pi$.}
    \label{fig: KW}
\end{figure}

Upon policy updates, the wave function representing the stochastic policy $\pi$ needs to be updated. For each state, one can compute the mean action $\mu(s)$ and the covariance matrix, $\Sigma(s)$, and the resulting Gaussian wave function within a quantum circuit. One option is to use a general-purpose wave function preparation (a.k.a. state preparation) techniques (e.g. \cite{Shende2006,Carvalho2024}). However, more special-purpose techniques for Gaussian wave function preparation, such as the technique proposed by \cite{Kitaev2008}, are available. To implement the technique by Kitaev and Webb for a given state $s \in \mathcal{S}$ and a single dimension, we use the circuit given in Fig.~\ref{fig: KW}. This implementation is based on \url{https://github.com/msohaibalam/gaussian-wavefunction}, which differs from the original Kitaev-Webb implementation in that no auxiliary registers are required. Our implementation requires $\sum_{n=0}^{Ak - 1} 2^n = \BigO(2^{Ak})$ controlled rotation gates, which implies a total gate complexity of $\BigO(2^{(S+A)k})$ considering it needs to be computed for each $s \in \mathcal{S}$.

First, note that amplitudes for a one-dimensional Gaussian with mean $m$ and standard deviation $v$ can be constructed based on integers as
\begin{align*}
c(a) = \frac{1}{\sqrt{F(m,v)}} e^{-\frac{1}{2v^2} (a - m)^2}
\end{align*} 
where $F(m,v) = \sum_{n=-\infty}^{\infty} e^{(n - m)^2}{v^2}$ which is related to the third Jacobi theta function, and which implies
\begin{align*}
\sum_a c^2(a) = \sum_a \frac{1}{F(m,v)} e^{-\frac{1}{v^2} (a - m)^2} = 1 \,.
\end{align*}
Starting from the least significant qubit, the rotation angles for the qubits $i=1,\dots,k$ are determined recursively as illustrated in Fig.~\ref{fig: KW}. The algorithm requires that $\mu \gg v$, $2^{k} \gg v$, and $v \gg 1$.

To extend this to the multi-dimensional Gaussian with diagonal covariance matrix, the state to prepare becomes 
\begin{align}
\label{eq: wavefun}
\ket{\psi_{\mu(s),\Sigma(s)}} &= C \sum_{a \in \mathcal{A}} e^{-\frac{1}{2}\tilde{a}^{\intercal} \Sigma(s)^{-1} \tilde{a}} \ket{a} \nonumber \\
     &= C \sum_{a \in \mathcal{A}} \prod_{i=0}^{A-1} e^{-\frac{1}{2} D_i \tilde{a}[i]^2 } \ket{a} \nonumber \\
 &= C \bigotimes_{i=1}^{A} \left( \sum_{a \in \mathcal{A}} e^{-\frac{1}{2} D_i \tilde{a}[i]^2 } \ket{a[i]} \right)
\end{align}
where $C^2 = \sqrt{\det \Sigma } \pi^{-A/2}$, $D_i = \Sigma(s)^{-1}_{ii}$ and $\tilde{a} = a - \mu(s)$. This can be implemented with a larger circuit where each of the dimensions is performed independently but completely analogous to Fig.~\ref{fig: KW}.

The Gaussian quantum kernel policy is parametrised classically. That is, the mean $\mu(s)$ is computed classically and then we apply controlled rotations based on $s \in \mathcal{S}$ to prepare the wave function $\ket{\psi_{\mu(s),\Sigma(s)}}$ according to Eq.~\ref{eq: wavefun}.\footnote{Note that there is no need for rounding $\mu(s)$ to the computational basis since the $\mu(s)$ only appears in the amplitude. In fact, it may be advantageous to have the mean function in between different eigenactions if it is not yet clear which of the actions is the best.} For state $s \in \mathcal{S}$, the observable $\langle P_a \rangle_{s} = \langle \psi_{\Sigma(s),\mu(s)} \vert P_a \vert \psi_{\Sigma(s), \mu(s)} \rangle$ defines the policy directly according to
\begin{equation}
\label{eq: raw GQK policy}
\pi(a \vert s) = \langle P_a \rangle_{s} \,,
\end{equation}
where $P_a$ is the projection associated to action $a$ such that $\sum_a P_{a} = \mathbb{I}$, $P_{a}P_{a'} = \delta_{a,a'} P_{a}$. 

\begin{figure}
    \centering
    \includegraphics[width=0.99\linewidth]{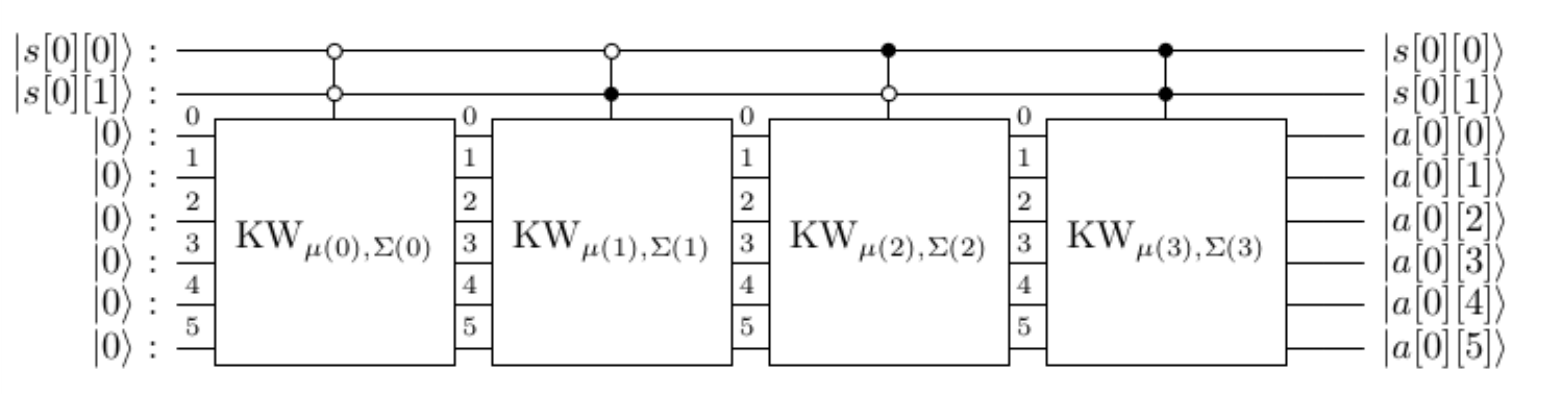}
    \caption{In the Gauss QKP, the policy evaluation oracle $\Pi$ is formed by calling multiple Gaussian wavefunction sub-circuits similar to that in Fig.~\ref{fig: KW}, each of which is controlled by a unique state. The figure illustrates this for a one-dimensional, two-qubit state space and a one-dimensional, six-qubit action space.}
    \label{fig: Pi}
\end{figure}

To define the oracle $\Pi$, which computes actions coherently, one needs to provide controls for all the states, which yield different $\mu(s)$ and potentially different $\Sigma(s)$, and therefore rotation angles. To this end, we formulate a circuit, shown in Fig.~\ref{fig: Pi}, with sub-circuits such as those in Fig.~\ref{fig: KW} each of which is controlled upon its respective state. 

\subsection{The number of representers}
In the Representer Raw-PQC and the Gauss-QKP, the mean function is given by $\mu(s)=\sum_{a \in \mathcal{A}} a \, \langle P_a \rangle_{s,\theta}$ is given by a representer formula with a number of $N$ representers, which is chosen based on a trade-off between expressiveness and efficiency. Extending the matching pursuit algorithm from \cite{Mallat1993} to vector-valued RKHS, the lemma below provides an efficient setting of $N$ which is logarithmic in the inverse of the error tolerance and allows to represent an arbitrary function in the Hilbert space of interest to a desired degree of accuracy.

\begin{lemma}
\label{lem: setting N}
\textbf{The number of representers for $\epsilon$-precise approximation.} Let $\epsilon > 0$, let $\kappa: \mathcal{X} \to \mathbb{R}$ be a scalar-valued kernel, and let $\mathcal{G}$ be an orthonormal basis in $\mathcal{H}_{\kappa}$. Moreover, let $\mu: \mathcal{X} \to \mathbb{C}$ be a squared-integrable function. Then applying scalar-valued kernel matching pursuit to form an estimate $\hat{\mu}$, the number of representers  
\begin{equation}
\label{eq: setting N}
N^* = \BigO\left(\log\left(\frac{\norm{\mu}{L_2}}{\epsilon}\right)\right)
\end{equation}
is sufficient to guarantee that
\begin{align*}
\norm{\hat{\mu} - \mu}{L_2} \leq \epsilon \,.
\end{align*}
For vector-valued kernel matching pursuit with kernel $K(x,y) = \kappa(x,y) \mathbb{I}_A$ for all $x,y \in \mathcal{X}$, and $\mu: \mathcal{X} \to \mathbb{C}^{A}$, the setting 
\begin{equation}
\label{eq: setting N vector}
N^* = \BigO\left(\log\left(\frac{\norm{\mu}{L_2(\mathcal{X}),p}}{\epsilon}\right)\right)
\end{equation}
is sufficient to guarantee that
\begin{align*}
\norm{\hat{\mu} - \mu}{L_2(\mathcal{X}),p} \leq \epsilon \,,
\end{align*}
where for function $f$ the notation $\norm{f}{L_2(\mathcal{X}),p}$ indicates the $L_2$ norm over the input space and the $p$-norm over the output dimensions of a function $f$.
\end{lemma}
The proof of this lemma is given in \refSI{7}.

\subsection{Learnability of Representer PQCs}
\label{sec: learnability RPQC}
To illustrate the learnability of Representer PQCs, we implement a Kronecker delta Representer PQC (see Eq.~\ref{eq: kronecker RPQC}) coherently as a policy evaluation oracle $\Pi$ in the state control example in Example~\ref{ex: quantum-accessible MDP} and perform gradient descent using classical estimates of the policy gradient. The experiments are conducted on a noiseless Qiskit simulator. 

\begin{figure}[htbp!]
    \centering
    \subfigure[Probability parametrisation]{
    \includegraphics[width=0.48\textwidth]{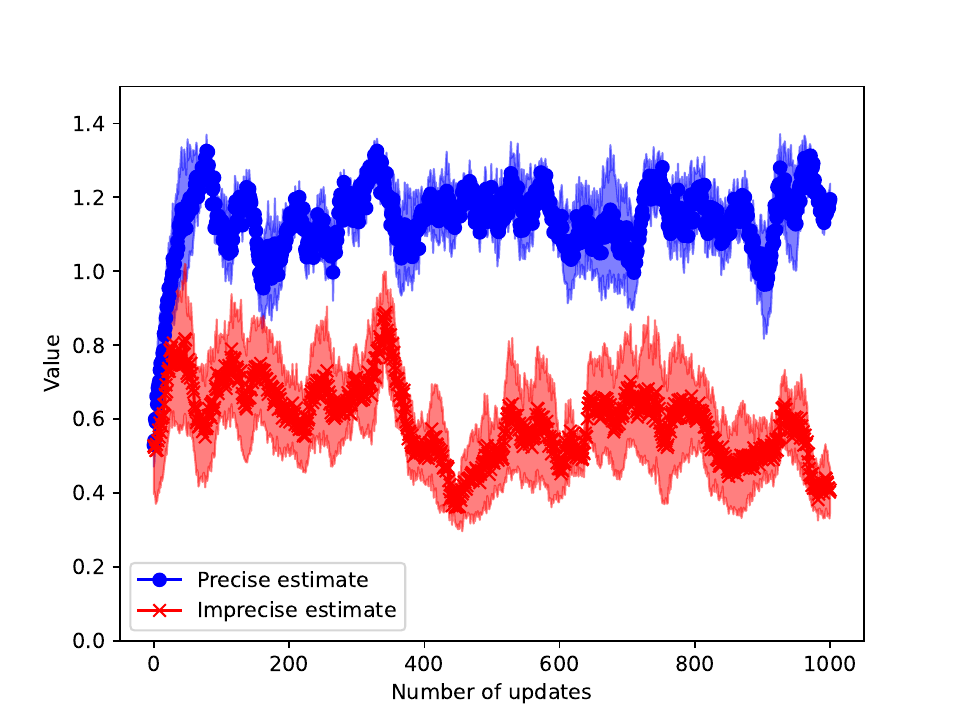}}
    \subfigure[Angle parametrisation]{
    \includegraphics[width=0.48\textwidth]{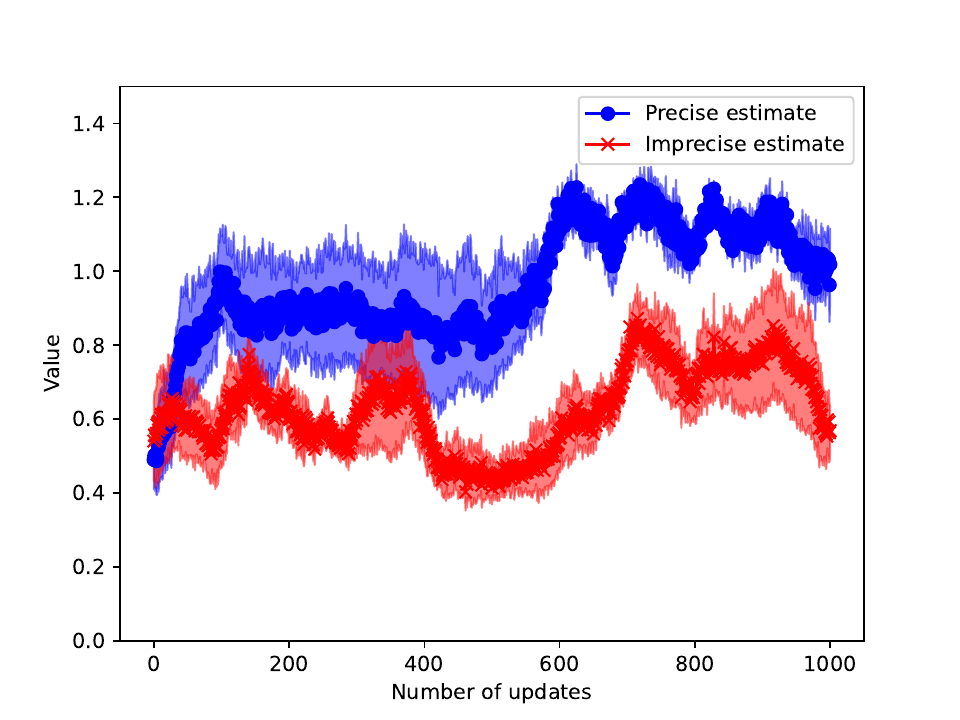}}
    \caption{Performance of classical central differencing based policy gradient algorithms on the state control example. The $y$-axis shows a value estimate based on 1,000 shots of the return oracle of the quantum-accessible MDP. The behavior policy $\pi$ is implemented based on the Kronecker delta policy evaluation oracle $\Pi_{\theta}^{\text{Kron}}$. The lines and shaded areas represent the means and standard errors across 5 independent runs of the algorithm.}
    \label{fig: kronecker delta gradient descent}
\end{figure}

To estimate the gradient of the quantum-accessible MDP, we use the central differencing scheme of \cite{Gilyen2019}, which is suitable for classical estimators, to estimate the gradient from repeated applications of the return oracle. The scheme reduces the approximation error with a rate $\BigO(e^{-m/2})$ where $m$ is the number of parameter perturbation pairs. While we use it here for classical gradient estimation, in the context of quantum gradient estimation it has been noted that this scheme allows for the query complexity scales logarithmically rather than linearly with the dimensionality $d$ as is the case when using an evenly spaced grid \citep{Jordan2005}. Moreover, we note that the scheme only differs from the scheme of Cornelissen (see Eq.~\ref{eq: CD scheme cornelissen}) by removing the estimation at zero perturbation (i.e. that coefficient is zero rather than one). Each perturbation in the central differencing scheme is applied for each dimension to obtain a high-precision estimate of each of the partial derivatives. After obtaining the gradient estimate, the policy parameters are updated based on the estimate of the policy gradient according to $\theta \gets \theta + \eta \nabla V(\theta)$ using the Adam optimiser \citep{Kingma2015} with learning rate $0.1$ and default running average parameters $0.9$ and $0.999$. An additional decay $\eta_t = \eta_{t-1} \alpha$ is applied where $\alpha = 0.998$. The experiments compare a high-precision setting, with $m=6$ and 1,000 shots, and a low-precision setting, with $m=1$ and 50 shots. Further, two distinct parametrisations are explored. In the (direct) angle parametrisation, the parameters that are being optimised range in $[-2\pi,2\pi]$ and are directly input to the controlled $R_Y$-gates of $\Pi$. In the probability-based parametrisation, the parameters that are being optimised range in $[0,1]$, and are first converted to the corresponding rotation angle $\theta' = 2 \sin^{-1}(\sqrt{\theta})$ before being input to the controlled $R_Y$-gates of $\Pi$. 

The results (see Fig.~\ref{fig: kronecker delta gradient descent}) indicate that it is indeed possible to learn high-performing policies with central differencing based policy gradient using the Kronecker delta policy. Under the high-precision regime, the probability parametrisation allows particularly rapid convergence to a high-performing policy while the angle parametrisation converges slower but also achieves a similarly high-performing policy. Under the low-precision regime, it is challenging to make any improvements over the initial policy. The results confirm the importance of obtaining high-precision estimates of the policy gradient, which is the subject of the following section.

\section{Numerical policy gradient}
\label{sec: numerical policy gradient}
For numerical gradient estimation, we use the central differencing algorithm by \cite{Cornelissen2019} as applied to the value function in \cite{Jerbi2022}, and evaluate the query complexity of Representer Raw-PQCs under this estimation scheme (Section~\ref{sec: numerical query complexity}).  The scheme can be used for optimising both the kernel parameters, as will be discussed further in Section~\ref{sec: parametric-kernel}, and the policy weights, as we have already demonstrated in the numerical experiments (Section~\ref{sec: learnability RPQC})

The central differencing technique is applied to the value function as a function of the parameters. We first illustrate the technique based on a one-dimensional parameter. To simplify the notation, we will use the shorthand $V(\theta) := V(d_0;\theta)$. The technique is based on formulating a Taylor expansion with the Lagrangian formulation of the remainder:
\begin{align*}
V(\theta + h) = V(\theta) + V'(\theta)h + \dots + \frac{V^{(k-1)}}{(k-1)!}(\theta) h^{k-1} +  V^{(k)}(\xi) h^k \,,
\end{align*}
for some $\xi \in [\theta,\theta+h]$ and $h > 0$. For $k=2$, such a formulation leads to first-order central differencing, where
\begin{align*}
V'(\theta) = \frac{V(\theta + h) - V(\theta -h)}{2h} +  \frac{V^{(k)}(\xi_1) - V^{(k)}(\xi_2)}{4} h \,,
\end{align*}
where $\xi_1 \in [\theta,\theta+h]$ and $\xi_2 \in [\theta-h,\theta]$.
For improving the estimates, the formulation extends the Taylor expansion to higher orders $k>2$ and applies function smoothing across several grid points, rather than just $\theta+h$ and $\theta-h$. A so-called central differencing scheme is defined according to
\begin{equation}
\label{eq: CD scheme cornelissen}
c_{l}^{(2m)} =
\begin{cases}
    1 & \text{for $l=1$} \\
    \frac{(-1)^{l+1} (m!)^2}{l(m+l)!(m-l)!} & \text{otherwise} 
\end{cases}
\end{equation}
for all $l=-m,-m+1,\dots,m-1,m$ where $m = \lfloor \frac{k-1}{2} \rfloor$. This leads to the following expression for the derivative (cf. Eq.74 in \cite{Jerbi2022}) 
\begin{align}
\label{eq: central differencing}
V'(\theta) = \underbrace{\sum_{l=-m}^{m} \frac{c_{l}^{(2m)} V(\theta + lh)}{h}}_{V_{(2m)}(\theta,h)}  +  \underbrace{\sum_{l=-m}^{m} c_{l}^{(2m)} \frac{V^{(k)}(\xi_l)}{k!} l^{k} h^{k-1}}_{R_V^k} \,,
\end{align}
where $h > 0$, $\xi_l \in [\theta,\theta+lh]$ for all $l$, $V_{(2m)}(\theta,h)$ indicates the estimate based on a smoothed function value, and $R_V^k$ is the Lagrangian remainder. The smoothed function value makes the different point estimates linear over the central differencing scheme such that its average is close to the true derivative.

\subsection{Quantum gradient estimation of Gevrey functions}
\label{sec: quantum gevrey}
The technique by \cite{Cornelissen2019} as applied to gradient of the value function can be summarised as follows:
\begin{enumerate}
\item Define $R$, the edge length of the grid, depending on Gevrey parameters $c$, $d$, and $\sigma$.
\item Repeat for $j=1,\dots,N_x = \BigO(\log(d))$:
\begin{enumerate}
\item Formulate a $d$-dimensional grid $G \subset [-R/2,R/2]^d$ within a hypercube with edge length $R$ centred around zero with $2^k$ evenly spaced grid points per dimension and form a uniform superposition,
\begin{equation}
\vert \psi_1 \rangle = \frac{1}{\sqrt{2^{kd}}} \sum_{\theta' \in G} \vert \theta' \rangle \,, 
\end{equation}
where $k$ is the number of qubits per dimension.
\item Apply a phase oracle for $V_{(2m)}$ over $G$, repeating $N_V = \frac{8 \pi d^{1/p}}{R\epsilon}$ times, such that 
\begin{align*}
O_{V_{(2m)},G}: \frac{1}{\sqrt{2^{kd}}} \sum_{\theta' \in G} \vert \theta' \rangle \to \frac{1}{\sqrt{2^{kd}}} \sum_{\theta' \in G} e^{i N_V V_{(2m)}(\theta,\theta')} \ket{\theta'} \,.
\end{align*}
Note that this oracle can be constructed from the phase oracle in Definition~\ref{def: phase oracle} as it follows from the definition of the smoothed function value that
\begin{align*}
e^{i n V_{(2m)}(\theta+\theta')} = \prod_{l=-m}^{m} e^{i N_V c_l^{(2m)} V(\theta+l\theta')} \,.
\end{align*}
Due to the linear approximation $V_{(2m)}(\theta,\theta') \approx V(\theta) + \nabla_{\theta} V(\theta) \theta'$, the state after step (b) is 
\begin{align*}
\ket{\psi_2} &\approx \frac{1}{\sqrt{2^{kd}}} \sum_{\theta' \in G} e^{i N_V (V(\theta) + \nabla_{\theta} V(\theta) \theta')} \ket{\theta'} \\
             &= \frac{1}{\sqrt{2^{kd}}} \sum_{\theta' \in G} e^{i N_V \nabla_{\theta} V(\theta) \theta'} \ket{\theta'} \,,
\end{align*}
where the term $V(\theta)$ is dropped since QFT is invariant to global phase factors.
\item Applying an inverse QFT to the state $\ket{\psi_2}$ separately for each dimension yields the slope of the phase as a function of the parameter:
\begin{align*}
\ket{\psi_3} \approx \ket{\text{round}\left(\frac{N_V R}{2\pi} \nabla_{\theta} V(\theta)\right)} \,.
\end{align*}
\item Measure and renormalise by factor $\frac{2\pi}{N_V R}$ to obtain $X_j \approx \nabla_{\theta} V(\theta)$.
\end{enumerate}
\item Define $\bar{X} = \text{mean}(X_1,\dots,X_{N_x})$ 
\end{enumerate}
The resulting query complexity of the algorithm is given by  $n = \tildeO(m N_X N_V)$, which depends on the construction of the smoothed value phase oracle $O_{V_{(2m)},G}$ which depends on the grid points according to $\BigO(m \log(\frac{m}{\delta}))$, the number of repetitions ($N_X$), and the number of applications of $O_{V_{(2m)},G}$ per repetition ($N_V$). Following the setting of $m$ and $N_X$, which are logarithmic in the Gevrey parameters and the precision, and a further derivation for $N_V$, this amounts to $n =  \tildeO(\frac{cd^{\sigma+1/p}}{\epsilon})$. The depth scales with $N_V = \BigO(\frac{d^{1/p}}{R\epsilon})$ since the phase oracle is applied repeatedly in sequence. In terms of space complexity, a first register of exactly $kd$ qubits is required for the grid point and the phase oracle for the smoothed value function $O_{V_{(2m)},G}$ requires $\Theta(kd)$ auxiliary qubits, amounting to $\Theta(kd)$ total qubits. In terms of gate complexity, applying $d$ parallel applications of the $k$-qubit QFT requires $\tildeO(kd)$ gates while the implementation of  $O_{V_{(2m)},G}$ is not known in general. However, in our specific numerical policy gradient algorithm, the trajectory oracle scales with $\BigO((S+A)kT)$ in terms of gate and space complexity. The return oracle implementation is not known exactly but should scale with $\BigO(T)$, and it is subsequently converted to a probability oracle via controlled $R_Y$-gates, followed by a phase oracle using a set of controlled phase gates.

\subsection{Quadratic improvements for numerical policy gradient}
\label{sec: numerical query complexity}
Since quantum gradient estimation of Gevrey functions scales in query complexity with the higher-order gradient of the policy, we first derive an upper bound on the higher-order gradient of the policy for the Representer Raw-PQC of Sec.~\ref{sec: representer PQC}.
\begin{lemma}
\label{lem: D}
\textbf{Bound on the higher-order gradient of the policy.} Let $\pi$ be a Representer Raw-PQC as in Def.~\ref{def: RRPQC policy}  implemented according to Fig.~\ref{fig: representer-pqc}b. Then 
\begin{align*}
D = \max_{p} D_p \,,
\end{align*}
where
\begin{align*}
D_p = \max_{s \in \mathcal{S}, \alpha \in [d]^p} \sum_{a \in \mathcal{A}} \vert \partial_{\alpha} \pi(a \vert s) \vert \,,
\end{align*}
is bounded by $D \leq 1$.
\end{lemma}
\begin{proof}
Noting it takes the form of a Raw-PQC, and the fact that the $R_Y$ gates have $\pm 1$ eigenvalues, the remainder of the proof is analogous to that of  \cite{Jerbi2022}. The full proof is given in \refSI{8}. \end{proof}

We apply the quantum Gevrey estimation as summarised in \ref{sec: quantum gevrey}. Using the upper bound $D$, we confirm the quadratic improvements for numerical policy gradient also hold in the context of Representer Raw-PQCs. 
\begin{theorem}
\textbf{Quadratic improvement for Representer Raw-PQCs under numerical policy gradient.} Let $\pi$ be the policy formed from a Representer Raw-PQC, let $\delta > 0$ be the upper bound on the failure probability, and let $\epsilon > 0$ be the tolerable $\ell_{\infty}$ error on the policy gradient. Then with probability at least $1-\delta$,  computing the policy gradient $\nabla_{\theta} V(\theta)$ numerically with quantum gradient estimation (Section~\ref{sec: quantum gevrey}) requires
\begin{equation}
\label{eq: numerical query complexity}
n = \tildeO\left(\sqrt{d} \left(\frac{r_{\text{max}}}{\epsilon(1-\gamma)} T^2\right)\right) 
\end{equation}
$\BigO(T)$ steps of interactions are required. This yields a quadratic improvement over classical estimators under general classical policy formulations (including but not limited to Gaussian and softmax policies).
\end{theorem}
\begin{proof}
The classical algorithm applies multivariate Monte Carlo to the above central differencing algorithm, independently for each parameter dimension. The resulting query complexity can be bounded using Theorem~3.4 in \cite{Cornelissen2019} and derivations in Appendix F and G of \cite{Jerbi2022} (see \refSI{9} for a summary); that is,
\begin{align*}
n = \tildeO\left(d \left(\frac{r_{\text{max}}}{\epsilon(1-\gamma)} DT^2\right)^2\right) \,.
\end{align*}

For the quantum algorithm, we use quantum Gevrey estimation as summarised in Section~\ref{sec: quantum gevrey}. In particular, we follow its application according to Theorem~3.1 of \cite{Jerbi2022}, where the phase oracle $O_V$ is constructed from a probability oracle $O_{PV}$ as defined in Definition~\ref{def: phase oracle}. To obtain the probability oracle, one rotates the last qubit proportional to the return, obtaining the state
\begin{align*}
\ket{\theta} \sum \sqrt{P(\tau)} \ket{\tau} \ket{R(\tau)} \left( \sqrt{\tilde{R}(\tau)} \ket{0} + \sqrt{1-\tilde{R}(\tau)}  \right) \,,
\end{align*}
which reduces to 
\begin{align*}
\ket{\theta} \sum \sqrt{\tilde{V}(d_0)} \ket{\varphi_0} \ket{0} + \sqrt{1 - \tilde{V}(d_0)} \ket{\varphi_1}\ket{0} \,,
\end{align*}
where $\tilde{R}(\tau) = \frac{(1-\gamma)R(\tau)}{r_{\text{max}}}$ and $\tilde{V}(d_0) = \frac{(1-\gamma)V(d_0)}{r_{\text{max}}}$. 
Due to the Gevrey value function parameters $c=DT^2$, $M=4\frac{r_{\text{max}}}{1-\gamma}$, and $\sigma=0$ (see Lemma~\ref{lem: Gevrey value}), we obtain
\begin{align*}
n &= \tildeO\left(\frac{M cd^{\max\{\sigma,1/2\}}}{\epsilon} \right) \\
n &= \tildeO\left(\sqrt{d} \left(\frac{r_{\text{max}}}{\epsilon(1-\gamma)} DT^2\right)\right) \,.
\end{align*}
For Representer Raw-PQCs, note that $D\leq 1$ following Lemma~\ref{lem: D}. Therefore, the factor $D$ vanishes in the query complexity, yielding Eq.~\ref{eq: numerical query complexity}. Since the classical policy was arbitrarily chosen, this represents a quadratic query complexity speedup as claimed.
\end{proof}

While the optimisation scheme comes with comparable quadratic improvement, a reduction in the number of parameters is further possible if the optimal deterministic policy $\mu^*$ is a Lipschitz continuous function.

\section{Analytical policy gradient}
\label{sec: analytical policy gradient}
As a second class of techniques, we use analytical policy gradient techniques based on quantum multivariate Monte Carlo \citep{Cornelissen2022}. The main policies analysed in this section are the Representer-Softmax-PQCs and Gauss-QKPs.

We first summarise how to use quantum multivariate Monte Carlo algorithm for computing the policy gradient before moving on to specific analytic quantum policy gradient algorithms.

As a warm-up example, we consider REINFORCE \citep{Jerbi2022,Peters2008a},
\begin{equation}
\nabla_{\theta} V(d_0) = \mathbb{E}\left[\sum_{t=0}^{T-1} \nabla_{\theta} \log(\pi(a_t \vert s_t)) \sum_{t'=0}^{T-1} \gamma^{t'} r_{t'} \right] \,,
\end{equation}
where we show two classes of QKPs that yield quadratic improvements over any classical policy, thereby extending Lemma~\ref{lem: analytical policy gradient}.

Following this example, we will prove the query complexity of two quantum actor-critic algorithms, which have oracles closely related to occupancy measures and which have different policy gradient updates.

\subsection{Quantum multivariate Monte Carlo}
The quantum multivariate Monte Carlo technique \citep{Cornelissen2022} generalises univariate techniques \citep{Montanaro2017} to multiple dimensions and the multivariate technique by \cite{VanApeldoorn2021} to compute the expected value over vectors depending on a random variable rather than over mutually exclusive unit vectors. The technique requires a binary oracle for $X$, which we denote $U_{X,\Omega}$ (see Definition~\ref{def: binary oracle}). The technique allows to estimate the expectation $\mathbb{E}[X]$ based on sampled trajectories, yielding an $\epsilon$-precise policy gradient. The basic algorithm, called QBounded (Theorem~3.3 in \cite{Cornelissen2022}), works under the condition of a bounded $\ell_2$ norm of $\mathbb{E}[\norm{X}{2}] \leq B$. It is the same algorithm as was used in Theorem`4.1 in \cite{Jerbi2022} and can be summarised for our purposes in the following steps:
\begin{enumerate}
\item Define a grid $G = \{ \frac{j}{m} - \frac{1}{2} + \frac{1}{2m}: j \in \{0,\dots,m-1\} \subset (-1/2,1/2)^{d}$, where $m=2^{\lceil \log(\frac{8\pi n}{\zeta \sqrt{B} \log(d/\delta)} ) \rceil}$ is the number of grid points per dimension and $d$ is the dimension of $X$. The grid represents vectors $x \in G$ to be used within the directional mean $\langle x, \mathbb{E}[X] \rangle$, where for example $\mathbb{E}[X] = \mathbb{E}\left[\sum_{t=0}^{T-1} \nabla_{\beta} \log(\pi(a_t \vert s_t)) \sum_{t'=0}^{T-1} \gamma^{t'} r_{t'} \right]$ for traditional REINFORCE.
\item For $j=1,\dots, N_x=\BigO(\log(d/\delta))$:
\begin{enumerate}
\item Compute a uniform superposition over the grid:
\begin{equation}
\vert \psi_1 \rangle = \frac{1}{m^{d/2}} \sum_{x \in G} \ket{x} \,.
\end{equation}
\item Compute the truncated directional mean oracle: within $\tildeO\left( m \sqrt{B} \log^2(1/\epsilon) \right)$ queries to $U_{X,\Omega}$, a state $\ket{\psi_2}$ is formed such that 
\begin{align*}
\norm{\ket{\psi_2(x)} - e^{im\mathbb{E}[\trunc{\zeta \langle x, X \rangle}{0}{1}]} \ket{0}}{2} \leq \epsilon 
\end{align*}
for some desirable $\epsilon > 0$ for a fraction at least  $1 - \zeta/2$ of grid points $x \in G$,  where $\zeta = \frac{1}{\sqrt{\log(400\pi n \sqrt{d})}}$. The technique is based on first computing a probability oracle for $\trunc{\zeta \langle x, X \rangle}{0}{1}$ by using controlled rotations, which encodes the directional mean for most values of $x \in G$, resulting in an amplitude that is close to the directional mean. Converting to a phase oracle then yields the desired state.
\item Apply inverse quantum Fourier transform $(\text{QFT}^{\dagger}_{G} \otimes \mathbb{I}) \ket{\psi_2}$, where
\begin{align*}
\text{QFT}_G: \ket{x} \to \frac{1}{m^{d/2}} \sum_{y \in G} e^{2i\pi m \langle x, y \rangle } \ket{y} 
\end{align*}
resulting in the state $\ket{y_j}$.
\item Measure $y_j$ and renormalise as $X_j=\frac{2\pi y_j}{\zeta}$.
\end{enumerate}
\item Obtain the estimate $\bar{X} = \text{median}(X_1,\dots,X_{N_x})$.
\end{enumerate}
With $N_x = \BigO(\log(d/\delta))$ preparations of the directional mean oracle (based on $\BigO\left( m \sqrt{B} \log^2(1/\epsilon) \right)$ queries to the binary oracle),  QBounded was shown to have a combined query complexity of $n = \tildeO(\frac{\sqrt{B}}{\epsilon})$. We expect that the depth scales similarly $\BigO(1/\epsilon)$ due to the sequential application of phase oracles. In terms of space complexity, one requires $d \log(m)$ qubits for the grid points and $\tildeO(d \log(m))$ for the QFT (assuming parallelisation is possible). The construction of the directional mean oracle relies on the binary oracle. In quantum-accessible MDPs, this oracle prepares the analytical expression of the gradient; we implement a specific instance of such an oracle within $\BigO(k(S + A + d)T)$ space complexity (see Lemma~\ref{lem: U_X}). Its gate complexity varies since it depends on the implementation of the policy evaluation oracle, the transition oracle, and how the policy gradient expression is obtained from the trajectory. For instance, a naive implementation uses $\BigO(T (2^{(S+A)k}))$ controlled X-gates for the gradient register and $\BigO(d k 2^{(S+A)k})$ controlled $R_Y$ gates for the gradient expression $X(s,a) \in \mathbb{R}^{d}$. The conversion to a probability oracle and then to a phase oracle again introduce additional controlled $R_Y$ and phase gates, respectively. 

The QEstimator algorithm (Theorem~3.4 in \cite{Cornelissen2022}) expands on QBounded based on a loop with additional classical and quantum estimators, each with logarithmic query complexity, thereby allowing improved query complexity as well as applicability to unbounded variables. The technique roughly goes as follows: 
\begin{enumerate}
    \item Run a classical sub-Gaussian estimator on $X$ (e.g. the polynomial-time estimator of Hopkins based on semi-definite programming with the sum of squares method \citep{Hopkins2020}) on $\log(1/\delta)$  $T$-step trajectories (e.g. from measurements of $U_{X,\Omega}$) to obtain an estimate $X'$ such that $\mathbb{P}(\norm{X' - \mathbb{E}[X]}{2} > \sqrt{\Tr(\Sigma_X)}) \leq \delta/2$ for failure probability $\delta > 0$, where $\Sigma_X$ is the covariance of $X$.
    \item For $j = 1,\dots, N_y=\BigO(\log(n/\log(d/\delta)))$:
    \begin{enumerate}
    \item Apply a univariate quantum quantile estimator \citep{Hamoudi2021}, which is based on sequential amplitude amplification, to estimate $q_j$, the $2^{-j}$'th order quantile of $\norm{X-X'}{2}$ based on $\BigO(\log(k/\delta)/\sqrt{2^{-j}})$ calls to $U_{X,\Omega}$. 
    \item Define the truncated random variable $Y_j = \frac{1}{q_j}\trunc{\norm{X-X'}{2}}{q_{j-1}}{q_j}$ and apply QBounded, obtaining the estimate $\bar{Y}_j$.
    \end{enumerate}
    \item Obtain the estimate $\bar{X} = X' + \sum_{j=1}^{N_y} q_j \bar{Y}_j$.
\end{enumerate}
Each estimate $\bar{Y}_j$, $j=1,\dots,N_y$, will have its error bounded by $2^{-(j-1)/2} \BigO(\log(kd/\delta))/n$ as guaranteed by QBounded, resulting in an overall error of at most $\frac{\sqrt{\norm{\mathbb{E}[X - X']}{2}^2}}{\sqrt{2}} \log(d/\delta)$. Due to the relation of $\norm{\mathbb{E}[X - X']}{2}^2$ with the covariance $\Sigma_X$, this amounts to a query complexity of $n = \tildeO(\sqrt{\Tr(\Sigma_X)}/\epsilon)$, where the logarithmic factors from the classical sub-Gaussian estimator and the quantum quantile estimator do not appear in the notation. The depth of the amplitude amplification circuit scales according to $T = \BigO(1/2^{-j})$ for each $j$ as the range of the random variable gets more and more restricted. A further requirement introduced by the quantum quantile estimator is a comparison oracle to distinguish the good state (i.e. when the variable is in the desired range). 

In our query complexity results, we will use QBounded for traditional REINFORCE (Section~\ref{sec: REINFORCE}) and Deterministic Compatible Quantum RKHS Actor-Critic \ref{sec: deterministic AC}, while using QEstimator for (stochastic) Compatible Quantum RKHS Actor-Critic \ref{sec: stochastic AC}. Using QEstimator allows query complexity bounds based on the variance, a quantity that we will reduce by considering a form of the policy gradient that subtracts a baseline from the critic prediction, similar to the objective in REINFORCE with baseline and the popular Advantage Actor-Critic (A2C) \citep{Mnih2016a}.

\subsection{Quadratic improvements for REINFORCE}
\label{sec: REINFORCE}
As a warm-up example, we first seek to establish that the quadratic improvements over classical Monte Carlo hold. Since the analytical expression of the policy gradient includes $\nabla_{\theta} \log(\pi(a \vert s))$, the $\ell_1$-norm of the gradient of the log-policy, denoted as $B_1$, appears in both the classical query complexity due to Hoeffding inequality in classical multivariate Monte Carlo (see \refSI{1}) and the quantum query complexity due to the bound from quantum multivariate Monte Carlo in Lemma~\ref{lem: analytical policy gradient}. Therefore, we first establish that under some conditions $B_1$ is bounded by a constant, which will enable a quadratic improvement in query complexity over any classical policy (not just kernel-based). The purpose of Lemma~\ref{lem: B} is to show that kernel policies in quantum oracles are bounded by $B_1 = \tildeO(1)$ which ensures a quadratic query complexity improvement over any classical policy with arbitrary $B_1$.
\begin{lemma}
\label{lem: B}
\textbf{$\ell_1$ bounds on the gradient of the log-policy.} Let $\kappa$ be a scalar-valued kernel such that $\vert \kappa(s,s') \vert \leq \kappa_{\text{max}}$ for all $s,s' \in \mathcal{S}$. The following statements hold for the $\ell_1$ upper bound on the gradient of the log-policy, $B_1 \geq \max_{s \in \mathcal{S},a \in \mathcal{A}} \norm{\nabla_{\theta} \log(\pi(a \vert s))}{1}$. \\
\textbf{a)} Then for any Gauss-QKP with $\theta=\beta$, $A$ action dimensions and $N$ representers, with probability $1-\delta$
\begin{align*}
B_1 \leq ANZ_{1-\frac{\delta}{2A}}\kappa_{\text{max}}
\end{align*}
where $Z_{1-\frac{\delta}{2A}}$ is the $1-\frac{\delta}{2A}$ quantile of the standard-normal Gaussian.\\
\textbf{b)} For any finite-precision Gauss-QKP with $\theta=\beta$, mean function $\mu: \mathcal{S} \to \mathcal{A}$, and number of representers  $N=\BigO(\frac{\sqrt{\Sigma_{\text{min}}}}{A\kappa_{\text{max}}})$, it follows that $B_1 = \BigO(1)$. \\
\textbf{c)} Any Representer Softmax-1-PQC satisfies $B_1=\BigO(1)$. \\
\textbf{d)} Any Representer Softmax-1-PQC B\&S satisfies $B_1=\tildeO(1)$ provided $N=\BigO(\kappa_{\text{max}}^{-1} N^*)$ with $N^*$ from Eq.~\ref{eq: setting N vector}.
\end{lemma}
\begin{proof}
\textbf{a)} For any Gauss-QKP, and noting the form of Eq.~\ref{eq:log-policy-grad} and applying union bound over the $1-\frac{\delta}{2A}$ quantile yields the desired result (see \refSI{10.1}).\\
\textbf{b)} The finite-precision Gauss-QKP will have bounded support and the variance is a fraction of this interval. Applying the settings to the vectorised form Eq.~\ref{eq:log-policy-grad} with parametrisation in the policy weights ($\beta$), and setting $N = \BigO(\frac{\sqrt{\Sigma_{\text{min}}}}{A \kappa_{\text{max}} })$ yields $B_1 = \BigO(1)$ (see \refSI{10.2}). \\
\textbf{c)} The Representer Softmax-1-PQC is an instance of Softmax-1-PQC, which yields $B_1 = \BigO(1)$ following Lemma~4.1 in \cite{Jerbi2022}. \\
\textbf{d)} Following the analytical form in \refSI{6}, the gradient of the log-policy of Representer Softmax-1-PQC B\&S is bounded by 
\begin{align*}
\norm{\mathcal{T} \left( \kappa((s,a),\colon) - \mathbb{E}_{a' \sim \pi(\cdot \vert s)} \kappa((s,a'),\colon) \right)}{1}  
&\leq 2\max_{s,a} \norm{\mathcal{T} \kappa((s,a),\colon)}{1} \\
& = \BigO(N \kappa_{\text{max}}) = \tildeO(1)  \,,
\end{align*}
where the last equality follows from setting $N = \BigO(\kappa_{\text{max}}^{-1} N^* )$ and $N^* = \tildeO(1)$.
\end{proof}

We note that the settings of $N$ in Lemma~\ref{lem: B} b) and d) are not restrictive in practice. For setting d), note that if $\kappa_{\text{max}} \leq 1$, as is the case for any quantum kernel and any orthonormal basis, then setting $N = \kappa_{\text{max}}^{-1} N^* \geq N^*$ ensures the kernel matching pursuit algorithm in Lemma~\ref{lem: setting N} can obtain an $\epsilon$-precise approximation $\hat{\mu}$ to the desired function $\mu$. For setting b), we will normalise the state space to a unit hypercube, in which case the $L_2$ distance is consistent with the root mean squared error. If $A$ is a small constant, as in many RL applications, and if $\kappa_{\text{max}} \leq 1$, $\sqrt{\Sigma_{i,i}} = \Theta(a_{\text{max}})$ for all $i \in A$, where $[-a_{\text{max}},a_{\text{max}}]_{i=1}^{A}$ is the action space with $a_{\text{max}} \geq 1$, it follows that setting $N = \kappa^{-1} N^*$ is sufficient since
\begin{align*}
N^* &= \BigO\left(\log\left(\frac{\norm{\mu}{L_2(\mathcal{S}),1}}{\epsilon} \right)\right)  = \BigO\left(\log\left(a_{\text{max}}\right)\right) = \BigO\left( \sqrt{\Sigma_{\text{min}}} \right)\,,
\end{align*}
which implies that $N = \kappa_{\text{max}}^{-1} N^* = \BigO\left( \frac{\sqrt{\Sigma_{\text{min}}}}{A \kappa_{\text{max}}} \right)$. Due to setting $N = N^*$, we again obtain a guarantee for an $\epsilon$-precise approximation $\hat{\mu}$ of the desired function $\mu$. To provide a guarantee over the root mean squared error in a discretised state space, one can consider the case where $2^{Sk}$ unique states are spread evenly across volume and each state represents a hypercube of volume of $2^{-Sk}$, such that $\sqrt{\int \vert \mu(s) - \hat{\mu(s)} \vert^2 ds} = \sqrt{\frac{1}{2^{Sk}} \sum_{i=1}^{2^{Sk}} \vert \mu(s_i) - \hat{\mu}(s_i) \vert^2}$.\footnote{Note that the exact equality follows regardless of the precision $k$ since a function $\mu(s)$ can be constructed to make the same prediction everywhere in the hypercube associated with $s$.} In conclusion, with settings as in Lemma~\ref{lem: B} b) and d), expressive policies can be represented with only limited resulting gradient norm $B_1 = \tildeO(1)$. 

Having defined the bounds on the gradient of the log-policy allows for a query complexity analysis of the QKPs. Below we analyse the above QKPs in the context of REINFORCE with quantum policy gradient.
\begin{theorem}
\label{th: REINFORCE quadratic}
\textbf{Quadratic improvements in REINFORCE.} Let $\delta \in (0,1)$ be the upper bound on the failure probability, and let $\epsilon > 0$ be an upper bound on the $\ell_{\infty}$  error of the policy gradient estimate. Moreover, let $\pi$ be a policy satisfying the preconditions of Lemma~\ref{lem: B}\textbf{b)} or \textbf{c)}. Then with probability at least $1-\delta$, applying QBounded (algorithm in Theorem 3.3 of \cite{Cornelissen2022} for quantum multivariate Monte Carlo) on a binary oracle for the policy gradient returns an $\epsilon$-correct estimate $\bar{X}$ of $ \mathbb{E}[X] = \nabla_{\theta} V(d_0)$ such that $\norm{\bar{X} - \mathbb{E}[X]}{\infty} \leq \epsilon$ within 
\begin{equation}
\label{eq: REINFORCE quadratic}
n=\tildeO\left(\dfrac{T r_{\text{max}}}{\epsilon(1-\gamma)} \right) \,,
\end{equation}
 $\BigO(T)$-step interactions with the environment. This represents a quadratic improvement compared to any policy evaluated with classical multivariate Monte Carlo yields query complexity
\begin{equation}
\label{eq: REINFORCE classic}
n = \tildeO\left(\left(\dfrac{B_1 T r_{\text{max}}}{\epsilon(1-\gamma)}\right)^2 \right)
\end{equation} 
 where $B_1 \geq \norm{\nabla_{\theta} \log(\pi(a \vert s))}{1}$.
\end{theorem}
\begin{proof}
We first construct the binary oracle used by \cite{Jerbi2022} which applies $U_P$ followed by $U_R$ and finally a simulation of the classical product of $\sum_{t'=0}^{T-1} \gamma^{t'} r_{t'}$ and $\sum_{t=0}^{T-1} \nabla_{\theta} \log(\pi(a_t \vert s_t))$. Defining the oracle in this manner yields $\BigO(T)$-step interactions with the environment, as it applies $\BigO(T)$ calls to policy evaluation ($\Pi$), transition ($O_P$), and reward ($O_R$) oracles. Note that $\sum_{t'=0}^{T-1} \gamma^{t}r_{t'} = \tildeO(\frac{r_{\text{max}}}{1-\gamma})$ due to the effective horizon of the MDP. Moreover, $\norm{\sum_{t=0}^{T-1} \nabla_{\theta} \log(\pi(a_t \vert s_t))}{1}$ is upper bounded by $\tildeO(T)$ since applying Lemma~\ref{lem: B}b, c, or d yields $\nabla_{\theta} \log(\pi(a_t \vert s_t)) = \tildeO(1)$. 

Now denote $\tilde{X} = \frac{(1-\gamma) X}{ T r_{\text{max}}}$. Since an $\ell_2$ bound $B_2 \leq B_1$ and $B_1 = \BigO(1)$, it follows that $\norm{\tilde{X}}{2} \leq 1$, and $\norm{\mathbb{E}[\tilde{X}]}{2} \leq 1$ as required by the QBounded algorithm. Applying QBounded (Theorem~3.3 of \cite{Cornelissen2022}) to $\tilde{X}$, we obtain an $\frac{(1-\gamma) \epsilon}{ T r_{\text{max}}}$-precise estimate of $\mathbb{E}[\tilde{X}]$ with probability $1-\delta$ within
\begin{align*}
n &= \tildeO\left(\dfrac{T r_{\text{max}}}{(1-\gamma)\epsilon} \right) 
\end{align*}
$\BigO(T)$-step interactions with the environment. Therefore, after renormalisation, an $\epsilon$-correct estimate of $\mathbb{E}[X]$ is obtained within the same number of queries.

By contrast, for classical multivariate Monte Carlo (see \refSI{1}) we note that $B_{\infty} \leq B_1$ and therefore bounding $X \in [-B,B]$ where $B = \frac{T B_1 r_{\text{max}}}{1-\gamma}$, we require
\begin{align*}
n &= \BigO\left(\left(\dfrac{B_1 T r_{\text{max}}\log(d/\delta)}{\epsilon(1-\gamma)}\right)^2 \right) \\
\end{align*}
$\BigO(T)$-step interactions with the environment.
\end{proof}

\subsection{Compatible Quantum RKHS Actor-Critic}
\label{sec: stochastic AC}
An alternative to REINFORCE is the Compatible RKHS Actor-Critic algorithm as proposed by \cite{Lever2015}, which reduces the variance of gradient estimates for improved sample efficiency. We briefly review the classical algorithm to help construct a suitable quantum policy gradient algorithm in Section~\ref{sec: compatible QRKHS-AC}, which we call Compatible Quantum RKHS Actor-Critic (CQRAC). 

\begin{figure}[htbp!]
    \centering
    \includegraphics[width=1.1\textwidth]{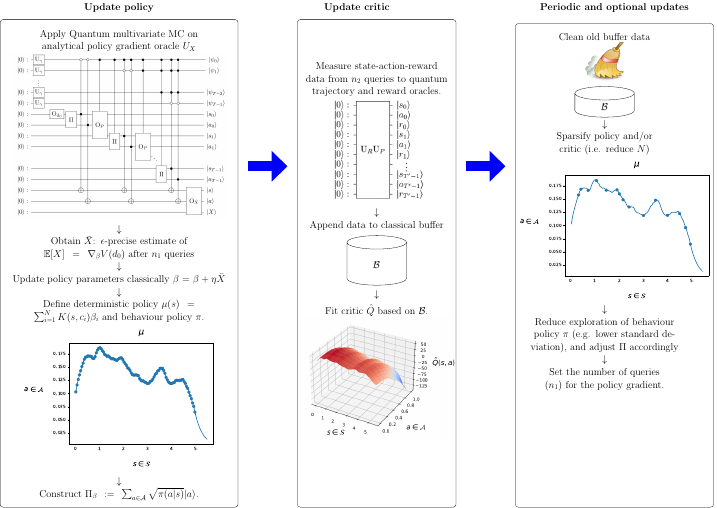}
    \caption{Overview of the algorithmic framework for Compatible Quantum RKHS Actor-Critic algorithms (see Algorithm~\ref{alg: quantum AC} and \ref{alg: quantum DAC}). The Quantum RKHS Natural Actor-Critic algorithm (see Algorithm~\ref{alg: quantum NAC}) derives the critic update from a binary oracle (similar to $U_X$) from which the natural policy gradient is then obtained without additional samples.}
    \label{fig: diagram}
\end{figure}

As illustrated in Fig.~\ref{fig: diagram}, our framework for implementing actor-critic algorithms is based on a quantum policy gradient and a classical critic. The algorithm repeats updates to the policy and the critic as follows. It updates the policy by making use of an occupancy oracle, which samples an analytic expression of the policy gradient according to its probability under $\Pi$ and $O_P$, based on the quantity $X(s,a) = \hat{Q}(s,a) \nabla_{\beta} \log(\pi(a \vert s))$, where $\hat{Q}(s,a)$ is the prediction from the critic and $\beta$ is the set of policy weights. The resulting policy gradient is estimated using quantum multivariate Monte Carlo. The critic is updated classically based on separate calls to the traditional trajectory and return oracles ($U_P$ and $U_R$) while setting the number of such classical samples such that there is no increase in query complexity. Additional periodic and optional steps include cleaning the trajectory data stored for replay, sparsifying the policy, and reducing the scale of the covariance. The full flow of the algorithm can be found in Algorithm~\ref{alg: quantum AC}, which uses the GaussQKP as a concrete implementation, such that $\nabla_{\beta} \log(\pi(a \vert s)) =  K(s, \cdot)\Sigma^{-1} (a - \mu(s))$.

\subsubsection{The classical algorithm}
For classical Gaussian kernel policies, the classical algorithm defines the policy gradient as
\begin{equation}
\label{eq: integral}
\nabla_{\mu} V(d_0) = \int \nu(z) Q(z) K(s, \cdot)\Sigma^{-1} (a - \mu(s))  dz
\end{equation}
where $z \in \mathcal{S}\times{A}$ and $\nu(z)$ is the occupancy measure (see Eq.~\ref{eq: occupancy measure}).  The integral in Eq.~\ref{eq: integral} can be approximated by sampling from the distribution formed from $(1- \gamma) \nu$ and computing the quantity based on \textit{iid} state-action pair samples:
\begin{align}
\label{eq: approximate-integral}
\nabla_{\mu} V(d_0) \approx \frac{1}{1-\gamma} \frac{1}{n} \sum_{i=0}^{n} \hat{Q}(z_i) K(s_i, \cdot)\Sigma^{-1} (a_i - \mu(s_i)) \,,
\end{align}
where $n$ is the total number of samples. Since direct knowledge of the occupancy measure is typically unrealistic, samples can be generated using a subroutine (see Algorithm~\ref{alg: occupancy}).

Lemma~\ref{lem: unbiased} states that sampling from Algorithm~\ref{alg: occupancy} provides unbiased estimates of the occupancy distribution. That is, it is equivalent to  $(s,a) \sim (1- \gamma) \nu$. This result provides the basis for kernel regression of the critic (e.g. based on kernel matching pursuit) as well as the sampling distribution for the policy gradient in Theorem~\ref{th: actor-critic query complexity}.
\begin{lemma}
\label{lem: unbiased}
\textbf{Occupancy distribution lemma.} 
Let $\gamma \in [0,1]$ be the discount factor and let $\tilde{\nu}(s,a)$ be a state-action sampler following Algorithm~\ref{alg: occupancy}. Then the sampling distribution is equivalent to the occupancy distribution, i.e. $\tilde{\nu}(s,a) = (1 - \gamma) \nu(s,a)$.
\end{lemma}
The proof is given in \refSI{11}.

\begin{algorithm}
\caption{Classical program for occupancy-based sampling in infinite horizon MDPs.} \label{alg: occupancy}
\begin{algorithmic}[1]
\Procedure{Classical occupancy-based sampling \citep{Agarwal2021}}{}
\State $s_0 \sim d_0$.
\State $a_0 \sim \pi(\cdot \vert s_0)$.
\For {$t=0,1,\dots,\infty$}
\State With probability $1 - \gamma$:
\State \indent \textbf{return} $(s_t,a_t)$
\State $s_{t+1} \sim P(\cdot \vert s_{t},a_t)$
\State $a_{t+1} \sim \pi(\cdot \vert s_{t+1})$
\EndFor
\EndProcedure
\end{algorithmic}
\end{algorithm}

\subsubsection{Compatible Quantum RKHS Actor-Critic (CQRAC)}
\label{sec: compatible QRKHS-AC}
In the quantum-accessible setting, the classical program is modified into suitable quantum oracle for occupancy-based sampling, which is formed from $\BigO(T)$ calls to the policy evaluation oracle $\Pi$ and the transition oracle $O_P$. A quantum multivariate Monte Carlo is then used to obtain reliable estimates of the policy gradient. The proposed algorithm, called Compatible Quantum RKHS Actor-Critic (CQRAC) is shown in Algorithm~\ref{alg: quantum AC}. Note that we now use vector-based gradients over $\beta$ rather than functional gradients over $\mu$ as this is more convenient in quantum circuits.

At each iteration, the algorithm computes the policy gradient based on $n_1$ queries to a quantum oracle, where $n_1$ is set according to Theorem~\ref{th: actor-critic query complexity}. The quantum oracle is a state-action occupancy, a particular binary oracle $U_{X,\mathcal{S} \times \mathcal{A}}$ defined in Definition~\ref{def: state-action occupancy oracle} which when measured yields the random variable $X(s,a) = (\hat{Q}(s,a) - b(s)) K(s, \cdot)\Sigma^{-1} (a - \mu(s))$ according to the occupancy measure. The resulting quantity $X$ has the policy gradient as its expectation up to a constant of $(1-\gamma)$, allowing an $\epsilon$-precise estimate of the policy gradient within $n_1$ queries via quantum Monte Carlo. Following $n_1$ calls to $U_{X,\mathcal{S} \times \mathcal{A}}$, one applies the trajectory oracle $U_P$ and $U_R$ $n_2$ times to measure the trajectories $\{\tau_i = \left(s_0,a_0,s_1,a_1,\dots,s_{T'-1},a_{T'-1}\right)\}_{i=1}^{n_2}$ and reward sequences $\{r_0,\dots,r_{T'-1}\}_{i=1}^{n_2}$, where $T' \leq 2T-1$. The setting of $T'$ can be based on bootstrapping (e.g. with 1-step return, it is $T'=T$) or based on a full $T$-step return (with or without bootstrapping), in which case $T'=2T-1$ environment interactions are needed to get $T$ return estimates. These classical data are then used to improve the critic by performing classical kernel ridge regression. The iteration concludes with periodic and optional updates (e.g. kernel matching pursuit, covariance shrinking). After many such iterations, the policy converges towards a near-optimal policy and the critic correctly estimates the state-action values of that policy. 

To further reduce the variance and improve the query complexity, we include a baseline $b(s)$ in the policy gradient according to 
\begin{align}
\label{eq: approximate-integral baseline}
\nabla_{\beta} V(d_0) &\approx \frac{1}{1-\gamma}\frac{1}{n} \sum_{i=0}^{n} (\hat{Q}(s_i,a_i) - b(s_i)) \nabla_{\beta} \log(\pi(a \vert s))  \\
                      &= \frac{1}{1-\gamma}\frac{1}{n} \sum_{i=0}^{n} (\hat{Q}(s_i,a_i) - b(s_i)) K(s_i,:)\Sigma^{-1} (a_i - \mu(s_i))   \,, \nonumber  
\end{align}
where the last line follows if $\Pi$ is the Gauss QKP. Note that the baseline $b(s) = \hat{V}_{\pi}(s) = \sum_{a \in \mathcal{A}} \pi(a \vert s) \hat{Q}(s,a)$ is a possible choice, in which case $\hat{Q}(s,a) - b(s)$ is the advantage function. The use of the baseline is common in algorithms such as REINFORCE with baseline and the popular Advantage Actor-Critic (A2C) \citep{Mnih2016a}. Including baselines such as these reduces the variance, since its maximum is reduced, and comes with no effect on the accuracy of the policy gradient \citep{Sutton2018b} due to the derivation $\sum_a b(s) \nabla_{\beta} \pi(a \vert s) = b(s) \nabla_{\beta} \sum_a  \pi(a \vert s) = 0$ (where we note that the gradient of a constant 1 is 0). Note that the term corresponding to the log-policy gradient is vectorised to yield parameters in $\mathbb{R}^{N \times A}$, according to \refSI{3}, where the $N$ may change after periodic calls to classical kernel matching pursuit.

A compatible critic can be formulated based on a linear function of the feature-map $\phi(s,a)= \nabla_{\beta} \log(\pi(a \vert s))$, as exemplified for the Gaussian policy in Section~\ref{sec: background-RKHS-AC}. For the Gauss QKP, this reduces to $\phi(s,a) = K(s_i,:)\Sigma^{-1} (a_i - \mu(s_i))$. The critic is trained by classical kernel regression to minimise the MSE on a buffer of previously collected data:
\begin{align}
\label{eq: KRR critic}
\hat{Q}(s,a) = \langle w, \phi(s,a) \rangle = \argmin_{\hat{Q} \in \mathcal{H}_{K_{\mu}}} \mathbb{E}_{(z,Q) \sim \mathcal{B}} \left[ \left(Q - \hat{Q}(z) \right)^2\right] + \lambda \norm{\hat{Q}}{\mathcal{H}_{K_{\mu}}}^2 \,.
\end{align}
The \textit{iid} replay of many trajectories collected from past behaviour policies provides more efficient convergence similar to a supervised learning setting, as it avoids to overfit on data from the new behaviour policy, and the batch of training data can be much larger than $n_2$ without affecting the query complexity.

In practice, the occupancy measure is not a distribution. However, a related occupancy distribution can be implemented by forming a quantum analogue of classical occupancy-based sampling (Algorithm~\ref{alg: occupancy}) in what we call a state-action occupancy oracle.
\begin{definition}
\label{def: state-action occupancy oracle}
\textbf{State-action occupancy oracle.} A state-action occupancy oracle $U_{X,\mathcal{S} \times \mathcal{A}}$ is a binary oracle that takes the form
\begin{align*}
U_{X,\mathcal{S} \times \mathcal{A}}: \vert 0 \rangle \to  \sum_{(s,a) \in \mathcal{S} \times \mathcal{A}} \sqrt{\tilde{\nu}(s,a)}   \ket{s} \ket{a} \ket{X(s,a)} \,,
\end{align*}
where $\ket{s} \ket{a}$ is the occupancy register (representing returned state-action pairs) and $\ket{X(s,a)}$ represents the contribution to the policy gradient, e.g. $X(s,a) = (\hat{Q}(s,a) - b(s)) K(s, \cdot)\Sigma^{-1} (a - \mu(s))$ for the Gauss QKP in CQRAC.
\end{definition}

\begin{figure}[htbp!]
    \centering
\includegraphics[width=1.1\textwidth]{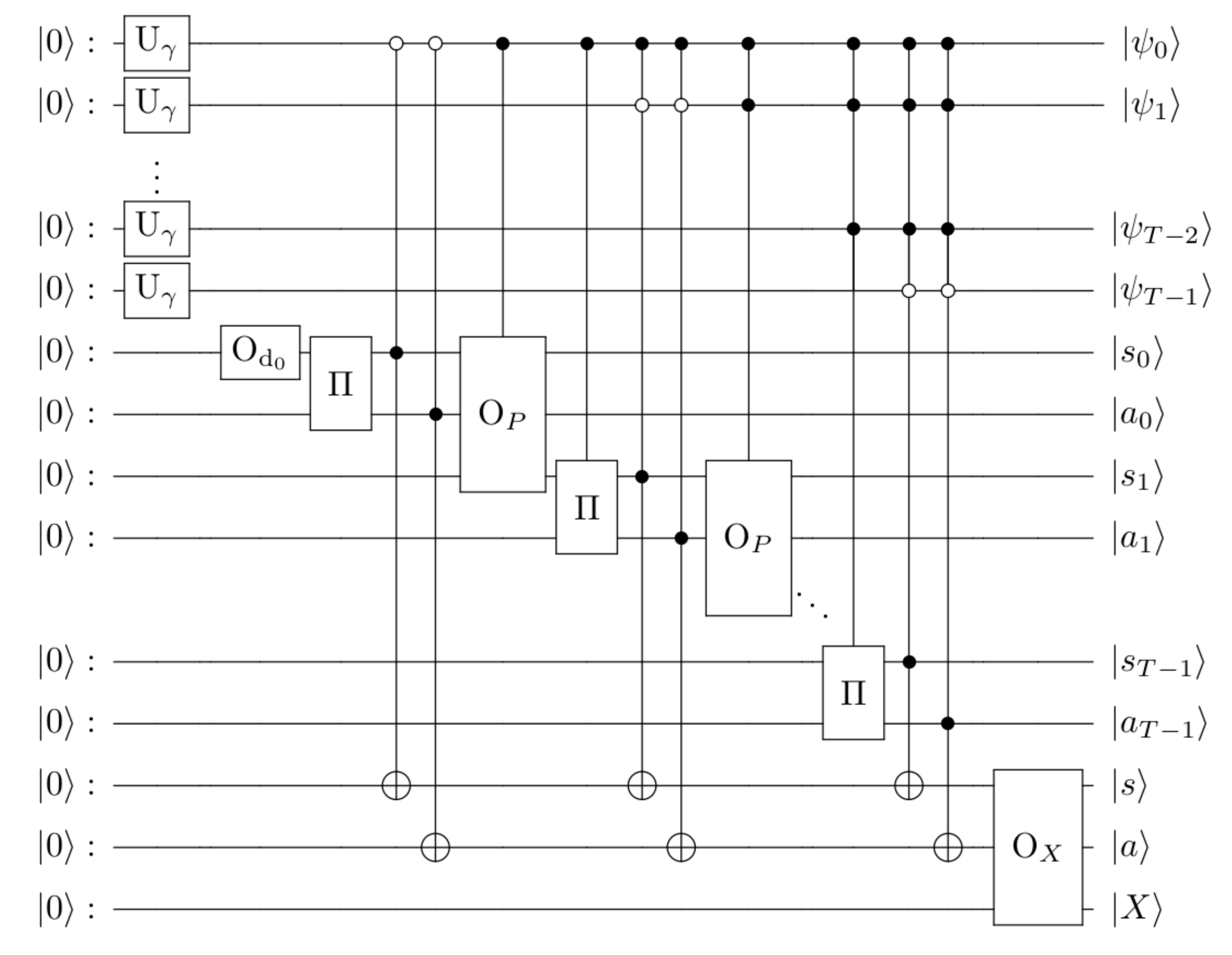}
    \caption{The circuit $U_{X,\mathcal{S}\times\mathcal{A}}$ for occupancy-based sampling to estimate the policy gradient within Compatible Quantum RKHS Actor-Critic. The unitary $U_{\gamma} \ket{0} = \sqrt{\gamma} \ket{1} + \sqrt{1-\gamma} \ket{0}$ is implemented based on multi-controlled $R_Y(2\sin^{-1}(\gamma))$ gates. $O_X$ denotes another unitary defined such that for any eigenstate $s \in \mathcal{S}$ and eigenaction $a \in \mathcal{A}$, $O_X \ket{s,a}\ket{0} = \ket{s,a}\vert X(s,a) \rangle$, where $X(s,a) = \hat{Q}(s,a) \nabla_{\beta} \log(\pi(a\vert s))$. Other oracles have the meanings as defined in Section~\ref{sec: quantum-classical setup}. The circuit $U_{X,\mathcal{S}}$ for DCQRAC is analogous but removes action controlled CNOT-gates and formulates the $O_X$ oracle such that for any eigenstate $s \in \mathcal{S}$, $O_X \ket{s}\ket{0} = \ket{s}\vert X(s) \rangle$, where $X(s) = \kappa(s,\cdot) \nabla_{a} \hat{Q}(s,a) \vert_{a=\mu(s)}$.}
    \label{fig: U_X}
\end{figure}

Note that implementing the oracle in Definition~\ref{def: state-action occupancy oracle} will typically require auxiliary registers (e.g. to encode the trajectories which terminate in particular state-action pairs). Moreover, we have assumed that Algorithm~\ref{alg: occupancy} runs with $T \to \infty$ such that it always returns, and $\tilde{\nu}$ is indeed a probability distribution summing to one. For finite $T$, the classical algorithm may not always return before time $T-1$. Considering these above points, we demonstrate the implementation of the quantum oracle and a suitable correction to yield the expected value over the the occupancy measure $\nu$. The proof is shown for a state-action occupancy oracle (Definition~\ref{def: state-action occupancy oracle}) but also applies to a state occupancy oracle (see Definition~\ref{def: state occupancy oracle}) by removing conditioning on actions.
\begin{lemma}
\label{lem: U_X}
\textbf{Occupancy oracle lemma.} An occupancy oracle $U_{X,\mathcal{S} \times \mathcal{A}}$ following Definition~\ref{def: state-action occupancy oracle} can be computed within $\BigO(T)$ calls to $O_P$ and $\Pi$ such that an estimator with $\mathbb{E}[\bar{X}] = \langle X \rangle$ has expectation equal to the analytical policy gradient after post-processing. The space complexity is $\BigO(k(S + A + d)T)$ where $S$ the action dimensionality, $A$ is the number of action qubits, $d$ is the number of dimensions of the policy gradient, and $k$ is the per-dimension precision.
\end{lemma}
\begin{proof}Fig.~\ref{fig: U_X} shows an example circuit to implement $U_{X,\mathcal{S}\times\mathcal{A}}$. The full proof is given in \refSI{12}.
\end{proof}

Now we turn to proving the query complexity of CQRAC, and more generally quantum actor-critic algorithms, using QEstimator. Theorem~\ref{th: actor-critic query complexity} states that shots from $U_{X,\mathcal{S} \times \mathcal{A}}$ can provide a sample-efficient estimate based on the approximation of Eq.~\ref{eq: integral} through Eq.~\ref{eq: approximate-integral}. Instead of a dependence on the maximal value as in QPG, the actor-critic has a dependence on the maximal deviation from the baseline $b(s)$. We provide the proofs in a generic way for all policies that can be prepared with state preparation and have a readily available binary form for $\nabla_{\theta} \log(\pi(a \vert s))$ and fit the other preconditions. In part \textbf{a)}, we derive an upper bound on the variance based on the range (e.g. $B_p=\BigO(1)$ for the Gauss-QKP), which leads to a query complexity that is comparable to QBounded. In part \textbf{b)}, we analyse a case where more information on the variance upper bound is known. This leads to an improvement over the range-based QBounded algorithm since the standard deviation is only a fraction of the range. For instance, for the Gauss-QKP, the improvement in query complexity for $p=1$ is at least $\Omega(\min_i (u_i -l_i))$, where $[l_i,u_i]_{i=1}^{d}$ is the support of the finite-precision Gaussian (see \refSI{13}).
\begin{theorem}
\label{th: actor-critic query complexity}
\textbf{Quantum actor critic theorem (CQRAC query complexity).} Let $\delta \in (0,1)$ be the upper bound on the failure probability, and let $\epsilon > 0$ be an upper bound on the $\ell_{\infty}$  error of the policy gradient estimate. Let $\Pi$ be a policy evaluation oracle parametrised by $\theta$ such that for any eigenstate $s \in \mathcal{S}$, $\Pi_{\theta} \ket{s} \ket{0} = \sum_{a \in \mathcal{A}} \sqrt{\pi_{\theta}(a\vert s)} \ket{s} \ket{a}$. Let $X(s,a) = \left(\hat{Q}(s,a) - b(s)\right) \nabla_{\theta} \log(\pi(a \vert s))$ and define $U_{X,\mathcal{S} \times \mathcal{A}}$ as a binary state-action occupancy oracle for $X$ based on Definition~\ref{def: state-action occupancy oracle} and Lemma~\ref{lem: U_X}. Moreover, let $\vert \hat{Q}(s,a) - b(s) \vert \leq \Delta_Q$ for all $(s,a) \in \mathcal{S} \times \mathcal{A}$, where $b$ is a baseline function. Then it follows that \\
\textbf{a)} with probability at least $1-\delta$, QEstimator (algorithm in Theorem 3.4 of \cite{Cornelissen2022} for quantum multivariate Monte Carlo) returns an $\epsilon$-correct estimate $\bar{X}$ such that $\norm{\bar{X} - \nabla_{\theta} V(d_0)}{\infty} \leq \epsilon$ within 
\begin{equation}
\label{eq: actor-critic query complexity}
n=\tildeO\left(\frac{d^{\xi(p)}\Delta_Q B_p}{(1-\gamma)\epsilon}\right) 
\end{equation}
 $\BigO(T)$ time steps of interactions with the environment, when there is an upper bound $B_p \geq \max_{s,a} \norm{\nabla_{\theta} \log(\pi(a \vert s))}{p}$ for some $p\geq 1$; and \\
 \textbf{b)} let the following assumptions hold for all gradient dimensions $i=1, \dots, d$: first, let $\mathrm{Var}_{\tilde{\nu}'}[\partial_i \log(\pi(a\vert s))] \leq \sigma_{\partial}(i)^2$ where $\tilde{\nu}'$ is the occupancy distribution before correction with $\gamma^T X(\mathbf{0},\mathbf{0})$; second, let $\mathrm{Cov}_{\tilde{\nu}'}((\hat{Q}(s,a) - b(s))^2, \partial_i \log(\pi(a \vert s))^2) \leq c \mathrm{Var}_{\tilde{\nu}'}(\hat{Q}(s,a) - b(s))  \mathrm{Var}_{\tilde{\nu}'}(\partial_i \log(\pi(a \vert s)))$ for some $c > 0$; third, let $\mathrm{Var}_{\tilde{\nu}'}(\partial_i \log(\pi(a \vert s))) ) \geq \mathbb{E}_{\tilde{\nu}'}[\partial_i \log(\pi(a \vert s)) ]^2$; finally, let $b(s) = \sum_{a \in \mathcal{A}} \pi(a \vert s) \hat{Q}(s,a)$. Under these conditions, QEstimator returns an $\epsilon$-correct estimate within 
 \begin{equation}
\label{eq: actor-critic query complexity b}
n=\tildeO\left(\frac{d^{\xi(p)} \sigma_Q \sigma_{\nabla_p}}{(1-\gamma)\epsilon}\right)
\end{equation}
 $\BigO(T)$ time steps of interactions with the environment, also with probability at least $1-\delta$, where $\sigma_{\nabla_p} = \norm{ \sigma_{\partial}(:)}{p}$ and $\sigma_Q^2 \geq \mathrm{Var}_{\tilde{\nu}'}( \hat{Q}(s,a) - b(s))$.
 \end{theorem}
\begin{proof}
We apply QEstimator (see start of this section) to $U_{X,\mathcal{S} \times \mathcal{A}}$ and the result follows from Theorem 3.4 of \cite{Cornelissen2022} after upper bound derivations. The full proof is given in \refSI{14}.
\end{proof}

We remark further on the assumptions in part b), which are mainly there to demonstrate that a significant improvement over QBounded is possible.\footnote{Note that we do not need assumption \textbf{b)} to hold throughout the algorithm since we can simply benefit from a reduced policy gradient error in some instances if we set the number of samples according to \textbf{a)}.} The first assumption, that of bounded variance of partial derivatives, is straightforward for smooth functions such as Gaussians. The second assumption states that squared deviations of values from their baseline do not have a strong positive covariance with the squared log-policy partial derivative. This is true when the mode of the policy does not correspond to the average Q-value; for instance, when the Q-distributions are skewed, the average Q-value (i.e. the baseline) does not correspond to a Gaussian policy's mode. The third assumption is true for Gaussians due to their peaked and symmetric nature; if the action chosen is frequently the same, it is the mode of the log-probability and it has expectation of the partial derivative near 
zero such that $\mathbb{E}[\partial_i \log(\pi(a \vert s))^2] \geq 2 \mathbb{E}[\partial_i \log(\pi(a \vert s))]^2$. The last assumption of the $b(s) = \sum_{a \in \mathcal{A}} \pi(a \vert s) \hat{Q}(s,a)$ is in general a recommended setting used in advantage-based algorithms. More generally, it is worth pointing out that $\sigma_Q^2 \geq  \frac{1}{4} \left(\max_{s,a} \hat{Q}(s,a) - b(s)  - ( \min_{s',a'} \hat{Q}(s',a') - b(s'))\right)^2$ provides a trivial upper bound via Popoviciu's inequality, but that generally more tight bounds can be available.

As shown in the corollary below, Theorem~\ref{th: actor-critic query complexity}\textbf{a} implies a quadratic query complexity speedup compared to its classical counterpart for $\Delta_Q \geq 1$. Such cut-off points are standard in big O notation to represent the asymptotic worst case, and indeed one may typically select $x \to \infty$ for terms in the enumerator and $x \to 0$ for terms in the denominator. The $\Delta_Q \geq 1$ case includes many settings where $T \to \infty$, and $\vert r_{\text{max}} \vert \to \infty$ but more generally settings where one action has a larger than 1 value benefit compared to others.
\begin{corollary}
\label{cor: speedup actor-critic}
\textbf{Quadratic query complexity speedup over classical sub-Gaussian estimator.} For any $\Delta_Q \geq 1$, any $p \geq 1$, upper bound $B_p \geq \max_{s,a} \norm{\nabla_{\theta} \log(\pi(a \vert s))}{p}$, and covariance matrix $\Sigma_X$ with operator norm (i.e. maximal eigenvalue) $\norm{\Sigma_X}{}$, Eq.~\ref{eq: actor-critic query complexity} provides a quadratic query complexity speedup in $\ell_{\infty}$ error compared to a comparable classical sub-Gaussian estimator, which yields
\begin{equation}
\label{eq: classical actor-critic query complexity}
n = \tildeO\left(\frac{d^{2\xi(p)}\Delta_Q^2 B_p^2 + \norm{\Sigma_X}{}}{(1-\gamma)^2\epsilon^2}\right) \,.
\end{equation}
Similarly, under the conditions of Theorem~\ref{th: actor-critic query complexity}\textbf{b)}, Eq.~\ref{eq: actor-critic query complexity b} provides a quadratic query complexity speedup since the classical sub-Gaussian estimator yields
\begin{equation}
\label{eq: classical actor-critic query complexity b}
n = \tildeO\left(\frac{d^{2\xi(p)}\sigma_Q^2 \sigma_{{\nabla}_p}^2 + \norm{\Sigma_X}{}}{(1-\gamma)^2\epsilon^2}\right) \,.
\end{equation}
\end{corollary}
The proof is given in \refSI{15}.

Since training the critic requires additional samples, below we analyse the total query complexity of CQRAC and compare it to the classical case. A first analysis uses a simple tabular average and disregards the role of replaying data from the buffer. We focus on part \textbf{a)} of Theorem~\ref{th: actor-critic query complexity} but note that part \textbf{b)} is completely analogous.
\begin{corollary}
\label{cor: total complexity}
\textbf{Total query complexity for CQRAC with a tabular averaging critic.}  Suppose the preconditions in Lemma~\ref{lem: KRR bounds} and Theorem~\ref{th: actor-critic query complexity}\textbf{a)}. Let $\delta > 0$ be the upper bound on the total failure probability (combining critic and policy gradient bounds) and let $\epsilon > 0$ be the upper bound on the $\ell_{\infty}$ error on the policy gradient. Let $Q(s,a)\in [-V_{\text{max}},V_{\text{max}}]$ 
and $\hat{Q}(s,a)$ be the state-action value and the prediction of the critic, respectively, for any state-action pair $(s,a) \in \mathcal{S} \times \mathcal{A}$. Moreover, let $\Delta_Q \geq 1$. Let $\epsilon' \geq \sqrt{\frac{(1-\gamma) \epsilon}{d^{\xi(p)}T\Delta_Q B_p}}V_{\text{max}}$  be a tolerable upper bound on the critic error, i.e. $\epsilon' \geq \max_{s,a} \vert \hat{Q}(s,a) - Q(s,a) \vert$. Then the total query complexity for CQRAC, combining queries for the policy gradient and the critic, is given by the same expression as in Eq.~\ref{eq: actor-critic query complexity},
\begin{align*}
n = \tildeO\left( \frac{d^{\xi(p)}\Delta_Q B_p}{(1-\gamma) \epsilon}\right)
\end{align*}
$\BigO(T)$ time steps of environment interaction, while the total query complexity for (classical) Compatible RKHS Actor-Critic is given by the same expression as in Eq.~\ref{eq: classical actor-critic query complexity},
\begin{align*}
n = \tildeO\left(  \frac{d^{2\xi(p)} \Delta_Q^2 B_p^2 \norm{\Sigma_X}{}}{(1-\gamma)^2 \epsilon^2}  \right)
\end{align*}
$\BigO(T)$ time steps of environment interaction. Therefore, a quadratic improvement holds for any $p \geq 1$.
\end{corollary}
The proof is given in \refSI{16.1}.

We now turn to analysing the total query complexity in the case where the critic is based on kernel ridge regression, focusing on the $L_2$ bound which is more common in the function approximation setting. The corollary imposes a requirement on the tolerable error such that the number of samples is limited compared to the number of samples for the policy gradient estimation.
\begin{corollary}
\label{cor: KRR}
\textbf{Total query complexity of CQRAC with a kernel ridge regression critic.} Suppose the preconditions in Lemma~\ref{lem: KRR bounds} and Theorem~\ref{th: actor-critic query complexity}\textbf{a)}. Moreover, let $\delta > 0$ be the upper bound on the total failure probability (combining critic and policy gradient bounds) and let $\epsilon > 0$ be the upper bound on the $\ell_{\infty}$ error on the policy gradient. Further, let $\Delta_Q \geq 1$. Further, let $\epsilon' \geq \left(\frac{(1-\gamma)\epsilon}{T d^{\xi(p)} \Delta_Q B_p} \right)^{1/4}$  
be a tolerable upper bound on the $L_2$ critic error such that $\epsilon' \geq \norm{\hat{Q} - Q}{L_2}$, let $m$ be the number of samples to estimate the critic, and let $n_2 = \frac{m}{T}$ denote the number of queries to the trajectory oracle. Then the total query complexity for CQRAC, combining queries for the policy gradient and the critic, is given by the same expression as in Eq.~\ref{eq: actor-critic query complexity},
\begin{align*}
n = \tildeO\left( \frac{d^{\xi(p)}\Delta_Q B_p}{(1-\gamma) \epsilon}\right)
\end{align*}
while the total query complexity for (classical) Compatible RKHS Actor-Critic is given by  the same expression as in Eq.~\ref{eq: classical actor-critic query complexity},
\begin{align*}
n = \tildeO\left(  \frac{d^{2\xi(p)}\Delta_Q^2 B_p^2 + \norm{\Sigma_X}{}}{(1-\gamma)^2 \epsilon^2}  \right) \,.
\end{align*}
Therefore, a quadratic improvement holds for any $p \geq 1$.
\end{corollary}
The proof is given in \refSI{16.2}.

\subsection{Deterministic Compatible Quantum RKHS Actor-Critic}
\label{sec: deterministic AC}
When seeking to learn the optimal deterministic policy, $\mu^*$, from samples of a behaviour policy, $\pi$, it is also possible to design an algorithm which uses a deterministic policy gradient to directly descend in a deterministic policy $\mu$, regardless of the form of the behaviour policy $\pi$. 
In this context, we analyse an off-policy actor-critic based on deterministic policy gradient algorithms \citep{Silverb}, where we define $\mu$ as a deterministic policy with the form of Eq.~\ref{eq: GQK mean}. The algorithm further applies experience replay, leading to a deep deterministic policy gradient (DDPG) \citep{Lillicrap2016} implementation. The algorithm is again implemented according to the framework in Fig.~\ref{fig: diagram}, making use of quantum policy gradient and a classical critic.

The algorithm, which we call Deterministic Compatible Quantum RKHS Actor-Critic (DCQRAC; Algorithm~\ref{alg: quantum DAC}), repeats iterations which are summarised as follows. At each iteration, the algorithm computes the policy gradient based on $n_1$ queries to a quantum oracle, where $n_1$ is set according to Theorem~\ref{th: dpg query complexity}. The quantum oracle is a binary oracle $U_{X,\mathcal{S}}$ (see Definition~\ref{def: state occupancy oracle}) which when measured yields the random variable $X =\nabla_a \hat{Q}(s_h,a) \vert_{a=\mu(s)} K(s, \cdot)$ according to the state-occupancy distribution. The resulting quantity provides an $\epsilon$-precise estimate of the policy gradient within $n_1$ queries via quantum multivariate Monte Carlo. In addition to calls to $U_{X,\mathcal{S}}$, the algorithm applies $n_2 = n_1$ calls to the trajectory oracle and return oracle, measuring the full trajectory with rewards $\{s_0,a_0,r_0\dots,s_{T-1},a_{T-1},r_{T-1}\}_{i=1}^{n_2}$, with $T$ interactions with environment per call. This is then followed by the estimation of the critic after which the iteration is concluded with some optional and periodic updates.

The policy gradient is based on a state-occupancy measure $\nu_{\pi}(s) := \sum_{t=0}^{T} \gamma^t \mathbb{P}_t(s \vert \pi)$. It substitutes the action from the trajectory by the action $\mu(s)$ of the deterministic policy, reformulating the value as
\begin{equation}
V_{\mu}(d_0) = \int \nu_{\pi}(s) Q_{\mu}(s,\mu(s))     \,,
\end{equation}
leading to an off-policy deterministic policy gradient of the form (see e.g. Eq.~15 in \cite{Silverb})
\begin{equation}
\nabla_{\beta} V_{\mu}(d_0) := \int \nu_{\pi}(s) \nabla_{\beta} \mu(s) \nabla_a Q_{\mu}(s,a) \vert_{a = \mu(s)} ds  \,,
\end{equation}
where $Q_{\mu}$ is the state-action value of the deterministic policy. The equality omits an approximation error, which is due to dropping a term which depends on $\nabla_{\beta} Q_{\mu}(s,a)$; since it is considered negligible \citep{Silverb}, it will be omitted in further analysis. 

With critic $\hat{Q}$ and $n$ samples from the occupancy measure $s_1, \dots, s_n \sim \tilde{\nu}_{\pi}$, using a representer formula $\mu$ for the deterministic policy leads to a convenient expression for the deterministic policy gradient
\begin{align}
\label{eq: approx DAC}
\nabla_{\theta} V(d_0) &\approx  \frac{1}{1-\gamma}\frac{1}{n} \sum_{i=0}^{n} \nabla_{\theta} \mu(s_i)  \nabla_{a} \hat{Q}(s_i,a) \vert_{a = \mu(s_i)} \nonumber \\
			  &=  \frac{1}{1-\gamma}\frac{1}{n} \sum_{i=0}^{n} K(s_i,\cdot)  \nabla_{a} \hat{Q}(s_i,a) \vert_{a = \mu(s_n)} \,,
\end{align}
for both the parametric and non-parametric settings ($\theta = \beta$ and $\theta = \mu$, respectively).

Following \cite{Silverb}, the compatible critic is now of the form
\begin{align}
\label{eq: compatible critic DPG}
\hat{Q}(s,a) &= \langle w, (a - \mu(s))^{\intercal} \nabla_{\theta} \mu(s) \rangle  + v^{\intercal} \phi(s) \nonumber  \\
			 &= \langle w, (a - \mu(s))^{\intercal} K(s,\cdot) \rangle + v^{\intercal} \phi(s) \,,
\end{align}
for some $d_f$-dimensional feature map $\phi(s) \in \mathbb{R}^{d_f}$ (not necessarily equal to the feature encoding of the kernel), and parameters $v \in \mathbb{R}^{d_f}$ and $w \in \mathbb{R}^{N \times A}$; one natural interpretation is of the second term as a state-dependent baseline and the first term as the advantage of the action in that state. 

Since the Q-value in Eq.~\ref{eq: compatible critic DPG} estimates the value of $\mu$ rather than the value of the behaviour policy $\pi$, a suitable off-policy technique should be used. A natural choice is to apply experience replay \citep{Mnih2015} with the DDPG style update \citep{Lillicrap2016}:
\begin{align}
L(\hat{Q}) = \mathbb{E}_{s,a,r,s' \sim \mathcal{B}} \left[ \left( r + \gamma \hat{Q}(s',\mu^{-}(s'); v^{-}, w^{-}) -  \hat{Q}(s,a) \right)^2 \right] \,,
\end{align}
where transitions are sampled \textit{iid} from a large buffer $\mathcal{B}$, and $\mu^{-}$,$v^{-}, w^{-}$ are updated infrequently (or with small increments using exponential averaging). The use of bootstrapping reduces the variance since the estimates only vary in one time step and make more efficient use of the trajectory (since all time steps can be used). The use of the slowly moving target network avoids the oscillatory behaviours and general poor convergence associated with frequently moving target values. Again, the \textit{iid} replay of many trajectories collected from past behaviour policies provides more efficient convergence.

Since Algorithm~\ref{alg: quantum DAC} requires a quantum oracle for state occupancy, we formalise this quantity below. Again $U_{X,\mathcal{S}}$ is based on a quantum analogue of Algorithm~\ref{alg: occupancy}, as shown in Lemma~\ref{lem: U_X} and \refSI{12}, but now one simply removes the action control qubits for the CX-gates. 
\begin{definition}
\label{def: state occupancy oracle}
\textbf{State occupancy oracle.} A state occupancy oracle $U_{X,\mathcal{S}}$ is a binary oracle that takes the form
\begin{align*}
U_{X,\mathcal{S}}: \vert 0 \rangle \to \sum_{s \in \mathcal{S}} \sqrt{\tilde{\nu}(s)} \ket{s} \ket{X(s)} \,,
\end{align*}
where $\ket{s}$ is the occupancy register (representing the returned state) and $\ket{X(s)}$ represents the contribution to the policy gradient, e.g. $X(s) = \nabla_a \hat{Q}(s,a) \vert_{a=\mu(s)} K(s, \cdot)$ for the Gauss QKP and a deterministic policy gradient.
\end{definition}

An analogue to Theorem~\ref{th: actor-critic query complexity} is now formulated using a binary oracle that returns after measurement the quantity $\kappa(s,\cdot)  \nabla_{a} \hat{Q}(s,a) \vert_{a = \mu(s)}$. With the critic available, the quantity $\nabla_{a} \hat{Q}(s,a) \vert_{a = \mu(s)}$ can be evaluated classically, before being input to the binary oracle.
\begin{theorem}
\label{th: dpg query complexity}
\textbf{Deterministic quantum actor critic theorem (DCQRAC query complexity).} Let $\delta \in (0,1)$ be the upper bound on the failure probability and let $\epsilon > 0$ be an upper bound on $\ell_{\infty}$ error of the policy gradient estimate. Further, let the critic be such that $C_p \geq \max_{s,a} \norm{\nabla_{a} \hat{Q}(s,a) \vert_{a = \mu(s)}}{p}$ for some $p \geq 1$. \\
\textbf{a)} Let $\mu_{\beta}$ be a deterministic policy parametrised by $\beta$ according to a representer formula (see Eq.~\ref{eq: representer}), let  $\Pi$ be a policy evaluation oracle parametrised by $\beta$ such that for any eigenstate $s \in \mathcal{S}$, $\Pi_{\beta} \ket{s} \ket{0} = \sum_{a \in \mathcal{A}} \sqrt{\pi_{\beta}(a\vert s)} \ket{s} \ket{a}$ for some behaviour policy $\pi_{\beta}$. Moreover, let $\kappa: \mathcal{S} \times \mathcal{S} \to \mathbb{C}$ be a kernel such that $\kappa_{p}^{\text{max}} \geq \max_{s} \norm{\kappa(s,:)}{p}$ where $:$ indicates vectorisation over the policy centres. Its policy gradient is given by Eq~\ref{eq: approx DAC}. Then with probability at least $1-\delta$, applying QBounded (algorithm in Theorem 3.3 of \cite{Cornelissen2022} for quantum multivariate Monte Carlo) on $U_{X,\mathcal{S}}$ returns an $\epsilon$-correct estimate $\bar{X}$ of $ \mathbb{E}[X] = \nabla_{\beta} V(d_0)$ such that $\norm{\bar{X} - \mathbb{E}[X]}{\infty} \leq \epsilon$ within 
\begin{equation}
\label{eq: dpg query complexity}
n=\tildeO\left(\frac{d^{\xi(p)}\kappa_{p}^{\text{max}}C_p}{(1-\gamma)\epsilon}\right) 
\end{equation}
 $\BigO(T)$-step interactions with the environment, where  $\xi(p) = \max\{0,1/2 - 1/p\}$. \\
 \textbf{b)} For a general deterministic policy $\mu_{\theta}$ and policy evaluation oracle $\Pi_{\theta}$, we obtain an $\epsilon$-correct estimate within 
 \begin{equation}
\label{eq: dpg query complexity general}
n=\tildeO\left(\frac{d^{\xi(p)}E_p C_p}{(1-\gamma)\epsilon}\right) \,,
\end{equation}
$\BigO(T)$-step interactions, where $E_p \geq \max_{s} \norm{\nabla_{\theta} \mu_{\theta}(s)}{p,\text{max}}$ for a norm defined as $\norm{\mathbf{A}}{p,\text{max}}:= \sum_{i} \max_{j} \vert a_{ij} \vert^p$ for any matrix $\mathbf{A}$.
\end{theorem}
\begin{proof}
We apply QBounded (see start of this section) to a normalised oracle $U_{\tilde{X},\mathcal{S}}$, which is based on the random variable $\tilde{X} = X /Z$ for some normalisation constant $Z$ to bound the $\ell_2$ norm. After upper bound derivations, the result follows from Theorem~3.3 of \cite{Cornelissen2022}. The full proof is given in \refSI{17}. 
\end{proof}
We have used QBounded rather than QEstimator for simplicity, even though in principle using QEstimator can improve the query complexity further, analogous to Theorem~\ref{th: actor-critic query complexity}. The difference between Eq.~\ref{eq: dpg query complexity} and Eq.~\ref{eq: dpg query complexity general} follows from the deterministic policy being defined in terms of a representer formula, which makes the gradient dependent on the kernel. The term $\kappa_{p}^{\text{max}}$ is often $\BigO(1)$; for instance, the Kronecker delta kernel yields $\kappa_{p}^{\text{max}} = 1$ since only a single state has non-zero value. Note that the policy gradient computed for the scalar-valued $\kappa$ can easily be converted to the matrix-valued kernel of the form $K(s,s') := \kappa(s,s') \mathbf{M}$ after the quantum oracle, since it follows immediately after matrix multiplication; in the worst case, this would only introduce a small multiplicative constant to the error.

Comparing Theorem~\ref{th: dpg query complexity}\textbf{a)} to Theorem~\ref{th: dpg query complexity}\textbf{b)} illustrates an advantage of kernel methods, namely that its query complexity depends on the number of representers rather than the parameter dimensionality. This is summarised in the informal corollary below.
\begin{corollary}
\textbf{Advantage of kernel policies.} Comparing a kernel policy with $N$ representers and $A$ action dimensions (and therefore $d_1 = \vert \beta \vert = N A$ parameters) to a general policy parametrised by $d_2 = \vert \theta \vert >N$ dimensions, the general policy has a higher query complexity since a norm is taken over $d_2$ rather than $N$ dimensions.
\end{corollary}

Theorem~\ref{th: dpg query complexity} implies a quadratic query complexity speedup compared to its classical counterpart.
\begin{corollary}
\label{cor: dpg quadratic speedup}
\textbf{Quadratic query complexity speedup over classical Hoeffding bounds.} For $p \in [1, 2]$, the results of Theorem~\ref{th: dpg query complexity}\textbf{a)} lead to a quadratic query complexity speedup over classical multivariate Monte Carlo. That is, classical multivariate Monte Carlo yields
\begin{equation}
\label{eq: classical dpg query complexity}
    n = \tildeO\left( \frac{(\kappa_{p}^{\text{max}} C_p)^2}{(1-\gamma)^2\epsilon^2}  \right) \,.
\end{equation}
\end{corollary}
The proof is given in \refSI{18}.

Theorem~\ref{th: dpg query complexity} implies a few key strategies for reducing the query complexity of DCQRAC, as summarised in the informal corollary below.
\begin{corollary}
\textbf{The importance of expressiveness control of $\mu$ and regularisation of $\hat{Q}$.} The results of Theorem~\ref{th: dpg query complexity}\textbf{a)} imply that controlling $N$ and $\nabla_{a} \hat{Q}$ are of critical importance for reducing query complexity. For the former, we propose the earlier-mentioned kernel matching pursuit technique (see Eq.~\ref{eq: matching-pursuit}). For the latter, regularisation techniques for $\hat{Q}$ are recommended.
\end{corollary}

Algorithm~\ref{alg: quantum DAC} formulates separate samples for the policy gradient and the critic estimation. Comparable to Corollary~\ref{cor: total complexity}, we compare the total query complexity of the algorithm to a classical variant thereof with a simple tabular critic (disregarding aspects of experience replay and the function approximator).
\begin{corollary}
\label{cor: total complexity deterministic}
\textbf{Total query complexity of DCQRAC with a tabular averaging critic.}  Suppose the preconditions in Theorem~\ref{th: dpg query complexity}\textbf{a)}. let $\epsilon > 0$ be an upper bound on the $\ell_{\infty}$  error of the policy gradient estimate. Let $Q(s,a)\in [-V_{\text{max}},V_{\text{max}}]$ and $\hat{Q}(s,a)$ be the state-action value and the prediction of the critic, respectively, for any state-action pair $(s,a) \in \mathcal{S} \times \mathcal{A}$. Moreover, let $\epsilon' \geq \sqrt{\frac{(1-\gamma) \epsilon}{T d^{\xi(p)} \kappa_{p}^{\text{max}} C_p}}V_{\text{max}}$ be the tolerable upper bound on the critic error, i.e. $\epsilon' \geq \max_{s,a} \vert \hat{Q}(s,a) - Q(s,a) \vert$. Then the total query complexity for DCQRAC, combining queries for the policy gradient and the critic, is given by the same expression as in Eq.~\ref{th: dpg query complexity}, i.e.
\begin{align*}
n = \tildeO\left(\frac{d^{\xi(p)} \kappa_{p}^{\text{max}} C_p}{(1-\gamma) \epsilon} \right)
\end{align*}
while the total query complexity for (classical) Deterministic Compatible RKHS Actor-Critic is given by the same expression as in Eq.~\ref{eq: classical dpg query complexity}, i.e.
\begin{align*}
n = \tildeO\left(  \frac{(\kappa_{p}^{\text{max}} C_p)^2}{(1-\gamma)^2\epsilon^2} \right) \,,
\end{align*}
yielding a quadratic improvement for any $p \in [1,2]$.
\end{corollary}
The proof is given in \refSI{19.1}.

We now turn to providing a similar total query complexity analysis when the critic is based on kernel ridge regression.
\begin{corollary}
\label{cor: KRR-deterministic}
\textbf{Total query complexity of DCQRAC with a kernel ridge regression critic.} Suppose the preconditions in Lemma~\ref{lem: KRR bounds} and Theorem~\ref{th: dpg query complexity}\textbf{a)}. let $\epsilon > 0$ be an upper bound on the $\ell_{\infty}$  error of the policy gradient estimate. Moreover, let $\epsilon'\geq \left( \frac{(1-\gamma)\epsilon}{T d^{\xi(p)} \kappa_{p}^{\text{max}} C_p}  \right)^{1/4}$ be a tolerable upper bound on the $\ell_{\infty}$ critic error and let $n_2 = \frac{m}{T}$ denote the number of queries to the trajectory oracle. Then the total query complexity for Compatible Quantum RKHS Actor-Critic, combining queries for the policy gradient and the critic, is given by the same expression as in Eq.~\ref{th: dpg query complexity}, i.e.
\begin{align*}
n = \tildeO\left(\frac{d^{\xi(p)} \kappa_{p}^{\text{max}} C_p}{(1-\gamma) \epsilon} \right)
\end{align*}
$\BigO(T)$ time steps of environment interaction, while the total query complexity for (classical) Deterministic Compatible RKHS Actor-Critic is given by the same expression as in Eq.~\ref{eq: classical dpg query complexity}
\begin{align*}
n = \tildeO\left(\frac{(\kappa_{p}^{\text{max}} C_p)^2}{(1-\gamma)^2\epsilon^2} \right)
\end{align*}
$\BigO(T)$ time steps of environment interaction. Therefore, a quadratic improvement holds for any $p \in [1,2]$.
\end{corollary}
The proof is given in \refSI{19.2}.

\subsection{Compatible Quantum RKHS Natural Actor-Critic}
\label{sec: natural AC}
While Section~\ref{sec: background-RKHS-AC} has highlighted the relation between the compatible critic and the natural policy gradient, the algorithms considered thus far do not fully exploit this relation. As a final contribution, we formulate a variant of our RKHS Actor-Critic framework based on natural actor critic \citep{Peters2008}, which ties in directly with the compatible critic from the CQRAC algorithm. The approach, which we call Compatible Quantum RKHS Natural Actor Critic (CQRNAC), is summarised in Algorithm~\ref{alg: quantum NAC}. The natural actor-critic is simple to implement in our approach, since the optimal solution to the critic,  $w^*$, is also the natural policy gradient. This allows implementing a natural policy gradient algorithm without the explicit computation of the Fisher information matrix.

As shown in \refSI{4.1} for the functional gradient with the feature-map of the Gaussian policy considered in CQRAC, note that if $\phi(s,a) = \nabla_{\beta} \log(\pi(a \vert s))$ and $\hat{Q} = \langle w,  \phi(s,a) \rangle$ for some parameter $w \in \mathbb{R}^d$, the optimal critic is given by
\begin{equation}
\label{eq: loss CQRNAC}
\hat{Q} = \argmin_{\hat{Q} \in \mathcal{H}_K} L(\hat{Q}) =  \frac{1}{2} \int \nu(z) \left( \hat{Q}(z) - Q(z) \right)^2 dz  \quad \in \mathcal{H}_{K}\,,
\end{equation}
and any optimal solution satisfies 
\begin{align*}
0 = \nabla_{w} L(\hat{Q}) &= \int \nu(z) (Q(z) - \hat{Q}(z)) \phi(z) dz \,.
\end{align*}
Following \refSI{4.2}, the optimum will correspond to the natural policy gradient, i.e.
\begin{equation}
\label{eq: weight as NPG}
w^* = \mathcal{F}(\beta)^{-1}\nabla_{\beta} V(d_0)\,,
\end{equation}
where $\mathcal{F}(\beta) = \int \nu(z) \nabla_{\beta} \log(\pi (a \vert s))  \nabla_{\beta}\log(\pi (a \vert s))^{\intercal}  dz$ is the Fisher information.

Classical natural actor-critic algorithms exploit this relation by considering the update based on the critic parameters. In our RKHS context, the set of parameter updates thus becomes
\begin{align}
\label{eq: updates CQRNAC}
&\text{critic update:} \quad \,\, w \gets w - \eta' \nabla_w L(\hat{Q}) \nonumber \\
&\text{policy update:} \quad \beta \gets \beta + \eta w \,,
\end{align}
where $\eta' > 0$ is the learning rate of the critic.

To ensure the desired analytical expression can be computed correctly for quantum multivariate Monte Carlo, a suitable binary oracle must be designed. Modifying the return oracle, the computation of the return for any trajectory $\tau$ starting from $(s,a)$ is now replaced by 
\begin{equation}
\label{eq: Rprime CQRNAC}
R'(\tau \vert s_0=s, a_0=a) = \left( R(\tau \vert s_0=s, a_0=a) - \hat{Q}(s,a) \right) \,, 
\end{equation}
such that 
\begin{equation}
\label{eq: expectation CQRNAC}
\mathbb{E}[R'(\tau \vert s_0=s, a_0=a) \phi(s,a)] = \left( Q(s,a)- \hat{Q}(s,a) \right) \phi(s,a) = \nabla_{w} L(\hat{Q}) \,.
\end{equation}
This allows to construct a binary oracle $U_{\nabla L}$ which computes Eq.~\ref{eq: Rprime CQRNAC} for state-action pairs sampled from the occupancy distribution. The first part of this construction is the binary state-action occupancy oracle $U_{X,\mathcal{S} \times \mathcal{A}}$ with $O_X \ket{s,a} \ket{0} = \ket{s,a} \ket{\phi(s,a)}$, which requires $T$ steps of environment interaction. The second part is to perform a $T-1$-step $U_P$ oracle, such that one has a $T$-step trajectory superposition starting from the state-action occupancy superposition ($\sum_{s,a} \sqrt{\tilde{\nu}(s,a)} \ket{s,a}$). The return is then computed based on the $T$-step $U_{R'}$ oracle over these last $T$ time steps. Finally, the classical product between $R'$ and $\phi(s,a)$ is computed. In total, this requires $M = \BigO\left((S+A)k T + N_{\text{aux}}\right)$ qubits which combines the $T$-step discount register, the $(2T-1)$-step trajectory register, the $T$-step reward register, and $N_{\text{aux}}$ auxiliary qubits. The auxiliary qubits are reserved for converting the rewards to the discounted return (Lemma~2.8 in \cite{Jerbi2022}) and computing the discounted return based on the classical product between $R'(s,a,\tau_{1:T-1})$ and $\phi(s,a)$ (as in Theorem~\ref{th: REINFORCE quadratic}, and Theorem~4.1 in \cite{Jerbi2022}). Formally, we have
\begin{align}
\label{eq: UnablaL}
\ket{0}^{\otimes M} & \overset{U_{X,\mathcal{S} \times \mathcal{A}}}{\to} \sum_{s,a} \sqrt{\tilde{\nu}(s,a)} \ket{\psi_1(s,a,\tau_{1:T-1})} \ket{s,a} \ket{\phi(s,a)} \ket{0}^{\otimes k} \ket{0}^{\otimes d} \nonumber \\
		&   \overset{U_P}{\to}  \sum_{s,a} \sqrt{\tilde{\nu}(s,a)} \sum_{\tau_{1:T-1}} \sqrt{P(\tau_{1:T-1})} \ket{\psi_2(s,a,\tau_{1:T-1})} \ket{s,a} \ket{\tau_{1:T-1}}   \ket{0}^{\otimes k}  \ket{0}^{\otimes d} \nonumber \\
		&  \overset{U_{R'}}{\to} \sum_{s,a} \sqrt{\tilde{\nu}(s,a)} \ket{s,a} \sum_{\tau_{1:T-1}} \sqrt{P(\tau_{1:T-1})} \ket{\psi_3(s,a,\tau_{1:T-1})} \ket{s,a} \ket{\tau_{1:T-1}} \nonumber \\
		& \hspace{5cm}  \ket{R'(s,a, \tau_{1:T-1})} \ket{0}^{\otimes d} \nonumber \\
		&  \overset{U_{*}}{\to} \sum_{s,a} \sqrt{\tilde{\nu}(s,a)} \ket{s,a} \sum_{\tau_{1:T-1}} \sqrt{P(\tau_{1:T-1})} \ket{\psi_4(s,a,\tau_{1:T-1})} \ket{s,a} \ket{\tau_{1:T-1}} \nonumber \\
		& \hspace{5cm} \ket{R'(s,a, \tau_{1:T-1})} \ket{R'(s,a, \tau_{1:T-1})\phi(s,a)}  \,,
\end{align}
where $\ket{\psi_1(s,a,\tau_{1:T-1})},\ket{\psi_2(s,a,\tau_{1:T-1})},\ket{\psi_3(s,a,\tau_{1:T-1})}$, and $\ket{\psi_4(s,a,\tau_{1:T-1})}$ refer to the combined, trajectory-dependent states of the auxiliary register, the reward register, and the discount register at the four different computational steps, and $U_{*}$ indicates the classical product unitary.  

Estimating the expectation in Eq.~\ref{eq: expectation CQRNAC} using a quantum algorithm, the query complexity comes from estimating the gradient of the critic, after which the policy gradient comes at no additional query complexity due to the critic update in Eq.~\ref{eq: updates CQRNAC}. The query complexity is summarised in the theorem below.
\begin{theorem}
\label{th: natural actor-critic query complexity}
\textbf{Quantum natural actor critic theorem (CQRNAC query complexity).} Let $\delta \in (0,1)$ be the upper bound on the failure probability and let $\epsilon > 0$ be an upper bound on the tolerable $\ell_{\infty}$  error of the compatible critic's gradient estimate. Further, let $\hat{Q}$ be a compatible RKHS critic such that $\hat{Q}(s,a) = \langle w, \phi(s,a) \rangle$ , with $B_p \geq \nabla_{\beta} \log(\pi(a \vert s)) = \phi(s,a)$ and $w \in \mathbb{R}^d$. Further, let $E \geq \sup_{\tau,z \in \mathcal{S} \times \mathcal{A}} \vert \hat{Q}(z) - R(z,\tau) \vert$ where $\tau$ is a $T-1$-step trajectory. Moreover, let $X(s,a) = \left(R(s,a,\tau) - \hat{Q}(s,a)\right) \phi(s,a)$ and define $U_{\nabla L}$ as in Eq.~\ref{eq: UnablaL}. Further, let $\vert Q(\mathbf{0},\mathbf{0}) - \hat{Q}(\mathbf{0},\mathbf{0}) \vert = \BigO\left(\frac{(1-\gamma)\epsilon}{\gamma^{T} \norm{\phi(\mathbf{0},\mathbf{0})}{\infty}}\right)$ bound the critic error for state-action pair $(\mathbf{0},\mathbf{0})$. Then with probability at least $1-\delta$, QBounded (algorithm in Theorem~3.3 of \cite{Cornelissen2022} for quantum multivariate Monte Carlo) returns an $\epsilon$-correct estimate $\bar{X}$ such that $\norm{\bar{X} - \nabla_{w} L(\hat{Q})}{\infty} = \BigO(\epsilon)$ within 
\begin{equation}
\label{eq: natural actor-critic query complexity}
n=\tildeO\left(\frac{d^{\xi(p)} E B_p}{(1-\gamma)\epsilon}\right)
\end{equation}
 $\BigO(T)$ time steps of interactions with the environment.
\end{theorem}
\begin{proof}
The proof applies QBounded on a normalised variant of the unitary $U_{\nabla L}$ of Eq.~\ref{eq: UnablaL}. The query complexity follows from Theorem~3.3 of \cite{Cornelissen2022}. The full proof is given in \refSI{20}.
\end{proof}

The quadratic speedup compared to classical also holds for CQRNAC as shown in the corollary below.
\begin{corollary}
\label{cor: nac quadratic speedup}
\textbf{Quadratic query complexity speedup over classical Hoeffding bounds.} For $p \in [1, 2]$, the results of Theorem~\ref{th: natural actor-critic query complexity} lead to a quadratic query complexity speedup over classical multivariate Monte Carlo. That is, classical multivariate Monte Carlo yields
\begin{equation}
\label{eq: classical nac query complexity}
    n = \tildeO\left( \frac{E^2 B_p^2}{(1-\gamma)^2 \epsilon^2}  \right) \,.
\end{equation}
\end{corollary}
\begin{proof}
The proof is analogous to that of Corollary~\ref{cor: dpg quadratic speedup}  (see \refSI{18}). After correcting for the discount factor, $\norm{X}{\infty} \leq \frac{E B_p}{1-\gamma}$, and computing Hoeffding's inequality for classical Monte Carlo in $[-B,B]^d$ with $B = \frac{E B_p}{1-\gamma}$, we obtain $n = \tildeO\left( \frac{E^2 B_p^2}{(1-\gamma)^2 \epsilon^2}  \right)$.
\end{proof}

\section{Discussion}
We now discuss the applicability of the techniques, future challenges, emerging trends, and further improvements.
\subsection{Applicability in the NISQ era}
The analysis of our algorithms has largely assumed fault-tolerant quantum computing capabilities, which is not realistic in near-term devices. It can be noted that despite the presence of noise, one has reason to be optimistic since its effect can be estimated. A useful tool in this regard is the diamond norm \citep{Aharonov1998}, which induces a distance that can be used to compare the desired channel to the channel that was actually implemented. While the structure of the RL setting with large depth leads to an accumulation of errors, the diamond distance can be bounded additively (i.e. layer by layer), and this can be directly applied to noisy channels \citep{Tan2025}. 

The quantum gradient estimation algorithms in this paper rely on amplitude estimation, which in its most general definition is the problem of estimating the amplitude $p = \vert P \ket{\psi} \vert$ for a given projector $P$ and state $\ket{\psi}$ \citep{Rall2023amplitudeestimation}. In general, such algorithms are expected to require depth of $\tildeO(1/\epsilon)$, where $\epsilon$ is the desired precision of the gradient estimate and the dependence on other parameters is suppressed. Although they asymptotically obtain the Heisenberg limited optimal complexity scaling, such depth requirements are well into the regime of fault-tolerant quantum computing and beyond the NISQ paradigm \citep{Meyer2022}. Although originally amplitude estimation was based on phase estimation \citep{brassard2000quantum}, there are recent alternatives with different underlying mechanisms (e.g. \cite{Rall2023amplitudeestimation,suzuki2020amplitude}). For instance, shallower depth algorithms for amplitude estimation have been regularly used for related quantum finance applications \citep{suzuki2020amplitude}; however, such algorithms tend to improve complexity only in the non-asymptotic regime and are still heavily reliant on error correction.  Another intriguing recent development is the log-depth in-place quantum Fourier transform \citep{kahanamoku2025log}, which promises to further reduce the depth of the quantum Fourier transform to scale logarithmically with the number of qubits. This could potentially reduce the depth of both the numerical and analytical policy gradient algorithms. 

Further overhead is encountered in policies that rely on preparing certain amplitude encoded quantum states. To overcome these issues and tailor a more explicitly NISQ oriented protocol, we introduce Representer PQCs which can be learned in a purely variational manner while allowing to be computed coherently within a quantum environment. In this approach, it is also possible to formulate a policy based on a shallow depth kernel (e.g. the Kronecker delta), which can be readily implemented within current architectures. For near-term use of such policies, classical policy gradient algorithms can already be used based on a simple central difference gradient computation that is NISQ friendly. By using central differencing schemes similar to those used in quantum central differencing algorithms, our experimental results confirm the benefit of high-quality central differencing estimates in simulated quantum-accessible environments. As indicated by the theoretical results, future implementations may provide quadratic improvements over these classical estimates which can improve convergence rates and scalability. To illustrate the potential impact, note that disregarding constants, state-of-the-art results indicate that approximate policy gradient algorithms require $\tildeO(\frac{1}{\epsilon^2})$ samples to find an $\epsilon$-optimal policy under a particular softmax parametrisation \citep{Cen2022}.

It is often challenging to obtain high-quality estimates of expectations and gradients in the context of quantum kernel methods and PQCs in general. Quantum kernel methods have been shown to suffer from the problem of exponential concentration \citep{Thanasilp2022}, where due to a variety of factors (inclusive of noise, expressiveness, and entanglement) the number of shots to accurately estimate the values of the quantum kernel scales exponentially with the number of qubits. In schemes where not just the policy weights but also the kernel function is adapted, re-estimation of the kernel matrix may be required which can potentially introduce significant computational costs. A related problem is the barren plateau phenomenon, where the gradient landscape becomes exponentially more flat with the number of qubits \citep{McClean2018}. However, there are at least four reasons to be optimistic for quantum policy gradient algorithms in RKHS. First, the highly expressive nature of PQCs suggests that quantum agents will be able to achieve complex tasks using kernels on low-dimensional Hilbert spaces. For instance, it has been shown that tasks executed by classical agents with high-dimensional state spaces, can be reproduced by quantum agents whose memory states lie in a much lower-dimensional Hilbert space \citep{elliott2022quantum}; similarly, even single-qubit systems are shown to be highly capable function approximators \citep{Perez2021}. Second, there may be novel types of quantum kernels on the horizon which can mitigate exponential concentration; for instance, recent work proposes quantum Fisher kernels, and in particular the anti-symmetric logarithmic derivative quantum Fisher kernel (Eq.~14 of  \cite{Suzuki}), as an alternative to fidelity-based quantum kernels. Third, we summarise the benefits of the kernel method in this context. The analytical expressions typically derive as a simple function of the kernel, which can be re-estimated only periodically, and with efficient placement of the policy centres, the parameter dimensionality is limited to $d = NA = \tildeO(A)$. The critic in CQRAC and DCQRAC is a linear function of the same kernel, which leads to a parameter efficient algorithm with a compatible function approximator. Finally, we note that the various policy formulations and algorithms presented in this paper come with their own advantages with respect to the above-mentioned problems but also with expressiveness trade-offs. For instance, with classically parametrised policy evaluation oracles and analytical gradient computation using quantum Monte Carlo, the barren plateau phenomenon can be avoided in some sense. Another example is that kernel matrix estimation or re-estimation is not always needed in our framework when the kernels are either classical or fixed.

Keeping the above in mind, we believe that our approach of using coherent policies for RL within quantum-accessible environments is especially applicable to closed-loop quantum control problems for state preparation and error correction but also Hamiltonian simulation of quantum systems on quantum devices or classical simulations thereof. By choosing the kernel, its associated representers, and regularisation techniques, the kernel policies provide smooth and convex optimisation landscapes, expressiveness control, parameter reduction, analytical gradients, and domain knowledge. These properties in turn also contribute to improved policy gradient estimates (or reduced query complexity), and therefore more sample-efficient learning.

\subsection{Optimising the kernel}
\label{sec: parametric-kernel}
In addition to optimising the policy weights, optimising the kernel may also possible with quantum policy gradient algorithms in RKHS. For instance, an interesting kernel in this respect is the bandwidth-based squared cosine kernel (Eq.~\ref{eq: bandwidth squaredcosine}) where only a single bandwidth factor $c > 0$ can impact the definition of the kernel in terms of expressivity, trainability, and generalisation. A disadvantage of such schemes is that gradient expressions and their norm bounds are more challenging to establish, which are essential for sample-efficient policy gradient estimates. Alternatively, central differencing techniques may be used to define inner products within the circuit, thereby formulating a new RKHS along with its unique function space and associated regulariser. Since redefining the RKHS may induce kernel estimation costs and also does not have known gradient norm bounds as is the case for policy weights (see Lemma~\ref{lem: B}), we would suggest to apply such updates in an alternating scheme where most of the updates are based on policy weights while infrequently the feature-maps of the RKHS are adapted. Another challenge that may need to be addressed is the potential instability changes in the kernel may induce in the critic. 

\subsection{Reducing the number of parameters}
\label{sec: QFIM-parameters}
While we propose kernel matching pursuit for analytical policy gradient algorithms, we note that it may also be possible to use techniques for pruning PQCs in the context of numerical policy gradient algorithms. For instance, using the quantum Fisher information matrix (QFIM), one can reduce the number of parameters in Representer PQCs following \cite{Haug2021b}. Noting that the QFIM for a state $\vert \psi(\theta)\rangle$ is given by
\begin{equation}
\mathcal{F}_{i,j} = 4 \Re \left[ \langle \partial_i \psi(\theta) \vert \partial_j \psi(\theta) \rangle - \langle \partial_i \psi(\theta) \vert \psi(\theta) \rangle \langle \psi(\theta)\vert \partial_j \psi(\theta) \rangle \right] \,
\end{equation}
where $\Re$ denotes the real part, the expressive capacity of a PQC can be determined by the rank of its QFIM. Although the QFIM for $\vert\psi(\theta)\rangle$ is a function of $\theta$ and hence is a local measure, its rank at random $\theta$ captures the global expressive power. Consequently, the QFIM for $\ket{\psi(\theta)}$ can be used to identify and eliminate redundant parameters, a process which involves calculating the eigenvectors of the QFIM that have zero eigenvalues. An iterative procedure can then be applied to remove parameters associated with zero components in the eigenvalues until all redundant gates are eliminated (see Algorithm~1 in \cite{Haug2021b}).

\section{Conclusion}
This paper presents optimisation techniques for quantum kernel policies for efficient quantum policy gradient algorithms, including numerical and analytical gradient computations as well as parametric and non-parametric  representations. We define various kernel-based policies based on representer theorem formalisms, which include the Representer Raw-PQC as a purely coherent PQC suitable for numerical policy gradient, as well as the Representer Softmax-PQC and the Gauss-QKP policy suitable for implementing with state preparation and analytical policy gradient. Empirical results indicate the learnability of such coherent PQCs. Theoretical results prove quadratic improvements of kernel-based policy gradient and actor-critic algorithms over their classical counterparts, across different formulations of stochastic and deterministic kernel-based policies. Quantum actor-critic algorithms are proposed that improve on quantum policy gradient algorithms under favourable conditions. Our approach results in improved query complexity results by reducing the number of parameters, improving the optimisation landscape by regularisation, lowering analytic bounds on deterministic gradients which are given by the kernel, and reducing the variance using baselines. Additionally, with our quantum natural actor-critic algorithm, it is possible to implement a quantum natural policy gradient algorithm without requiring the computation of the Fisher information matrix. Compared to traditional parametrised quantum circuit policies, the proposed quantum kernel policies allow convenient analytical forms for the gradient and techniques for expressiveness control, and are suitable for vector-valued action spaces.

\begin{appendices}

\section{Algorithm pseudo-code}
\label{app: CQRAC algorithm}
This appendix provides the pseudo-code for CQRAC, DCQRAC, and CQRNAC.
\begin{algorithm}[htbp]
\caption{CQRAC algorithm} \label{alg: quantum AC}
\begin{algorithmic}[1]
\State \textbf{Input:} error tolerance for policy gradient $\epsilon > 0$, learning rate $\eta > 0$, regularisation parameter $\lambda \geq 0$, covariance shrinkage $\alpha \in (0,1)$, discount factor $\gamma \in [0,1)$, failure probability $\delta \in (0,1)$, upper bound on deviation from baseline $\Delta_Q$, upper bound on the $2$-norm of the partial derivatives of the log-policy $B_2$,  number of policy centres $N$, action dimensionality $A$, parameter dimensionality $d=NA$, horizon $T$, number of iterations $N_{\text{it}}$.
\State \textbf{Output:} near-optimal policy $\pi$
\State Define $n_1 = \BigO\left(\frac{\Delta_Q B_2 \log(d/\delta)}{(1-\gamma) \epsilon}\right)$ (Theorem~\ref{th: actor-critic query complexity}a)
\State $\mathcal{Z} = \emptyset; \mathcal{Q} = \emptyset$.
\For {$i = 1,\dots, N_{\text{it}}$}
\LineComment{Estimate policy gradient (Eq.~\ref{eq: approximate-integral}) and update policy}
\State Define binary oracle $U_{X,\mathcal{S}\times\mathcal{A}}: \vert 0 \rangle \to \sum_{s,a} \sqrt{\tilde{\nu}(s,a)} \vert (\hat{Q}(s,a) - b(s)) K(s, \cdot)\Sigma^{-1} (a- \mu(s))\rangle$ according to Lemma~\ref{lem: U_X} and Fig.~\ref{fig: U_X} ($T$ interactions with environment per call).
\State Perform quantum multivariate Monte Carlo with $X(s,a) = (\hat{Q}(s,a) - b(s)) K(s,:)\Sigma^{-1} (a - \mu(s))$ based on $n_1$ queries of $U_{X,\mathcal{S} \times \mathcal{A}}$, following Theorem~\ref{th: actor-critic query complexity}a.
\State Obtain the final estimate $\bar{X} \approx \mathbb{E}[X] = \nabla_{\beta}V(d_0)$ from quantum multivariate Monte Carlo.
\State Compute update: $\beta \inc \eta \bar{X}$; $\mu = \sum_{i=1}^{N} \beta_i K(\cdot,c_i)$.
\LineComment{Update critic classically from measured trajectories}
\State Apply $n_2 = n_1$ calls to $U_P$ and $U_R$, measuring trajectories $\{\tau = s_0,a_0,s_1,a_1,\dots,s_{T'-1},a_{T'-1}\}_{i=1}^{n_2}$ and reward sequences $\{r_0,\dots,r_{T'-1}\}_{i=1}^{n_2}$ ($T' = 2T-1$ interactions with the environment)
\State Add trajectories $\mathcal{Z} \gets \mathcal{Z} \cup \{\tau\}_{i=1}^{n_2}$.
\State Add reward sequences $\mathcal{Q} \gets \mathcal{Q} \cup \{r_0,\dots,r_{T'-1}\}_{i=1}^{n_2}$.
\State Define an occupancy-based distribution $\mathcal{B}$ over $z$ and $R(\tau \vert z)$ by applying Algorithm~\ref{alg: occupancy} to randomly selected trajectories in $\mathcal{Z}$ and their associated $T$-step returns in $\mathcal{Q}$.
\NoteComment{An alternative with $T'=T$ is experience replay with bootstrapping (analogous to l.14 of Algorithm~\ref{alg: quantum DAC}). }
\State Classical kernel regression for the critic based on random samples from $\mathcal{B}$:
\begin{align*}
\hat{Q}(s,a) = \langle w, K(s, \cdot)\Sigma^{-1} (a - \mu(s))  \rangle = \argmin_{\hat{Q} \in \mathcal{H}_{K_{\mu}}} \mathbb{E}_{(z,Q) \sim \mathcal{B}} \left[ \left(Q - \hat{Q}(z) \right)^2\right] + \lambda \norm{\hat{Q}}{\mathcal{H}_{K_{\mu}}}^2 \,.
\end{align*}
\State Update baseline (e.g. $b(s) = \sum_{a \in \mathcal{A}} \pi(a \vert s) \hat{Q}(s,a) \quad \forall s \in \mathcal{S}$).
\LineComment{Optional and periodic updates} 
\State Remove proportion of old data in $\mathcal{Z}$ and $\mathcal{Q}$ (periodically).
\State Sparsify policy (periodically, optional): kernel matching pursuit, with tolerance $\epsilon_{\mu}$.
\State Update number of queries (optional): $n_1 = \BigO\left(\frac{\Delta_Q B_2 \log(d/\delta)}{(1-\gamma) \epsilon}\right)$ based on new $\Delta_Q$ and $d$.
\State Shrink covariance matrix (optional): $\Sigma \gets \Sigma * \alpha$.
\EndFor
\end{algorithmic}
\end{algorithm}

\clearpage

\begin{algorithm}[htbp]
\caption{DCQRAC algorithm} \label{alg: quantum DAC}
\begin{algorithmic}[1]
\State \textbf{Input:} error tolerance for policy gradient $\epsilon > 0 $, learning rate $\eta > 0$, regularisation parameter $\lambda \geq 0$, discount factor $\gamma \in [0,1)$, failure probability $\delta \in (0,1)$, upper bounds $C_2 \geq \max_{s,a} \norm{\nabla_a \hat{Q}(s,a)  \vert_{a=\mu(s)}}{2}$ and $\kappa_{2}^{\text{max}} \geq \max_{s} \norm{\kappa(s,:)}{2}$, the number of representers $N$, action dimensionality $A$, parameter dimensionality $d=NA$, horizon $T$, number of iterations $N_{\text{it}}$.
\State \textbf{Output:} near-optimal policy $\pi$.
\State Define $n_1 = \BigO\left(\frac{\kappa_{2}^{\text{max}} C_2\log(d/\delta)}{(1-\gamma) \epsilon}\right)$
\State $\mathcal{B} = \emptyset$.
\For {$i = 1,\dots, N_{\text{it}}$}
\LineComment{Estimate policy gradient (Eq.~\ref{eq: approx DAC}) and update policy}
\State Define binary oracle $U_{X,\mathcal{S}}: \vert 0 \rangle \to \sum_{s} \sqrt{\tilde{\nu}(s)} \vert \nabla_a \hat{Q}(s,a) \vert_{a=\mu(s)} K(s,:) \rangle$ according to Lemma~\ref{lem: U_X} and Fig.~\ref{fig: U_X} ($T$ interactions with environment per call).
\State Perform quantum multivariate Monte Carlo with $X(s) = \nabla_a \hat{Q}(s,a) \vert_{a=\mu(s)} K(s,:) $ based on $n_1$ queries of $U_{X,\mathcal{S}}$, according to Theorem~\ref{th: dpg query complexity}.
\State Obtain the final estimate $\bar{X} \approx \mathbb{E}[X] = \nabla_{\beta}V(d_0)$ from quantum multivariate Monte Carlo.
\State Compute update: $\beta \inc \eta \bar{X}$; $\mu = \sum_{i=1}^{N} \beta_i K(\cdot,c_i)$.
\LineComment{Update critic (Eq.~\ref{eq: critic MSE})}
\State Apply $n_2 = n_1$ calls to the trajectory oracle and return oracle, measuring the full trajectory with rewards $\{s_0,a_0,r_0\dots,s_{T-1},a_{T-1},r_{T-1}\}_{i=1}^{n_2}$ ($T$ interactions with environment per call).
\State Add trajectory and rewards to buffer $\mathcal{B}$.
\State Define distribution over buffer (e.g. uniform, occupancy-based or prioritised).
\State Classical kernel regression for the critic based on buffer $\mathcal{B}$ and bootstrapping with target critic $\hat{Q}(s,\mu^{-}(s); v^{-},w^{-})$:
\begin{align*}
\hat{Q}(s,a) &= \langle w, (a - \mu(s))^{\intercal} K(s, \cdot)  \rangle + v^{\intercal} \phi(s) \\
             &= \argmin_{\hat{Q} \in \mathcal{H}_{K_{\mu}}} \mathbb{E}_{s,a,r,s' \sim \mathcal{B}} \left[\left(\hat{Q}(s,a) -  (r + \gamma \hat{Q}(s',\mu^{-}(s'); v^{-},w^{-})) \right)^2 \right] + \lambda \norm{\hat{Q}}{\mathcal{H}_{K_{\mu}}}^2 \,.
\end{align*}
\LineComment{Optional and periodic updates} 
\State Remove proportion of old data in $\mathcal{B}$ (periodically).
\State Sparsify policy (periodically, optional): kernel matching pursuit, with tolerance $\epsilon_{\mu}$.
\State Update number of queries (periodically, optional) $n_1 = \BigO\left(\frac{\kappa_{2}^{\text{max}} C_2 \log(d/\delta)}{(1-\gamma) \epsilon}\right)$ based on new parameter dimension and kernel vector upper bound.
\EndFor
\end{algorithmic}
\end{algorithm}
\clearpage
\begin{algorithm}[htbp]
\caption{CQRNAC algorithm} \label{alg: quantum NAC}
\begin{algorithmic}[1]
\State \textbf{Input:} initial critic $\hat{Q}$ and upper bound $E \geq \max_{s,a,\tau} \vert \hat{Q}(s,a) - R(s,a,\tau)  \vert$, error tolerance for policy gradient $\epsilon > 0$, learning rates $\eta,\eta' > 0$, covariance shrinkage $\alpha \in (0,1)$, discount factor $\gamma \in [0,1)$, failure probability $\delta \in (0,1)$, upper bound on the $2$-norm of the partial derivatives of the log-policy $B_2$,  number of policy centres $N$, action dimensionality $A$, parameter dimensionality $d=NA$, horizon $T$, number of iterations $N_{\text{it}}$.
\State \textbf{Output:} near-optimal policy $\pi$
\State Define $n = \BigO\left(\frac{E B_2 \log(d/\delta)}{\epsilon}\right)$ (Theorem~\ref{th: natural actor-critic query complexity})
\For{$i = 1,\dots, N_{\text{it}}$}
\LineComment{Estimate the critic's gradient (Eq.~\ref{eq: expectation CQRNAC}), then update critic and policy}
\State Define binary oracle $U_{\nabla L}: \vert 0 \rangle \to \sum_{s,a} \sqrt{\tilde{\nu}(s,a)}\ket{R'(s,a, \tau_{1:T-1})\phi(s,a)}$ according to Eq.~\ref{eq: UnablaL} ($2T-1$ interactions with environment per call).
\State Perform quantum multivariate Monte Carlo with $X(s,a) = R'(s,a, \tau_{1:T-1})\phi(s,a)$ based on $n$ queries of $U_{\nabla L}$, following Theorem~\ref{th: natural actor-critic query complexity}.
\State Obtain the final estimate $\bar{X} \approx \mathbb{E}[X] = \nabla_{w} L(\hat{Q})$ from quantum multivariate Monte Carlo.
\State Critic update: $ w \gets w - \eta' \bar{X}$; $\hat{Q}(s,a) \gets \langle w, K(s,:)\Sigma^{-1} (a- \mu(s)) \rangle$ for all $(s,a) \in \mathcal{S} \times \mathcal{A}$.
\State Natural policy gradient update: $\beta \gets \beta + \eta w$; $\mu = \sum_{i=1}^{N} \beta_i K(\cdot,c_i)$.
\LineComment{Optional and periodic updates} 
\State Update number of queries (optional): $n = \BigO\left(\frac{E B_2 \log(d/\delta)}{\epsilon}\right)$ based on new upper bound $E$.
\State Shrink covariance matrix (optional): $\Sigma \gets \Sigma * \alpha$.
\EndFor
\end{algorithmic}
\end{algorithm}

\end{appendices}

\bibliographystyle{apalike}
\bibliography{library}

\begin{thebibliography}{}

\bibitem[Agarwal et~al., 2021]{Agarwal2021}
Agarwal, A., Kakade, S.~M., Lee, J.~D., and Mahajan, G. (2021).
\newblock {On the theory of policy gradient methods: Optimality, approximation,
  and distribution shift}.
\newblock {\em Journal of Machine Learning Research}, 22:1--76.

\bibitem[Aharonov et~al., 1998]{Aharonov1998}
Aharonov, D., Kitaev, A., and Nisan, N. (1998).
\newblock Quantum circuits with mixed states.
\newblock In {\em Proceedings of the Thirtieth Annual ACM Symposium on Theory
  of Computing (STOC 98)}, page 20–30, New York, NY, USA. Association for
  Computing Machinery.

\bibitem[Bagnell and Schneider, 2003]{Bagnell2003}
Bagnell, J.~A. and Schneider, J. (2003).
\newblock {Policy Search in Kernel Hilbert Space}.
\newblock {\em Tech. Rep. RI–TR-03–45}.

\bibitem[Brassard et~al., 2000]{brassard2000quantum}
Brassard, G., Hoyer, P., Mosca, M., and Tapp, A. (2000).
\newblock Quantum amplitude amplification and estimation.
\newblock {\em arXiv preprint quant-ph/0005055}.

\bibitem[Canatar et~al., 2022]{Canatar2022}
Canatar, A., Peters, E., Pehlevan, C., Wild, S.~M., and Shaydulin, R. (2022).
\newblock Bandwidth enables generalization in quantum kernel models.
\newblock {\em Transactions on Machine Learning Research}, pages 1--31.

\bibitem[Cen et~al., 2022]{Cen2022}
Cen, S., Cheng, C., Chen, Y., Wei, Y., and Chi, Y. (2022).
\newblock Fast global convergence of natural policy gradient methods with
  entropy regularization.
\newblock {\em Operations Research}, 70(4):2563--2578.

\bibitem[Chen, 2023]{Chen2023}
Chen, S. Y.~C. (2023).
\newblock {Asynchronous training of quantum reinforcement learning}.
\newblock {\em Procedia Computer Science}, 222:321--330.

\bibitem[Cornelissen, 2019]{Cornelissen2019}
Cornelissen, A. (2019).
\newblock {Quantum gradient estimation of Gevrey functions}.
\newblock {\em arXiv preprint arXiv:1909.13528}, pages 1--48.

\bibitem[Cornelissen et~al., 2022]{Cornelissen2022}
Cornelissen, A., Hamoudi, Y., and Jerbi, S. (2022).
\newblock Near-optimal quantum algorithms for multivariate mean estimation.
\newblock In {\em {Annual ACM SIGACT Symposium on Theory of Computing (STOC
  2022)}}, pages 33--43, New York, NY, USA. Association for Computing
  Machinery.

\bibitem[de~Carvalho et~al., 2024]{Carvalho2024}
de~Carvalho, J.~A., Batista, C.~A., de~Veras, T. M.~L., Araujo, I.~F., and
  da~Silva, A.~J. (2024).
\newblock Quantum multiplexer simplification for state preparation.

\bibitem[Dong et~al., 2008]{Dong2008}
Dong, D., Chen, C., Li, H., and Tarn, T.~J. (2008).
\newblock {Quantum reinforcement learning}.
\newblock {\em IEEE Transactions on Systems, Man, and Cybernetics, Part B:
  Cybernetics}, 38(5):1207--1220.

\bibitem[Dunjko et~al., 2017]{Dunjko2017}
Dunjko, V., Liu, Y.-K., Wu, X., and Taylor, J.~M. (2017).
\newblock {Exponential improvements for quantum-accessible reinforcement
  learning}.
\newblock {\em arXiv preprint arXiv:1710.11160}, pages 1--27.

\bibitem[Elliott et~al., 2022]{elliott2022quantum}
Elliott, T.~J., Gu, M., Garner, A.~J., and Thompson, J. (2022).
\newblock Quantum adaptive agents with efficient long-term memories.
\newblock {\em Physical Review X}, 12(1):011007.

\bibitem[Gily{\'{e}}n et~al., 2019]{Gilyen2019}
Gily{\'{e}}n, A., Arunachalam, S., and Wiebe, N. (2019).
\newblock {Optimizing quantum optimization algorithms via faster quantum
  gradient computation}.
\newblock In {\em {Annual ACM-SIAM Symposium on Discrete Algorithms (SODA
  2019)}}, pages 1425--1444.

\bibitem[Hamoudi, 2021]{Hamoudi2021}
Hamoudi, Y. (2021).
\newblock {Quantum Sub-Gaussian Mean Estimator}.
\newblock In Mutzel, P., Pagh, R., and Herman, G., editors, {\em {Annual
  European Symposium on Algorithms (ESA 2021)}}, volume 204 of {\em {Leibniz
  International Proceedings in Informatics (LIPIcs)}}, pages 50:1--50:17.
  Schloss Dagstuhl -- Leibniz-Zentrum f{\"u}r Informatik.

\bibitem[Haug et~al., 2021]{Haug2021b}
Haug, T., Bharti, K., and Kim, M. (2021).
\newblock Capacity and quantum geometry of parametrized quantum circuits.
\newblock {\em PRX Quantum}, 2:040309.

\bibitem[Hopkins, 2020]{Hopkins2020}
Hopkins, S.~B. (2020).
\newblock {Mean estimation with sub-Gaussian rates in polynomial time}.
\newblock {\em Annals of Statistics}, 48(2):1193--1213.

\bibitem[Jerbi et~al., 2023a]{Jerbi2022}
Jerbi, S., Cornelissen, A., Ozols, M., and Dunjko, V. (2023a).
\newblock {Quantum policy gradient algorithms}.
\newblock In {\em {Conference on the Theory of Quantum Computation,
  Communication and Cryptography (TQC 2023)}}, pages 1--24.

\bibitem[Jerbi et~al., 2023b]{Jerbi2023}
Jerbi, S., Fiderer, L.~J., {Poulsen Nautrup}, H., K{\"{u}}bler, J.~M., Briegel,
  H.~J., and Dunjko, V. (2023b).
\newblock {Quantum machine learning beyond kernel methods}.
\newblock {\em Nature Communications}, 14(1):1--8.

\bibitem[Jerbi et~al., 2021]{Jerbi2021}
Jerbi, S., Gyurik, C., Marshall, S.~C., and Briegel, H.~J. (2021).
\newblock {Parametrized Quantum Policies for Reinforcement Learning}.
\newblock In {\em {Advances in Neural Information Processing (NeurIPS 2021)}},
  pages 1--14.

\bibitem[Jordan, 2005]{Jordan2005}
Jordan, S.~P. (2005).
\newblock Fast quantum algorithm for numerical gradient estimation.
\newblock {\em Physical Review Letters}, 95:050501.

\bibitem[Kahanamoku-Meyer et~al., 2025]{kahanamoku2025log}
Kahanamoku-Meyer, G.~D., Blue, J., Bergamaschi, T., Gidney, C., and Chuang,
  I.~L. (2025).
\newblock A log-depth in-place quantum fourier transform that rarely needs
  ancillas.
\newblock {\em arXiv preprint arXiv:2505.00701}.

\bibitem[Kakade, 2002]{Kakade2002}
Kakade, S. (2002).
\newblock {A Natural Policy Gradient}.
\newblock In {\em {Advances in Neural Information Processing Systems
  (NeurIPS2002)}}, pages 1057--1063.

\bibitem[Kandala et~al., 2017]{Kandala2017}
Kandala, A., Mezzacapo, A., Temme, K., Takita, M., Brink, M., Chow, J.~M., and
  Gambetta, J.~M. (2017).
\newblock {Hardware-efficient variational quantum eigensolver for small
  molecules and quantum magnets}.
\newblock {\em Nature}, 549(7671):242--246.

\bibitem[Kingma and Ba, 2015]{Kingma2015}
Kingma, D.~P. and Ba, J.~L. (2015).
\newblock {Adam: A method for Stochastic Optimisation}.
\newblock In {\em {International Conference on Learning Representations (ICLR
  2015)}}, pages 1--15.

\bibitem[Kitaev and Webb, 2008]{Kitaev2008}
Kitaev, A. and Webb, W.~A. (2008).
\newblock {Wavefunction preparation and resampling using a quantum computer}.
\newblock {\em arXiv preprint arXiv:0801.0342}, pages 1--8.

\bibitem[Lan, 2021]{Lan2021}
Lan, Q. (2021).
\newblock Variational quantum soft actor-critic.
\newblock {\em arXiv preprint arXiv:2112.11921}, pages 1--8.

\bibitem[Lever and Stafford, 2015]{Lever2015}
Lever, G. and Stafford, R. (2015).
\newblock {Modelling policies in MDPs in reproducing kernel Hilbert space}.
\newblock In {\em {Proceedings of the International Conference on Artificial
  Intelligence and Statistics (AISTATS 2015)}}, volume~38, pages 590--598.

\bibitem[Lillicrap et~al., 2016]{Lillicrap2016}
Lillicrap, T.~P., Hunt, J.~J., Pritzel, A., Heess, N., Erez, T., Tassa, Y.,
  Silver, D., and Wierstra, D. (2016).
\newblock Continuous control with deep reinforcement learning.
\newblock In Bengio, Y. and LeCun, Y., editors, {\em {International Conference
  on Learning Representations ({ICLR} 2016)}}.

\bibitem[Mallat and Zhang, 1993]{Mallat1993}
Mallat, S. and Zhang, Z. (1993).
\newblock {Matching Pursuits with Time-Frequency Dictionaries}.
\newblock {\em IEEE Transactions on Signal Processing}, 41(12):3397--3415.

\bibitem[Markov et~al., 2022]{Markov2022}
Markov, V., Stefanski, C., Rao, A., and Gonciulea, C. (2022).
\newblock {A Generalized Quantum Inner Product and Applications to Financial
  Engineering}.
\newblock {\em arXiv preprint arXiv:2201.09845}, pages 1--17.

\bibitem[McClean et~al., 2018]{McClean2018}
McClean, J.~R., Boixo, S., Smelyanskiy, V.~N., Babbush, R., and Neven, H.
  (2018).
\newblock {Barren plateaus in quantum neural network training landscapes}.
\newblock {\em Nature Communications}, 9(1):1--6.

\bibitem[Meyer et~al., 2023]{Meyer2023}
Meyer, N., Scherer, D.~D., Plinge, A., Mutschler, C., and Hartmann, M.~J.
  (2023).
\newblock {Quantum Natural Policy Gradients: Towards Sample-Efficient
  Reinforcement Learning}.
\newblock In {\em {IEEE International Conference on Quantum Computing and
  Engineering (QCE 2023)}}, volume~2, pages 36--41.

\bibitem[Meyer et~al., 2022]{Meyer2022}
Meyer, N., Ufrecht, C., Periyasamy, M., Scherer, D.~D., Plinge, A., and
  Mutschler, C. (2022).
\newblock {A Survey on Quantum Reinforcement Learning}.
\newblock {\em arXiv preprint arXiv:2211.03464}, pages 1--83.

\bibitem[Mnih et~al., 2016]{Mnih2016a}
Mnih, V., Badia, A.~P., Mirza, M., Graves, A., Lillicrap, T.~P., Harley, T.,
  Silver, D., and Kavukcuoglu, K. (2016).
\newblock {Asynchronous Methods for Deep Reinforcement Learning}.
\newblock In {\em {International Conference on Machine Learning (ICML 2016)}},
  volume~48, New York, NY, USA.

\bibitem[Mnih et~al., 2015]{Mnih2015}
Mnih, V., Kavukcuoglu, K., Silver, D., Rusu, A.~A., Veness, J., Bellemare,
  M.~G., Graves, A., Riedmiller, M., Fidjeland, A.~K., Ostrovski, G., Petersen,
  S., Beattie, C., Sadik, A., Antonoglou, I., King, H., Kumaran, D., Wierstra,
  D., Legg, S., and Hassabis, D. (2015).
\newblock {Human-level control through deep reinforcement learning}.
\newblock {\em Nature}, 518(7540):529--533.

\bibitem[Montanaro, 2017]{Montanaro2017}
Montanaro, A. (2017).
\newblock {Quantum speedup of Monte Carlo methods}.
\newblock {\em Proceedings of the Royal Society A}, 471:20150301.

\bibitem[Nakaji and Yamamoto, 2021]{Nakaji2021}
Nakaji, K. and Yamamoto, N. (2021).
\newblock {Expressibility of the alternating layered ansatz for quantum
  computation}.
\newblock {\em Quantum}, 5:1--20.

\bibitem[P{\'{e}}rez-Salinas et~al., 2020]{Perez-Salinas2020}
P{\'{e}}rez-Salinas, A., Cervera-Lierta, A., Gil-Fuster, E., and Latorre, J.~I.
  (2020).
\newblock {Data re-uploading for a universal quantum classifier}.
\newblock {\em Quantum}, 4:226.

\bibitem[P\'erez-Salinas et~al., 2021]{Perez2021}
P\'erez-Salinas, A., L\'opez-N\'u\~nez, D., Garc\'{\i}a-S\'aez, A.,
  Forn-D\'{\i}az, P., and Latorre, J.~I. (2021).
\newblock One qubit as a universal approximant.
\newblock {\em Phys. Rev. A}, 104:012405.

\bibitem[Peters and Schaal, 2008a]{Peters2008}
Peters, J. and Schaal, S. (2008a).
\newblock Natural actor-critic.
\newblock {\em Neurocomputing}, 71(7):1180--1190.
\newblock Progress in Modeling, Theory, and Application of Computational
  Intelligenc.

\bibitem[Peters and Schaal, 2008b]{Peters2008a}
Peters, J. and Schaal, S. (2008b).
\newblock {Reinforcement learning of motor skills with policy gradients}.
\newblock {\em Neural Networks}, 21(4):682--697.

\bibitem[Puterman, 1994]{Puterman1994}
Puterman, M. (1994).
\newblock {\em {Markov Decision Processes: Discrete Stochastic Dynamic
  Programming}}.
\newblock Wiley.

\bibitem[Rall and Fuller, 2023]{Rall2023amplitudeestimation}
Rall, P. and Fuller, B. (2023).
\newblock Amplitude {E}stimation from {Q}uantum {S}ignal {P}rocessing.
\newblock {\em {Quantum}}, 7:937.

\bibitem[Sch{\"{o}}lkopf and Smola, 2003]{Scholkopf2003}
Sch{\"{o}}lkopf, B. and Smola, A.~J. (2003).
\newblock {\em {Learning With Kernels: Support Vector Machines, Regularization,
  Optimization, and Beyond}}.
\newblock MIT Press.

\bibitem[Schuld, 2021]{Schuld2021}
Schuld, M. (2021).
\newblock {Supervised quantum machine learning models are kernel methods}.
\newblock {\em arXiv preprint arxiv:2101.11020}, pages 1--25.

\bibitem[Schuld et~al., 2019]{Schuld2019a}
Schuld, M., Bergholm, V., Gogolin, C., Izaac, J., and Killoran, N. (2019).
\newblock {Evaluating analytic gradients on quantum hardware}.
\newblock {\em Physical Review A}, 99(3):032331.

\bibitem[Schuld and Killoran, 2019]{Schuld2019}
Schuld, M. and Killoran, N. (2019).
\newblock Quantum machine learning in feature hilbert spaces.
\newblock {\em Physics Review Letters}, 122:040504.

\bibitem[Sequeira et~al., 2023]{Sequeira2023a}
Sequeira, A., Santos, L.~P., and Barbosa, L.~S. (2023).
\newblock {Policy gradients using variational quantum circuits}.
\newblock {\em Quantum Machine Intelligence}, 5(18):18.

\bibitem[Sequeira et~al., 2024]{Sequeira2024}
Sequeira, A., Santos, L.~P., and Barbosa, L.~S. (2024).
\newblock {On Quantum Natural Policy Gradients}.
\newblock {\em arXiv preprint arXiv:2401.08307}, pages 1--14.

\bibitem[Shende et~al., 2006]{Shende2006}
Shende, V., Bullock, S., and Markov, I. (2006).
\newblock Synthesis of quantum-logic circuits.
\newblock {\em IEEE Transactions on Computer-Aided Design of Integrated
  Circuits and Systems}, 25(6):1000–1010.

\bibitem[Silver et~al., 2014]{Silverb}
Silver, D., Lever, G., Heess, N., Degris, T., Wierstra, D., and Riedmiller, M.
  (2014).
\newblock {Deterministic Policy Gradient Algorithms}.
\newblock In {\em {International Conference on Machine Learning (ICML 2014)}},
  pages 1--9, Bejing, China.

\bibitem[Sutton and Barto, 2018]{Sutton2018b}
Sutton, R.~S. and Barto, A.~G. (2018).
\newblock {\em {Reinforcement learning: an introduction}}.
\newblock MIT Press, Cambridge, MA.

\bibitem[Suzuki et~al., 2024]{Suzuki}
Suzuki, Y., Kawaguchi, H., and Yamamoto, N. (2024).
\newblock {Quantum Fisher kernel for mitigating the vanishing similarity
  issue}.
\newblock {\em Quantum Science and Technology}, 9(3):035050.

\bibitem[Suzuki et~al., 2020]{suzuki2020amplitude}
Suzuki, Y., Uno, S., Raymond, R., Tanaka, T., Onodera, T., and Yamamoto, N.
  (2020).
\newblock Amplitude estimation without phase estimation.
\newblock {\em Quantum Information Processing}, 19(2):75.

\bibitem[Tan et~al., 2025]{Tan2025}
Tan, K.~C., Liu, D., Bharti, K., Bhowmick, D., and Sengupta, P. (2025).
\newblock Probing quantum phase transitions via short-depth quantum circuits
  for estimating quantum coherence and discrete berry phases.
\newblock {\em Physical Review B}, 112:014108.

\bibitem[Thanasilp et~al., 2022]{Thanasilp2022}
Thanasilp, S., Wang, S., Cerezo, M., and Holmes, Z. (2022).
\newblock {Exponential concentration and untrainability in quantum kernel
  methods}.
\newblock {\em Nature Communications}, 15(1):5200.

\bibitem[Tuo et~al., 2020]{Tuo2020}
Tuo, R., Wang, Y., and {Jeff Wu}, C.~F. (2020).
\newblock {On the Improved Rates of Convergence for Mat{\'{e}}rn-Type Kernel
  Ridge Regression with Application to Calibration of Computer Models}.
\newblock {\em {SIAM/ASA Journal on Uncertainty Quantification}},
  8(4):1522--1547.

\bibitem[van Apeldoorn, 2021]{VanApeldoorn2021}
van Apeldoorn, J. (2021).
\newblock {Quantum probability oracles \& multidimensional amplitude
  estimation}.
\newblock In {\em {Leibniz International Proceedings in Informatics (LIPIcs
  2021)}}, volume 197, pages 9:1--9:11. Schloss Dagstuhl – Leibniz-Zentrum
  f{\"{u}}r Informatik, Dagstuhl Publishing, Germany.

\bibitem[Vincent and Bengio, 2002]{Vincent2002}
Vincent, P. and Bengio, Y. (2002).
\newblock {Kernel matching pursuit}.
\newblock {\em Machine Learning}, 48(1-3):165--187.

\bibitem[Wang and Jing, 2022]{Wang2022}
Wang, W. and Jing, B.-y. (2022).
\newblock {Gaussian process regression: Optimality, robustness, and
  relationship with kernel ridge regression}.
\newblock {\em Journal of Machine Learning Research}, 23(193):1--67.

\bibitem[Wu et~al., 2020]{Wu2020}
Wu, S., Jin, S., Wen, D., Han, D., and Wang, X. (2020).
\newblock {Quantum reinforcement learning in continuous action space}.
\newblock {\em arXiv preprint arXiv:2012.10711}, pages 1--15.

\end{thebibliography}


\newcommand{\etalchar}[1]{$^{#1}$}
\begin{thebibliography}{SBG{\etalchar{+}}19}

\bibitem[CC21]{Cerezo2021}
M~Cerezo and Patrick~J Coles.
\newblock {Higher order derivatives of quantum neural networks with barren
  plateaus}.
\newblock {\em Quantum Science and Technology}, 6(3):035006, 2021.

\bibitem[CHJ22]{Cornelissen2022}
Arjan Cornelissen, Yassine Hamoudi, and Sofiene Jerbi.
\newblock Near-optimal quantum algorithms for multivariate mean estimation.
\newblock In {\em {Annual ACM SIGACT Symposium on Theory of Computing (STOC
  2022)}}, pages 33--43, New York, NY, USA, 2022. Association for Computing
  Machinery.

\bibitem[Cor19]{Cornelissen2019}
Arjan Cornelissen.
\newblock {Quantum gradient estimation of Gevrey functions}.
\newblock {\em arXiv preprint arXiv:1909.13528}, pages 1--48, 2019.

\bibitem[Hop20]{Hopkins2020}
Samuel~B. Hopkins.
\newblock {Mean estimation with sub-Gaussian rates in polynomial time}.
\newblock {\em Annals of Statistics}, 48(2):1193--1213, 2020.

\bibitem[JCOD23]{Jerbi2022}
Sofiene Jerbi, Arjan Cornelissen, Māris Ozols, and Vedran Dunjko.
\newblock {Quantum policy gradient algorithms}.
\newblock In {\em {Conference on the Theory of Quantum Computation,
  Communication and Cryptography (TQC 2023)}}, pages 1--24, 2023.

\bibitem[LM19]{Lugosi2019}
G{\'{a}}bor Lugosi and Shahar Mendelson.
\newblock {Mean Estimation and Regression Under Heavy-Tailed Distributions: A
  Survey}.
\newblock {\em Foundations of Computational Mathematics}, 19(5):1145--1190,
  2019.

\bibitem[LS15]{Lever2015}
Guy Lever and Ronnie Stafford.
\newblock {Modelling policies in MDPs in reproducing kernel Hilbert space}.
\newblock In {\em {Proceedings of the International Conference on Artificial
  Intelligence and Statistics (AISTATS 2015)}}, volume~38, pages 590--598,
  2015.

\bibitem[MZ93]{Mallat1993}
Stephane Mallat and Zhifeng Zhang.
\newblock {Matching Pursuits with Time-Frequency Dictionaries}.
\newblock {\em IEEE Transactions on Signal Processing}, 41(12):3397--3415,
  1993.

\bibitem[SBG{\etalchar{+}}19]{Schuld2019a}
Maria Schuld, Ville Bergholm, Christian Gogolin, Josh Izaac, and Nathan
  Killoran.
\newblock {Evaluating analytic gradients on quantum hardware}.
\newblock {\em Physical Review A}, 99(3):032331, 2019.

\bibitem[SBM06]{Shende2006}
V.V. Shende, S.S. Bullock, and I.L. Markov.
\newblock Synthesis of quantum-logic circuits.
\newblock {\em IEEE Transactions on Computer-Aided Design of Integrated
  Circuits and Systems}, 25(6):1000–1010, June 2006.

\bibitem[Sta23]{stack}
Stackoverflow.
\newblock
  https://stats.stackexchange.com/questions/15978/variance-of-product-of-dependent-variables,
  2023.

\end{thebibliography}

\end{document}


\maketitle

\section{Classical multivariate Monte Carlo}
\label{app: classical MC}
For Multivariate Monte Carlo, an estimate $\bar{X} = \frac{1}{n} \sum_{i=1}^{n} X^{i}$ of the expectation $\mathbb{E}[X]$ is obtained by taking $n$ \textit{iid} samples of a $d$-dimensional random variable $X$. For error tolerance $\epsilon > 0$, an erroneous estimate can be defined by having at least one dimension with error greater than $\epsilon$, leading to the following upper bound:
\begin{align*}
\mathbb{P}(\vert \vert \bar{X} - \mathbb{E}[X] \vert \vert_{\infty} \geq \epsilon) 
&\leq \sum_{i=1}^{d} \mathbb{P}( \vert \bar{X}_i -\mathbb{E}[X_i] \vert \geq \epsilon)   \tag{union bound} \\
&\leq d \times \max_{j} \mathbb{P}( \vert \bar{X}_j - \mathbb{E}[X_j]  \vert \geq \epsilon) \\
&\leq 2 d \exp\left( - \frac{2n^2\epsilon^2}{4nB^2}\right)  \text{(Hoeffding inequality and $X_i \in [-B,B]$ for all $i \in [d]$} \\
&= \delta \,.
\end{align*}
Therefore the number of samples required for an error at most $\epsilon$ and failure rate at most $\delta$ is
\begin{align*}
 \delta &= 2 d \exp\left(-\frac{n\epsilon^2}{2B^2}\right) \\
n &= \frac{2B^2}{\epsilon^2}\log(2d/\delta) \\
  &= \BigO\left(\frac{B^2}{\epsilon^2} \log(d/\delta)\right) \\
  &= \tildeO\left(\frac{B^2}{\epsilon^2}\right) \,.
\end{align*} \qed

\section{Functional log-policy gradient} 
\label{app: log-policy grad}
The gradient of the log-policy of the Gaussian RKHS policy,
\begin{align*}
\pi(a \vert s) &= \mathcal{N}(\mu(s),\Sigma)  \nonumber \\
               &= \frac{1}{Z} \exp\left(-\frac{1}{2} (\mu(s) - a)^{\intercal} \Sigma^{-1} (\mu(s) - a)\right) \,,
\end{align*}
with respect to the parameters can be derived following the proof in Lever and Stafford \cite{Lever2015}. 

Define $g: \mathcal{H}_K \to \mathbb{R} : \mu \to \log(\pi(a \vert s))$ and the Fr\'{e}chet derivative as a bounded linear map $Dg\vert_{\mu}: \mathcal{H} \to \mathbb{R}$ according to 
\begin{equation}
\label{eq: frechet}
\lim_{\norm{h}{} \to 0} \dfrac{\norm{g(\mu + h) - g(\mu)}{\mathbb{R}} - Dg\vert_{\mu}(h)}{\norm{h}{\mathcal{H}_K}} = 0 \,.
\end{equation}
We want to find the gradient of $g$ with respect to $\mu$. To do so, we note that the Fr\'{e}chet derivative relates to the gradient through the inner product 
\begin{equation}
Dg\vert_{\mu} = \langle \nabla g , h(\cdot) \rangle \,.
\end{equation}
In our setting, this gives
\begin{align} 
\label{eq: Dg}
Dg\vert_{\mu}: & h \to (a - \mu(s) ) \Sigma^{-1} h(s)  \nonumber  \\
             &= \langle K(s, \cdot)\Sigma^{-1} (a - \mu(s)) , h(\cdot) \rangle  \,,
\end{align}
two forms which are equivalent due to the reproducing property.

Expanding $g$, we get
\begin{align*} 
g(\mu + h) &= \log\left(\frac{1}{Z} \exp\left[-\frac{1}{2} (\mu(s) + h(s) - a)^{\intercal} \Sigma^{-1} (\mu(s)+h(s) - a)  \right]\right) \\
		   &= -\log(Z) - \frac{1}{2} (\mu(s) + h(s) - a)^{\intercal} \Sigma^{-1} (\mu(s) + h(s) - a) \,,
\end{align*}
where $Z$ is the normalisation constant, and 
\begin{align*} 
g(\mu) &= \log\left(\frac{1}{Z} \exp\left[ \frac{1}{2} (\mu(s) - a)^{\intercal} \Sigma^{-1} (\mu(s) - a)  \right]\right) \\
		   &= -\log(Z) - \frac{1}{2} (\mu(s) - a)^{\intercal} \Sigma^{-1} (\mu(s) - a) \,. 
\end{align*}

Note that due to cancelling $\log(Z)$ and $\mu(s) -a$ terms,
\begin{align*}
g(\mu + h) - g(\mu) = h(s)^{\intercal} \Sigma^{-1} h(s) \,.
\end{align*}

Now evaluate the criterion for Fr\'{e}chet differentiability:
\begin{align*} 
 &\lim_{\norm{h}{} \to 0} \dfrac{\norm{g(\mu + h) - g(\mu) - Dg\vert_{\mu}(h)}{\mathbb{R}}}{\norm{h}{\mathcal{H}_K}} \\
 &= \lim_{\norm{h}{} \to 0} \dfrac{\norm{g(\mu + h) - g(\mu) - \langle K(s, \cdot)\Sigma^{-1} (a - \mu(s)) , h(\cdot) \rangle}{\mathbb{R}}}{\norm{h}{\mathcal{H}_K}} \tag{definition of $Dg\vert_{\mu}(h)$ in Eq.~\ref{eq: Dg}} \\
&=\lim_{\norm{h}{} \to 0} \dfrac{\norm{-h(s)^{\intercal}\Sigma^{-1} h(s) - 2h(s)^{\intercal} \Sigma^{-1} (\mu(s) - a) - 2\langle K(s, \cdot)\Sigma^{-1} (a - \mu(s)) , h(\cdot) \rangle}{\mathbb{R}}}{2\norm{h}{\mathcal{H}_K}} \tag{subtracting $g(\mu + h) - g(\mu)$} \\
&= \lim_{\norm{h}{} \to 0} \dfrac{\norm{\langle K(s,\cdot) \left(\Sigma^{-1}h(s) - 2\Sigma^{-1} (\mu(s) - a) - 2\Sigma^{-1} (a - \mu(s))\right) , h(\cdot) \rangle_{\mathcal{H}_K}}{\mathbb{R}}}{2\norm{h}{\mathcal{H}_K}}  \tag{reproducing property}\\ 
&=\lim_{\norm{h}{} \to 0} \dfrac{\langle K(s,\cdot) \Sigma^{-1} h(s),h(\cdot) \rangle }{2\norm{h}{\mathcal{H}_K}} \tag{cancelling out terms}\\
&\leq \lim_{\norm{h}{} \to 0}  \dfrac{(\Sigma^{-1} h(s))^{\intercal} K(s,s) \Sigma^{-1} h(s) \, \norm{h}{\mathcal{H}_K}}{2\norm{h}{\mathcal{H}_K}} \tag{Cauchy-Schwarz} \\
&= \lim_{\norm{h}{} \to 0} (\Sigma^{-1} h(s))^{\intercal} K(s,s) \Sigma^{-1} h(s)  \\
&= 0 \,.
\end{align*}

Thus the gradient of the log-policy is indeed 
\begin{align*}
\nabla_{\mu} \log(\pi(a \vert s)) =  K(s, \cdot)\Sigma^{-1} (a - \mu(s)) \,.
\end{align*} 
\qed

\section{Vectorised gradient of log-policy} 
\label{app: analytical grad}
First note that
\begin{align*}
\nabla_{\beta} \mu(s) = (\partial_{\beta_1} \mu(s), \dots,\partial_{\beta_d} \mu(s) )  = (K(s,c_1),\dots,K(s,c_N)) \,. \\
\end{align*}
If the policy centres exhaust the state-space, i.e. $\{c\}_{i=1}^{N} = \mathcal{S}$, then this can be written as $K(s,\cdot)$. However, this is clearly not tractable in high-dimensional, continuous state spaces. Instead, we use the notation $K(s,:)$ below to denote vectorisation across the $N$ policy centres.
Therefore
\begin{align*}
\nabla_{\beta} \log(\pi(a \vert s)) &= \nabla_{\mu} \log(\pi(a \vert s))  \nabla_{\beta} \mu(s) \tag{chain rule} \nonumber \\
							  &= \nabla_{\mu} \left( \log(C e^{-\frac{1}{2}(a-\mu(s))^{\intercal} \Sigma^{-1} (a - \mu(s))}) \right)  K(s,:) \tag{product rule} \nonumber  \\
                            &= \left(\frac{1}{2}(a-\mu(s))^{\intercal} \Sigma^{-1} \mathbf{1} +\frac{1}{2} \mathbf{1}^{\intercal} \Sigma^{-1}*(a-\mu(s)) \right) K(s,:) \nonumber  \\
							  &= \left((a-\mu(s))^{\intercal} \Sigma^{-1}  \right)  K(s,:)   \in \mathbb{C}^{A \times N} \,.
\end{align*}
 \qed

\section{Compatible function approximation and the natural policy gradient}
\label{app: NPG}
Define a feature-map of the form $\phi: (s,a) \mapsto K(s,\cdot) \Sigma^{-1} (a - \mu(s)) \in \mathcal{H}_K$ and an associated scalar-valued kernel 
\begin{align*}
K_{\mu}((s,a),(s',a')) = K(s,s') \Sigma^{-1} (a - \mu(s)) \Sigma^{-1} (a' - \mu(s')) \,.
\end{align*}
Given that the kernel satisfies $K_{\mu}((s,a),(s',a')) = \langle \phi(s,a),\phi(s',a')\rangle$, its associated Hilbert space $\mathcal{H}_{K_{\mu}}$ has the reproducing property.

\textbf{1. There exists a $w^* \in \mathcal{H}_K$ such that 
\begin{align*}
\hat{Q}(s,a) = \langle w^*, K(s,\cdot) \Sigma^{-1} (a - \mu(s)) \rangle \in \mathcal{H}_{K_{\mu}}
\end{align*} is a compatible approximator.
} \\
Suppose that $\hat{Q}$ is defined as
\begin{align*}
\hat{Q} &= \argmin_{\hat{Q} \in \mathcal{H}_K} L(\hat{Q}) =  \frac{1}{2} \int \nu(z) \left( \hat{Q}(z) - Q(z) \right)^2 dz  \quad \in \mathcal{H}_{K_{\mu}}\,.
\end{align*}
Due to the reproducing property of $\mathcal{H}_{K_{\mu}}$, there exists a $w^*$ in the feature space $\mathcal{H}_K$ such that $\hat{Q}_{\pi_{\mu}}(z) = \langle w^*, \phi(z) \rangle$. It follows that 
\begin{align*}
\nabla_{w^*} \hat{Q}_{\pi_{\mu}}(s,a) = \phi(s,a) = K(s,\cdot) \Sigma^{-1} (a - \mu(s))\,.
\end{align*}
It follows that at a (possibly local) optimum, $w^*$ satisfies
\begin{align*}
0 = \nabla_{w} L(\hat{Q}) &= \int \nu(z) (Q(z) - \hat{Q}(s,a)) \nabla_{w} \hat{Q}(s,a) dz \\
					&= \int \nu(z) (Q(z) - \hat{Q}(s,a)) K(s,\cdot) \Sigma^{-1} (a - \mu(s)) dz \,.
\end{align*}
From the policy gradient theorem, the above equality, and the analytical gradient for $\nabla_{\mu} \log(\pi(a \vert s))=K(s,\cdot) \Sigma^{-1} (a - \mu(s))$ (see Appendix~\ref{app: log-policy grad}), it follows that the gradient of the value function allows substituting the state-action value with the critic prediction:
\begin{align*}
\nabla_{\mu} V(d_0) &= \int \nu(z) Q(z) \nabla_{\mu} \log(\pi(a \vert s))   dz \\
					&= \int \nu(z) \hat{Q}(s,a) \nabla_{\mu} \log(\pi(a \vert s))   dz \,.
\end{align*}
\qed

\textbf{2. $w^*$ is the natural policy gradient, i.e. $w^* = \mathcal{F}(\mu)^{-1}\nabla_{\mu} V(d_0)$, where $\mathcal{F}(\mu)$ is the Fisher information.}\\
Note that the Fisher information is given by 
\begin{align*}
\mathcal{F}(\mu) &= \mathbb{E}\left[ \nabla_{\mu}  \log(\pi (a \vert s)) \nabla_{\mu} \log(\pi (a \vert s))^{\intercal} \right]   \\
	   &= \int \nu(z) \nabla_{\mu} \log(\pi (a \vert s))  \nabla_{\mu}\log(\pi (a \vert s))^{\intercal}  dz \,. \\
\end{align*}
Due to the compatible function approximation property of $\hat{Q}$ and the feature-map of the critic being related to the gradient of the log-policy, $\phi(s,a) = \nabla_{\mu} \log(\pi (a \vert s)) = K(s,\cdot) \Sigma^{-1} (a - \mu(s))$, it follows that 
\begin{align*}
&\int \nu(s,a) \nabla_{\mu}  \log(\pi (a \vert s)) (\langle w^*,  K(s,\cdot) \Sigma^{-1} (a - \mu(s)) \rangle - Q(z) ) dz  \tag{Appendix~\ref{app: NPG}.1} \\
&\int \nu(s,a) \nabla_{\mu}  \log(\pi (a \vert s)) (\langle w^*, \nabla_{\mu} \log(\pi (a \vert s))^{\intercal}) \rangle - Q(z) ) dz 
\end{align*}
and therefore
\begin{align*}
\mathcal{F}(\mu) w^* = \int \nu(s,a) Q(z) \nabla_{\mu} \log(\pi (a \vert s)) dz = \nabla_{\mu} V(d_0)\,.
\end{align*}
In other words, $w^* = \mathcal{F}(\mu)^{-1}\nabla_{\mu} V(d_0)$. \qed

\section{GIP Representer PQC proof}
\label{app: GIP RPQC}
We apply the full circuit with random centre indices $l \in \{0,\dots,N-1\}$ on the first register, actions on the second register, and inner products on the remaining registers. Note that each subsequent rotation $l$ will be controlled on $\ket{j}\ket{0}$ where the second ket is read from the $l$'th inner product register. Therefore, each rotation $l=0,\dots,N-1$ corresponds to a mutually exclusive quantum experiment outcome (and a unique quantum policy weight $\ket{\beta_l'}$), and corresponds to a state of the form
\begin{align}
\label{eq: controlled rotation}
\ket{l} \ket{\psi_{\text{GIP}}(l)} := \ket{l} \left( \ket{\beta_l'} \bigotimes_{j<l} \ket{J_j} \ket{J_l^+} \bigotimes_{j=l+1}^{N} \ket{J_j}+  \ket{0}^{Ak}  \bigotimes_{j<l} \ket{J_j} \ket{J_l^-} \bigotimes_{j=l+1}^{N} \ket{J_j} \right) \,
\end{align}
where $\ket{J_j}$ is the $j$'th inner product register, $\ket{J_l^+} = \langle \phi(s) \vert \phi(c_l) \rangle \ket{0}^{n_f}$ and $\ket{J_l^-} = \sum_{i=1}^{2^{n_f}-1} c_{G_l}(i) \ket{i}$, where $c_{G_l}(:)$ is the $l$'th set of garbage amplitudes. Since the zero action does not contribute to the expectation, observe that
\begin{align*}
\sum_{a \in \mathcal{A}} a \langle \psi_{\text{GIP}}(l) \vert P_a \vert  \psi_{\text{GIP}}(l) \rangle &= \sum_{a \in \mathcal{A} \setminus \{\mathbf{0}\}} a \,  \langle \psi_{\text{GIP}}(l) \vert P_a \vert  \psi_{\text{GIP}}(l) \rangle \\
			&= \sum_{a \in \mathcal{A} \setminus \{\mathbf{0}\}} a \,  \psi_l(a)^2 \vert \langle \phi(s) \vert \phi(c_l) \rangle \vert^2  
\end{align*}
The main part to prove now is that the circuit prepares the state $\frac{1}{\sqrt{N}} \sum_{l=1}^{N} \ket{l} \ket{\psi_{\text{GIP}}(l)}$ and to derive the associated classical policy weight.

 Denoting $\ket{J_l} = \ket{J_l^+} + \ket{J_l^-}$ and applying the quantum circuit on eigenstate $\ket{s}$ yields
\begin{align*}
\ket{0}^{\otimes \log(N)} \ket{0}^{\otimes Ak}  \ket{0}^{\otimes N n_f} 	&\overset{H^{\otimes \log(N)}}{\to} \frac{1}{\sqrt{N}} \sum_{j=1}^{N} \ket{j-1}  \ket{0}^{\otimes Ak} \ket{0}^{\otimes N n_f}    \\
				&\overset{\mathbf{B}_1^{\dagger} \mathbf{A} \dots \mathbf{B}_N^{\dagger} \mathbf{A}}{\to} \frac{1}{\sqrt{N}} \sum_{j=1}^{N}   \ket{j-1}  \ket{0}^{\otimes Ak} \bigotimes_{l=1}^{N} \left( \langle \phi(s) \vert \phi(c_l) \rangle \ket{0}^{\otimes n_f} + \sum_{i=1}^{2^{n_f}-1} c_{G_l}(i) \ket{i} \right) \\
&\overset{CR_Y(\theta_{1,0}),\dots,CR_Y(\theta_{Ak,0})}{\to}  \frac{1}{\sqrt{N}} \left( \ket{\psi_{\text{GIP}}(0)} + \sum_{j=2}^{N}   \ket{j-1}  \ket{0}^{\otimes Ak}  \bigotimes_{l=1}^{N} \ket{J_l}  \right)  \tag{see Eq.~\ref{eq: controlled rotation})} \\
&\overset{CR_Y(\theta_{1,1}),\dots,CR_Y(\theta_{Ak,1})}{\to}   \frac{1}{\sqrt{N}}  \left( \sum_{j=1}^{2} \ket{\psi_{\text{GIP}}(j-1)}  + \sum_{j'=3}^N \ket{j'-1} \ket{0}^{Ak} \bigotimes_{l=1}^{N} \ket{J_l} \right)   \\ \\
				& \vdots \\
				&\overset{CR_Y(\theta_{1,N-1}),\dots,CR_Y(\theta_{Ak,N-1})}{\to} \frac{1}{\sqrt{N}}  \sum_{j=1}^{N} \ket{\psi_{\text{GIP}}(j-1)} = \Pi_{\theta}^{\text{GIP}} \ket{0}^{\otimes \log(N)} \ket{0}^{\otimes Ak}  \ket{0}^{\otimes N n_f} \,.
\end{align*}

Applying measurement in the computational basis on the action register, the expectation of the circuit on the action qubits can be written in terms of a representer formula depending on the kernel $\kappa(s,s') = \vert \langle \phi(s) \vert \phi(c_j) \rangle \vert^2$. That is,
\begin{align*}
\langle P \rangle_{s,\theta} &= \sum_{a \in \mathcal{A}} a \, \langle P_a \rangle_{s,\theta} \\ 
							&= \frac{1}{N}  \sum_{a \in \mathcal{A} \setminus \{\mathbf{0}\}} a \, \left(\sum_{j=1}^{N} \psi_l(a)^2 \vert \langle \phi(s) \vert \phi(c_j) \rangle \vert^2  \right) \\
                            &= \frac{1}{N}  \sum_{a \in \mathcal{A} \setminus \{\mathbf{0}\}} a \left(\sum_{j=1}^{N} \kappa(s,c_j) \psi_j(a)^2 \right) \tag{fidelity kernel} \\
                            &= \sum_{j=1}^{N} \kappa(s,c_j) \left( \frac{1}{N}   \sum_{a \in \mathcal{A} \setminus \{\mathbf{0}\}} \psi_j(a)^2  a  \right) \tag{rearranging} \\
                            &= \sum_{j=1}^{N} \kappa(s,c_j) \beta_j \tag{definition of the associated classical policy weight} \,.
\end{align*}
One may then convert between the real vector space and $\mathcal{A}$ with some loss of precision analogous to the proof of Lemma~4.2.
\qed

\section{Analytical policy gradient for softmax RKHS policy}
\label{app: gradient softmax RKHS policy}
The functional policy gradient of the policy gradient of the softmax RKHS policy is given by 
\begin{align*}
\nabla_f \log(\pi(a \vert s)) &= \nabla_f \log( \frac{1}{Z} e^{\mathcal{T} f(s,a)}) \\
									&= \nabla_f \left(\log(e^{\mathcal{T} f(s,a)}) -\log(Z) \right) \\
									&= \nabla_f \left(\mathcal{T} f(s,a) - \log(\int e^{\mathcal{T} f(s,a')} da')  \right) \\
								    &= \mathcal{T} \kappa((s,a),\cdot) - \nabla_f \left( \int e^{\mathcal{T} f(s,a')} da' \right) / Z \\
								    &= \mathcal{T} \kappa((s,a),\cdot) - \frac{1}{Z}  \int \nabla_f e^{\mathcal{T} f(s,a')} da' \\
                                   &= \mathcal{T} \kappa((s,a),\cdot) -   \frac{1}{Z}  \int e^{\mathcal{T} f(s,a')}  \nabla_f \mathcal{T} f(s,a')  da' \\
                                   &= \mathcal{T} \kappa((s,a),\cdot) -  \frac{1}{Z}  \int e^{\mathcal{T} f(s,a')}  \nabla_f \mathcal{T} f(s,a')  da' \\
                                   &= \mathcal{T} \kappa((s,a),\cdot) - \mathcal{T}  \int \pi(a' \vert s) \kappa((s,a'),\cdot)  da'  \\
                                   &= \mathcal{T} \kappa((s,a),\cdot) - \mathcal{T}  \mathbb{E}_{a' \sim \pi(\cdot \vert s)}[\kappa((s,a'),\cdot)] \\
                                   &= \mathcal{T} \left( \kappa((s,a),\cdot) - \mathbb{E}_{a' \sim \pi(\cdot \vert s)}[\kappa((s,a'),\cdot)] \right) \,.
\end{align*}
\qed

\section{The number of representers}
\label{app: setting N}

\textbf{Proof for the scalar-valued case.} 
The matching pursuit algorithm minimises residuals by successively adding new orthonormal basis functions $g \in \mathcal{G} \subset \mathcal{H}_{\kappa}$ until the stopping criterion is met. In RKHS, this amounts to choosing centres $c_j \in \mathcal{X}$ and then defining $g_j = \kappa(c_j,\cdot)$. Note that initially, the residual is defined as $R^0 = \mu$. At each iteration $j=0,\dots,N-1$, the weight $w_j = \langle R_j , g_j \rangle$ is based on the correlation between the basis function and the residual. The basis function $g_j \in \mathcal{H}_{\kappa}$ is selected based on the objective
\begin{align*}
\max_{g_j \in \mathcal{G}} \alpha \vert  \langle R_j , g_j \rangle \vert \,,
\end{align*}
which seeks to maximise the correlation between the residual and the feature function, and $\alpha \in [0,1]$ accounts for a suboptimal solution to the maximisation problem. 

The estimated function is defined as 
\begin{align*}
\hat{f}_{N} = \sum_{j=0}^{N-1} w_j g_j = f - R_N \,,
\end{align*}
and the residual $R_j \in \mathcal{H}$ is defined recursively as
\begin{align*}
R_j &= R_{j-1} -  w_{j-1} g_{j-1} \,.
\end{align*}
Note that the above implies that the efficiency of the algorithm is dependent on the correlation ratio, defined for any $f \in \mathcal{H}_{\kappa}$ as
\begin{align*}
\lambda(f) = \sup_{g \in \mathcal{G}} \frac{\langle f, g \rangle }{\norm{f}{L_2}} \,,
\end{align*}
which can be used to upper bound the decay rate of the residual. 

Due to $\mathcal{G}$ being an orthogonal basis, the residual $R_j$ and the feature $g_{j-1}$ will be orthogonal, such that as in Eq.13 of \cite{Mallat1993}, 
\begin{align*}
\norm{R_j}{L_2}^2 &= \norm{R_{j-1}}{L_2}^2 -  \norm{\langle R_{j-1}, g_{j-1}  \rangle g_{j-1}}{L_2}^2\,.
\end{align*}
Using the derivations from Lemma~2 in \cite{Mallat1993},
\begin{align*}
\norm{R_N}{L_2} &=  \norm{R_{N-1}}{L_2} \left( 1 - \alpha^2 \lambda^2(R_j) \right)^{1/2}  \,,
\end{align*}
and
\begin{align*}
\norm{R_N}{L_2} &\leq \norm{\mu}{L_2} \left( 1 - \alpha^2 \text{Inf}(\lambda)^2 \right)^{N/2} \,,
\end{align*}
where $\text{Inf}(\lambda) = \inf_{f \in \mathcal{H}_{\kappa}}  \lambda(f)$.

Requiring $\norm{\mu}{L_2}\left( 1 - \alpha^2 \text{Inf}(\lambda)^2 \right)^{N/2} \leq \epsilon$, it follows that 
\begin{align*}
N &\leq 2 \frac{\log\left(\frac{\epsilon}{\norm{\mu}{L_2}}\right)}{\log(1 - \alpha^2 \text{Inf}(\lambda))} \\
  &= 2 \log_{\frac{1}{1 - \alpha^2 \text{Inf}(\lambda)}}\left(\frac{\norm{\mu}{L_2}}{\epsilon}\right) \\
  &= \BigO\left(\log\left(\frac{\norm{\mu}{L_2}}{\epsilon}\right)\right) \,.
\end{align*}

\textbf{Proof for the vector-valued case.} 
Perform the same algorithm but note that the residuals now represent vector-valued functions in the operator-valued RKHS $\mathcal{H}_K$. Defining $L_j = g_j \mathbb{I}_A$, note that $L_j$ is equivalent to applying the operator-valued kernel on the centre $c_j$. Denote then also $L_j(x)$ as its application to $x \in \mathcal{X}$. 

The weight is computed as $w_j = \int_{x \in \mathcal{X}} L_j(x) R_j(x) dx$, which computes a correlation between the kernel and the residual. The optimal setting of $w_j$ is the one that maximally reduces the norm of the residual. In other words, we can select the basis function based on the objective
\begin{align*}
\max_{g_j \in \mathcal{G}} H(R_j,g_j) = \alpha \norm{w_j}{p} \,.
\end{align*}

The estimated function is defined as 
\begin{align*}
\hat{f}_{N} = \sum_{j=0}^{N-1} w_j g_j = f - R_N \,,
\end{align*}
where $R_j \in \mathcal{H}_K$ is defined as
\begin{align*}
R_j = R_{j-1} - w_{j-1} g_{j-1} \,.
\end{align*}

One can define for any residual $f \in \mathcal{H}_K$ a correlation ratio
\begin{align*}
\lambda(f) = \sup_{g \in \mathcal{G}} \frac{H(f,g_j)}{\norm{f}{L_2(\mathcal{X}),p}} \,,
\end{align*}
and $\text{Inf}(\lambda) = \inf_{f \in \mathcal{H}_{K}}  \lambda(f)$. It is straightforward to see that as the ratio $\lambda(f)$ grows closer to 1, the residual will converge more rapidly to zero in the $\norm{}{L_2(\mathcal{X}),p}$

Note that the orthogonality of $R_j$ and $L_{j-1}$ is preserved since the basis functions are orthonormal. It follows that for all $i \in [A]$,
\begin{align*}
\norm{R_N^i}{L_2(\mathcal{X}),p} &=  \norm{R_{N-1}^{i}}{L_2(\mathcal{X}),p} \left( 1 - \alpha^2 \lambda^2(R_j) \right)^{1/2} \\
                                 &\leq \norm{\mu}{L_2(\mathcal{X}),p} \left( 1 - \alpha^2 \text{Inf}(\lambda)^2 \right)^{N/2} \,,
\end{align*}
it follows that 
\begin{align*}
N &= \BigO\left(\log\left(\frac{\norm{\mu}{L_2(\mathcal{X}),p}}{\epsilon}\right)\right) \,.
\end{align*}

\qed 

\section{Raw-PQC and bound on $D$ (numerical gradient)}
\label{app: bound on D} 
The proof is analogous to Lemma~3.1 of Jerbi et al. \cite{Jerbi2022}, which generalises the parameter shift rule \cite{Schuld2019a} to higher-order derivatives using the formulation from Cerezo et al. \cite{Cerezo2021}. The approach requires eigenvalues $\pm 1$, which is true for the C-$R_Y$ gates in the Representer Raw-PQCs of Figure~1a--b; this can be seen by constructing analogous circuits with uncontrolled $R_Y$ gates (see e.g. \cite{Shende2006}).
      
The gradients of Representer Raw-PQCs are given by the parameter shift rule with one qubit rotations as
\begin{align*}
 \partial_{i} \pi_{\theta}(a \vert s) =  \partial_{i} \langle P_a \rangle_{s,\theta} = \dfrac{\langle P_a \rangle_{s,\theta + \frac{\pi}{2} e_i}  - \langle P_a \rangle_{s,\theta - \frac{\pi}{2} e_i} }{2} \,,
\end{align*}
which can be generalised to higher-order derivatives (see Eq.~50 \cite{Jerbi2022}; Eq.~19 \cite{Cerezo2021})
\begin{align*}
\partial_{\alpha} \pi_{\theta}(a \vert s) = \frac{1}{2^p} \sum_{\omega} c_{\omega} \langle P_a \rangle_{s,\theta + \omega} \,,
\end{align*}
where $\alpha \in [d]^p$, $\omega \in \{0,\pm \pi/2, \pm \pi, \pm 3\pi/2 \}^p$, and $c_{\omega} \in \mathbb{Z}$ are integer (negative or non-negative) coefficients such that $\sum_{\omega} \vert c_{\omega} \vert = 2^p$.

The quantity $D_p$ will be bounded by
\begin{align*}
D_p &= \max_{s \in \mathcal{S}, \alpha \in [d]^p} \sum_{a \in \mathcal{A}} \left\vert \frac{1}{2^p} \sum_{\omega} c_{\omega} \langle P_a \rangle_{s,\theta + \omega} \right\vert  \\
&\leq \max_{s \in \mathcal{S}, \alpha \in [d]^p} \sum_{a \in \mathcal{A}} \frac{1}{2^p} \sum_{\omega} \vert c_{\omega} \vert  \vert \langle P_a \rangle_{s,\theta + \omega} \vert \\
&= \max_{s \in \mathcal{S}, \alpha \in [d]^p} \frac{1}{2^p} \sum_{\omega} \vert c_{\omega} \vert \sum_{a \in \mathcal{A}}  \vert \langle P_a \rangle_{s,\theta + \omega} \vert \\
&= 1 \,,
\end{align*}
where the last line follows from $\sum_{\omega} \vert c_{\omega} \vert = 2^p$ and $\sum_a P_a = I$. Since this result holds for all $p \in \mathbb{N}$, it also hold that $D \leq 1$. \qed

\section{Classical central differencing}
\label{app: classical central differencing}
We briefly summarise the proof of Jerbi et al. \cite{Jerbi2022}. 

Recall that the derivative,
\begin{align*}
V'(\theta) = \underbrace{\sum_{l=-m}^{m} \frac{c_{l}^{(2m)} V(\theta + lh)}{h}}_{V_{(2m)}(\theta)}  +  \underbrace{\sum_{l=-m}^{m} c_{l}^{(2m)} \frac{V^{(k)}(\xi_l)}{k!} l^{k} h^{k-1}}_{R_V^k} \,,
\end{align*}
can be decomposed into the $k$'th order central differencing estimator $V_{(2m)}(\theta)$ and the remainder $R_V^{k}$. In the proof below, we seek to bound the error on both terms, each to $\epsilon/2$. 

Let $G_k$ be an upper bound for $\vert V^{(k)}(\xi_l) \vert$ for $\xi_l \in [\theta-mh,\theta+mh]$. Following properties of the central differencing scheme, the remainder $R_V^k$ is bounded in absolute value by 
\begin{align*}
\vert R_V^{k} \vert &= \Big\vert \sum_{l=-m}^{m} \frac{c_{l}^{(2m)} V^{(k)}(\xi_l)}{k!}  l^k h^{k-1} \Big\vert \\
                    &\leq  \Big\vert \sum_{l=-m}^{m} c_{l}^{(2m)} l^{k} \Big \vert\frac{G_k}{k!}  h^{k-1}   \\
                    &\leq 2m^{k} \frac{G_k}{k!} h^{k-1} \,.  \tag{$\vert \sum_{l=-m}^{m} c_{l}^{(2m)} l^{k} \vert \leq 2m^k$, Theorem~3.4 \cite{Cornelissen2019}} 
\end{align*}

To obtain $\vert R_V^{k} \vert$, we need a finite difference
\begin{align*}
h &\leq \left( \frac{k!\epsilon}{4m^k G_k}  \right)^{\frac{1}{k-1}} \,.
\end{align*} 
The setting of 
\begin{align}
\label{eq: setting h}
h = \frac{2}{e} \left(\frac{\epsilon}{4G_k}\right)^{\frac{1}{k}}
\end{align} 
satisfies this requirement.

Applying classical multivariate Monte Carlo (Appendix~\ref{app: classical MC}) with a zero'th order bound $G_0$ and precision $\epsilon$, we note that for each $l=-m,\dots,m$, the required precision for $V(\theta+lh)$ is given by $\frac{\epsilon h }{k c_{l}^{(2m)}}$ (such that the errors sum to $\epsilon$). Therefore, the query complexity is given by
\begin{align*}
n &= \tildeO\left( \sum_{l=-m}^{m} \left( \frac{k c_{l}^{(2m)} G_0}{ \epsilon h} \right)^2 \right) \\
  &= \tildeO\left( \left( \frac{G_0 k}{\epsilon h} \right)^2 \sum_{l=-m}^{m} \vert c_{l}^{(2m)} \vert \right) \\ 
  &= \tildeO\left( \left( \frac{G_0 k}{\epsilon h} \right)^2  \left( 1 + 2 \sum_{l=1}^{m} \frac{1}{l} \right)\right)  \tag{$\vert c_{l}^{(2m)} \vert \leq 1/l$, Theorem~3.4 \cite{Cornelissen2019}}  \\
&= \tildeO\left( \left( \frac{G_0 k}{\epsilon h} \right)^2  \left( 3 + 2 \log(m) \right) \right)\tag{upper bound on harmonic numbers} \,. 
\end{align*}
Dropping the logarithm and applying Eq.~\ref{eq: setting h}, the query complexity is given by
\begin{equation}
\label{eq: interm query classical central differencing}
n  = \tildeO\left( \left(\frac{G_0 k}{\epsilon} \left( \frac{G_k}{\epsilon} \right)^{\frac{1}{k}} \right)^2\right)  \,.
\end{equation}
Following general combinatorial arguments for MDPs (see Lemmas F.2--F.4 in \cite{Jerbi2022}), the value function as a function of the parameters satisfies the Gevrey condition with $\sigma = 0$, $M = \frac{4 r_{\text{max}}}{1 - \gamma}$, and $c = DT^2$ (as claimed in Lemma~3.1). Therefore, the $k$'th partial derivative of the value function is bounded by 
\begin{align*}
\vert \partial_{\alpha} V(\theta) \vert &\leq \frac{M}{2} c^k (k!)^{\sigma} \\ 
                                        &\leq \frac{2 r_{\text{max}}}{1-\gamma} (D T^2)^k \,.
\end{align*}
for any $k \geq 0$. This implies $G_k :=  \frac{2 r_{\text{max}}}{1-\gamma} (D T^2)^k$ is a suitable upper bound for $\vert V^{(k)}(\xi_l) \vert$. Defining $x=\frac{2 r_{\text{max}}}{\epsilon(1-\gamma)}$ and $k =\log\left(x\right)$, we have 
\begin{align*}
n &= \tildeO\left( \left(\frac{G_0 k}{\epsilon} \left( \frac{G_k}{\epsilon} \right)^{\frac{1}{k}} \right)^2\right) \\
  &= \tildeO\left( \left(x k \left(x\right)^{\frac{1}{k}} \right)^2\right) \\
  &= \tildeO\left( \left(x \log(x) e DT^2\right)^2\right) \tag{$k=\log(x)$, $x^{1/\log(x)}=e$, and $(D T^2)^{k/k}$)} \\
  &= \tildeO\left(\left(x DT^2\right)^2\right) \\
  &= \tildeO\left(\left( \frac{r_{\text{max}} DT^2}{\epsilon(1-\gamma)} \right)^2\right)  \tag{definition of $x$}
\end{align*}
for a single partial derivative. For the full $d$-dimensional gradient,  classical central differencing therefore obtains a query complexity of 
\begin{align}
\label{eq: query classical central differencing}
n = \tildeO\left(d \left(\frac{r_{\text{max}}}{\epsilon(1-\gamma)} DT^2\right)^2\right) \,,
\end{align}
which is indeed a quadratic slowdown compared to $n = \tildeO\left(\sqrt{d} \left(\frac{r_{\text{max}}}{\epsilon(1-\gamma)} T^2\right)\right) $
\qed

\section{Proof of bound $B_1$}
\subsection{Proof for analytic Gaussian: $B_1 \leq  A N Z_{1-\frac{\delta}{2A}} \Sigma_{\text{min}}^{-1/2} \kappa_{\text{max}}$ with probability $1 - \delta$}
\label{app: bound on B analytic Gaussian}
The gradient is defined as
\begin{align*}
\nabla_{\beta} \log(\pi(a \vert s))&=  \left((a - \mu(s)) \Sigma^{-1}  \right)  \kappa(s,:)  \in \mathbb{C}^{A \times N} \,,
\end{align*}
implying that for every $j \in [A]$, $(a[j] - \mu(s)[j])/\sqrt{\Sigma_{j,j}^{-1}} \in \mathcal{N}(0,1)$.

Let $Z_j = (a[j] - \mu(s)[j])/\sqrt{\Sigma_{j,j}}$ and define $\delta > 0$. Then by distributing failure probability $\delta$ across dimensions by allocating for each dimension the one-sided failure probability $1-\frac{\delta}{2A}$, this leads to a two-sided bound based on comparing the absolute $Z$-score to the quantile $Z_{1-\frac{\delta}{2A}}$. That is, the probability of observing any action dimension with a more extreme $Z$-score is bounded by
\begin{align*}
P( \cup_{j=1}^{A} \vert  Z_j \vert > Z_{1-\frac{\delta}{2A}}) &\leq \sum_{j=1}^{A}  P(\vert  Z_j \vert > Z_{1-\frac{\delta}{2A}})   \quad (\text{union bound})\\
		&= 2A (1 - \Phi(Z_{1-\frac{\delta}{2A}})) \quad (\text{symmetry and definition of cumulative density function})\\
		&= \delta \,.
\end{align*}
Therefore with probability at least $1 - \delta$, we have
\begin{align*}
B_1 & := \vert\vert \nabla_{\beta} \log(\pi(a \vert s)) \vert\vert_1 \\
&= \vert\vert \left((a - \mu(s)) \Sigma^{-1}  \right)  \kappa(s,:) \vert\vert_1 \\
 &= \sum_{i,j} \vert Z_j \Sigma_{j,j}^{-1/2}  \kappa(s,c_i) \vert \\
&\leq A N Z_{1-\frac{\delta}{2A}} \Sigma_{\text{min}}^{-1/2} \kappa_{\text{max}} \,,
\end{align*}
where $\kappa_{\text{max}} \geq \kappa(s,s')$ for all $s,s' \in \mathcal{S}$.  \qed

\subsection{Proof for finite-precision Gauss-QKP: $B_1 = \BigO(1)$}
\label{app: bound on B bounded Gaussian}
Note that the finite-precision Gaussian will have support over some interval $[l_i,u_i]$ and $u_i - l_i = \BigO(\sqrt{\Sigma})$ for all $i=1,\dots,A$. It follows that 
\begin{align*}
\vert\vert \nabla_{\beta} \log(\pi(a \vert s)) \vert\vert_1 &= \vert\vert \left((a - \mu(s)) \Sigma^{-1}  \right)  \kappa(s,:) \vert\vert_1 \\
            &\leq \sum_{i,j} \left\vert \frac{u_i -l_i}{\Sigma_{i,i}}  \kappa(s,c_j) \right\vert \\
            &= \BigO(\frac{N A  \kappa_{\text{max}}}{\sqrt{\Sigma_{\text{min}}}})   \tag{$\Sigma_{\text{min}} = \min_i \Sigma_{i,i}$} \\
            &= \BigO\left(1\right) \tag{setting of $N=\BigO(\frac{\sqrt{\Sigma_{\text{min}}}}{A \kappa_{\text{max}}})$)} \,.
\end{align*}
\qed

\section{Occupancy distribution lemma proof}
\label{app: occupancy distribution lemma}
By definition of Algorithm~1, 
\begin{align*}
\tilde{\nu}(s,a) &= \sum_{t=0}^{\infty} P(\text{algorithm 1 returns $(s,a)$ at step } t) \\
		           &=   \sum_{t=0}^{\infty} (1-\gamma) \gamma^t  \mathbb{P}_t(s,a \vert \pi) \tag{definition of $\mathbb{P}_t$} \\
		          &=  (1-\gamma) \nu(s,a)  \,.
\end{align*}
\qed

\section{Occupancy oracle lemma proof}
\label{app: occupancy-proof}
Implement $U_{X,\mathcal{S} \times \mathcal{A}}$ according to Fig.~5, which requires $T-1$ calls to $O_P$, and $T$ calls to $\Pi$. To simplify the proof, we formulate the circuit to first compute the trajectory superposition before doing the controls, which yields an equivalent final state. The first step of the circuit is to compute $U_{\gamma} \ket{0} = \sqrt{\gamma} \ket{1} + \sqrt{1-\gamma} \ket{0}$ independently on discount registers $\ket{\psi_0,\dots,\psi_{T-1}}$. Note that 
\begin{align}
\label{eq: p_t equivalence}
\mathbb{P}_t(s',a' \vert \pi) &= \sum_{\tau_{0:t}} P(\tau_{0:t}) \mathbb{P}_t(s',a' \vert \tau_{0:t}) \nonumber  \\
							&= \sum_{\tau_{0:t-1}} P(\tau_{0:t-1}) P(s',a' \vert \tau_{t-1}) \,.
\end{align}
Denote $\tau_t$ as the state-action pair at time $t$, $\tau_{j:t}$ as a trajectory from time step $j$ to time step $t$, $\psi_{j:t}$ as the discount register from time step $j$ to time step $t$, and
\begin{align}
\label{eq: J_t}
\ket{J_t} &= \sum_{(s',a') \in \mathcal{S}\times\mathcal{A}} \sum_{\tau_{0:t-1}} \sum_{\tau_{t+1:T-1}}  \sqrt{P(\tau_{0:t-1})}   \ket{\tau_{0:t-1}} \sqrt{P(s',a' \vert \tau_{t-1})}   \ket{s',a'} \ \sqrt{P(\tau_{t+1:T-1})}  \nonumber \\
& \ket{\tau_{t+1:T-1}} \sqrt{(1-\gamma) \gamma^t} \ket{1}^{\otimes t-1} \ket{0,\psi_{t+1:T-1}} \ket{s',a'} \ket{0,\dots,0}\,.
\end{align}
 With the first register indicating the trajectory superposition, the second register indicating the discount registers, the third register indicating the occupancy register, and the fourth circuit indicating the gradient register, the circuit develops as follows: 
\begin{align*}
& \ket{0} \ket{0,\dots,0}  \ket{\mathbf{0},\mathbf{0}} \ket{0,\dots,0}  \\
& \to \sum_{\tau} \sqrt{P(\tau)} \ket{\tau} \ket{\psi_{0:\psi_{T-1}}}  \ket{\mathbf{0},\mathbf{0}} \ket{0,\dots,0}\\
& \overset{CX(0,0),\dots,CX(0,Ak-1)}{\to} \sum_{(s',a') \in \mathcal{S}\times\mathcal{A}} \sum_{\tau_{1:T-1}}  \sqrt{P(s',a' \vert d_0, \pi)} \ket{s',a'}  \sqrt{P(\tau_{1:T-1})} \ket{\tau_{1:T-1}}  \sqrt{(1-\gamma)} \ket{0,\psi_{1:T-1}}  \ket{s',a'} \ket{0,\dots,0} \\ 
&\hspace{4cm}  +  \sqrt{\gamma}  \sum_{\tau} \sqrt{P(\tau)} \ket{\tau} \ket{1,\psi_{1:T-1}} \ket{\mathbf{0},\mathbf{0}} \ket{0,\dots,0}  \\
& = \ket{J_0} + \sqrt{\gamma} \sum_{\tau} \sqrt{P(\tau)} \ket{\tau} \ket{1,\psi_{1:T-1}} \ket{\mathbf{0},\mathbf{0}}  \ket{0,\dots,0} \tag{definition in Eq.~\ref{eq: J_t}} \\
& \overset{CX(1,0),\dots,CX(1,Ak-1)}{\to} \ket{J_0} + \\
&\sqrt{\gamma} \Bigg( \sum_{(s',a') \in \mathcal{S}\times\mathcal{A}} \sum_{\tau_{0}}  \sum_{\tau_{2:T-1}}  \sqrt{P(\tau_{0})} \sqrt{P(s',a' \vert \tau_{0})}  \ket{\tau_0} \ket{s',a'} \sqrt{P(\tau_{2:T-1})} \ket{\tau_{2:T-1}} \sqrt{(1-\gamma)} \ket{1}\ket{0,\psi_{2:T-1}} \ket{s',a'} \ket{0,\dots,0} \\
& \hspace{3cm}    + \sqrt{\gamma} \sum_{\tau} \sqrt{P(\tau)} \ket{\tau}  \ket{1,1,\psi_{2:T-1}} \ket{\mathbf{0},\mathbf{0}} \ket{0,\dots,0} \Bigg)  \\
& = \sum_{t=0}^{1} \ket{J_t} + \sqrt{\gamma^{2}} \sum_{\tau}  \sqrt{P(\tau)} \ket{\tau} \ket{1,1,\psi_{2:T-1}} \ket{\mathbf{0},\mathbf{0}} \ket{0,\dots,0} \\
& \vdots \\
\end{align*}
\begin{align*}
& \overset{CX(T-1,0),\dots,CX(T-1,Ak-1)}{\to}  \sum_{t=0}^{T-1} \ket{J_t} + \sqrt{\gamma^T}  \sum_{\tau}  \sqrt{P(\tau)} \ket{\tau} \ket{1}^{\otimes T} \ket{\mathbf{0},\mathbf{0}} \ket{0,\dots,0}  \\
& =  \sum_{t=0}^{T-1} \sum_{(s',a') \in \mathcal{S}\times\mathcal{A}} \sum_{\tau_{0:t-1}} \sum_{\tau_{t+1:T-1}}  \sqrt{P(\tau_{0:t-1})}   \ket{\tau_{0:t-1}} \sqrt{P(s',a' \vert \tau_{t-1})}   \ket{s',a'} \ \sqrt{P(\tau_{t+1:T-1})}  \nonumber \\
&\ket{\tau_{t+1:T-1}} \sqrt{(1-\gamma) \gamma^t} \ket{1}^{\otimes t} \ket{0,\psi_{t+1:T-1}} \ket{s',a'} \ket{0,\dots,0}\\
& + \sqrt{\gamma^T}  \sum_{\tau}  \sqrt{P(\tau)} \ket{\tau} \ket{1}^{\otimes T} \ket{\mathbf{0},\mathbf{0}} \ket{0,\dots,0} \\
&\overset{O_X}{\to} \sum_{t=0}^{T-1} \sum_{(s',a') \in \mathcal{S}\times\mathcal{A}} \sum_{\tau_{0:t-1}} \sum_{\tau_{t+1:T-1}}  \sqrt{P(\tau_{0:t-1})}   \ket{\tau_{0:t-1}} \sqrt{P(s',a' \vert \tau_{t-1})}   \ket{s',a'} \ \sqrt{P(\tau_{t+1:T-1})} \\
&\ket{\tau_{t+1:T-1}} \sqrt{(1-\gamma) \gamma^t} \ket{1}^{\otimes t} \ket{0,\psi_{t+1:T-1}} \ket{s',a'} \ket{X(s',a')} \\
&  \hspace{4cm}  + \sqrt{\gamma^T}  \sum_{\tau}  \sqrt{P(\tau)} \ket{\tau} \ket{1}^{\otimes T} \ket{\mathbf{0},\mathbf{0}} \ket{X(\mathbf{0},\mathbf{0})} \\
&= \sum_{(s',a') \in \mathcal{S}\times\mathcal{A}} \sum_{t=0}^{T-1}  \left( \sqrt{\mathbb{P}_t(s',a' \vert \pi)} \ket{\chi_{1,t}(s',a')}  \ket{s',a'}  \ket{\chi_{2,t}(s',a')} \sqrt{(1-\gamma) \gamma^t} \ket{1}^{\otimes t} \ket{0,\psi_{t+1:T-1}} \ket{s',a'} \ket{X(s',a')}\right)   \tag{via Eq.~\ref{eq: p_t equivalence}}\\
&  \hspace{4cm} +\sqrt{\gamma^T}  \sum_{\tau}  \sqrt{P(\tau)} \ket{\tau} \ket{1}^{\otimes T} \ket{\mathbf{0},\mathbf{0}} \ket{X(\mathbf{0},\mathbf{0})} \,,
\end{align*}
where $\ket{\chi_{1,t}(s',a')}$ and $\ket{\chi_{2,t}(s',a')}$ are (unnormalised) trajectory superpositions of $t$ steps and $T- t -1$ steps which precede and follow, respectively, $(s',a')$.

Consequently, the gradient register has expectation
\begin{align*}
\langle X \rangle &=  \sum_{(s,a) \in \mathcal{S}\times\mathcal{A}} \left( \sum_{t=0}^{T-1} (1-\gamma)\gamma^t  \mathbb{P}_t(s',a' \vert \pi) \right) X(s,a) + \gamma^T X(\mathbf{0},\mathbf{0}) \\
				  &= \sum_{(s,a) \in \mathcal{S}\times\mathcal{A}} (1 - \gamma) \nu(s,a) X(s,a) + \gamma^T X(\mathbf{0},\mathbf{0}) \,.
\end{align*}
Since $\mathbb{E}[\bar{X}] =  \langle X \rangle$ and  the quantities $X(s,a)$ are analytically known for all $(s,a) \in \mathcal{S} \times \mathcal{A}$, it is straightforward to post-process $\bar{X} \gets \frac{\bar{X} - \gamma^T X(\mathbf{0},\mathbf{0})}{(1-\gamma)}$ to obtain an estimator with expectation
\begin{align*}
\mathbb{E}[\bar{X}] = \sum_{s,a} \nu(s,a) X(s,a) \,.
\end{align*}
In total, the circuit required $(S+A)k T$ state-action qubits, $d k T$ qubits for the gradient register, and $T$ qubits for the discount register, yielding $\BigO(k(S + A + d)T)$ qubits.\footnote{We have ignored auxiliary qubits as e.g. may be required for policy evaluation. However, it is reasonable to further assume that the auxilary qubits relate linearly to the main registers.} 
\qed

\section{Improvement over QBounded}
\label{app: improvement over QBounded}
First we note that $\sigma_Q \leq \Delta_Q$. Let $[m,M]$ be the range of the Q-values as predicted by the critic. Note that $\sigma_Q^2 \leq \frac{(M - m)^2}{4} \leq \frac{(2\Delta_Q)^2}{4} = \Delta_Q^2$ following Popoviciu's inequality and noting that the range $M-m$ is limited by two times the maximal deviation from the baseline. 

Now we show the relation between $\sigma_{\nabla_1}$ and $B_1$. Note that
\begin{align*}
\sigma_{\partial}(i,j) &=\mathrm{SD}_{(s,a) \sim \tilde{\nu}'}(\partial_{i,j} \log(\pi(a \vert s)))  \\
                  &= \mathrm{SD}_{(s,a) \sim \tilde{\nu}'}\left( Z_i \kappa(s,c[j]) /\sqrt{\Sigma_{ii}}) \right) \\
                  &\leq \max_{s}\mathrm{SD}_{a \sim \pi(a \vert s)}\left(Z_i  \kappa(s,c_j) /\sqrt{\Sigma_{ii}})\right) \\
                  &= \max_{s} \kappa(s,c_j)/\sqrt{\Sigma_{ii}} \tag{standard deviation of $Z$-score is 1} \\
                  &\leq \kappa_{\text{max}}/\sqrt{\Sigma_{ii}} \,.
\end{align*}
where we use $Z_i = (a[i] - \mu(s)[i]) \Sigma_{ii}^{-1} \sim \mathcal{N}(0,1/\sqrt{\Sigma_{ii}})$ because the policy is Gaussian, and the upper bound on $\kappa$ (e.g. for quantum kernels, $\kappa(s,c_j) \leq 1$ for all state-pairs). Therefore, we set the upper bound according to
\begin{align*}
\sigma_{{\nabla}_{1}} &= \norm{ \sigma_{\partial}(:)}{1}  \\
                      &\leq \frac{\kappa_{\text{max}} NA}{\sqrt{\Sigma_{\text{min}}}} \,,
\end{align*}
where $\Sigma_{\text{min}} = \min_i \Sigma_{ii}$. This bound yields the same big O complexity as $B_1$ in the finite-precision Gaussian (Appendix~\ref{app: bound on B bounded Gaussian}). However, the hidden constant is much larger for $B_1$. To illustrate the hidden constant, observe that
\begin{align*}
\frac{\sum_{i=1}^{A} \sum_{j=1}^{N} \frac{u_i -l_i}{\sqrt{\Sigma_{i,i}}}  \kappa(s,c_j)}{\sum_{i=1}^{A} \sum_{j=1}^{N} \kappa(s,c_j) / \sqrt{\Sigma_{ii}}} \geq  \min_{i}  \frac{A  \frac{u_i -l_i}{\sqrt{\Sigma_{i,i}}} \sum_{j=1}^{N} \kappa(s,c_j) }{A \sum_{j=1}^{N} \kappa(s,c_j) / \sqrt{\Sigma_{ii}}} = \min_i u_i -l_i \,.
\end{align*}

Combining both $\sigma_Q$ and $\sigma_{\nabla_1}$, the improvement is at least a factor $2 \min_i (u_i - l_i)$. \qed

\section{CQRAC query complexity}
\label{app: actor-critic query complexity}
\textbf{a)} Performing QEstimator on $U_{X,\mathcal{S} \times \mathcal{A}}$, the random variable of interest is the $d$-dimensional $X(s,a) = (\hat{Q}(s,a) - b(s)) \nabla_{\theta} \log(\pi(a \vert s))$. Each query takes  $T=\BigO(T)$ time steps of interactions with the environment. It follows that
\begin{align*}
\sqrt{\Tr(\Sigma_{X})} &= \sqrt{\sum_{i=1}^{d} \mathrm{Var}(X_i)} \\
					 & \leq \max_{s,a} \sqrt{\sum_{i=1}^{d} \left( X_i - \mathbb{E}[X_i]\right)^2} \tag{definition of variance}  \\
                    &\leq \max_{s,a}  \norm{\tilde{X} - \mathbb{E}[\tilde{X}]}{2}  \quad  \text{(definition of $\ell_2$ norm}  \\
                    &\leq \max_{s,a}  \left( \norm{\tilde{X}}{2} + \norm{- \mathbb{E}[\tilde{X}]}{2} \right) \quad  \text{(triangle inequality}  \\
                    &\leq \max_{s,a} 2 \norm{\tilde{X}}{2}  \quad  \text{(drop the sign and $\norm{\mathbb{E}[X]}{2} \leq \max_{s,a} \norm{X}{2}$}  \\
                    &= \max_{s,a} 2 \norm{(\hat{Q}(s,a) - b(s)) \nabla_{\theta} \log(\pi(a\vert s))}{2}  \quad  \text{(definition of $\tilde{X}$}  \\
                    &\leq  2 \max_{s,a} \norm{\hat{Q}(s,a) - b(s)}{2} \max_{s',a'} \norm{ \nabla_{\theta} \log(\pi(a'\vert s'))}{2}   \tag{Cauchy-Schwarz and maximum} \\
                    &= 2 \max_{s,a} \vert \hat{Q}(s,a) - b(s) \vert \max_{s',a'} \norm{ \nabla_{\theta} \log(\pi(a'\vert s'))}{2}   \tag{$\hat{Q}(s,a) - b(s)$ is one-dimensional} \\
                    &\leq 2 \max_{s,a} \vert \hat{Q}(s,a) - b(s) \vert \max_{s',a'}  d^{\xi(p)} \norm{\nabla_{\theta} \log(\pi(a'\vert s'))}{p}   \tag{H\"older's inequality} \\
                    &\leq 2 d^{\xi(p)}\Delta_Q B_p   \tag{definition of $\Delta_Q$ and $B_p$} \,.
\end{align*}
Following Theorem~3.4 in \cite{Cornelissen2022}, the QEstimator algorithm will return an $\epsilon$-correct estimate $\bar{X}$ such that
\begin{align}
\label{eq: Qestimator-result}
\norm{\bar{X} - \mathbb{E}[X]}{\infty} &\leq \frac{\sqrt{\Tr(\Sigma_{X})} \log(d/\sqrt{\delta})}{n} \nonumber \\
									   &\leq \frac{2 d^{\xi(p)}\Delta_Q B_p \log(d/\sqrt{\delta})}{n}   
\end{align}
with probability at least $1-\delta$. 

Following the reasoning of Lemma~6.3, we post-process $\bar{X} \gets \frac{\bar{X} - \gamma^T X(\mathbf{0},\mathbf{0})}{(1-\gamma)}$ to convert the state occupancy distribution to the occupancy measure, $\nu(s,a) = \frac{1}{1-\gamma} \tilde{\nu}(s,a)$. It follows from Eq.~\ref{eq: Qestimator-result} that an $\epsilon$-correct estimate for $\mathbb{E}[X]$ is obtained within
\begin{align*}
n &\leq \frac{2d^{\xi(p)}\Delta_Q B_p \log(d/\sqrt{\delta})}{(1-\gamma)\epsilon} \\
  &= \tildeO\left(\frac{d^{\xi(p)}\Delta_Q B_p}{(1-\gamma)\epsilon}\right) 
\end{align*}
$\BigO(T)$ time steps of interactions with the environment. 
\qed

\textbf{b)} Performing QEstimator on $U_{X,\mathcal{S} \times \mathcal{A}}$, the random variable of interest is the $d$-dimensional $X(s,a) = (\hat{Q}(s,a) - b(s)) \nabla_{\theta} \log(\pi(a \vert s))$. Denote $Y(s,a)=(\hat{Q}(s,a) - b(s))$ and $Z(s,a)= \nabla_{\theta} \log(\pi(a \vert s))$. Then for any $i=1,\dots,d$, the random variables $Y_i$ and $Z_i$ obtained under $\tilde{\nu}'$ satisfy the following:
\begin{align*}
\mathrm{Var}(Y_i Z_i) &= \mathrm{Cov}(Y_i^2,Z_i^2)+(\mathrm{Var}(Y_i)+\mathbb{E}[Y_i]^2)(\mathrm{Var}(Z_i)+\mathbb{E}[Z_i]^2) - (\mathrm{Cov}(X_i Y_i) + \mathbb{E}[X_i]\mathbb{E}[Y_i])^2  \\  & \quad \tag{variance of product of dependent random variables, e.g. \cite{stack})} \\ 
            &\leq  \mathrm{Cov}(Y_i^2,Z_i^2)+(\mathrm{Var}(Y_i)+\mathbb{E}[Y_i]^2)(\mathrm{Var}(Z_i)+\mathbb{E}[Z_i]^2) \tag{drop negative term } \\
            &= \mathrm{Cov}(Y_i^2,Z_i^2)+\mathrm{Var}(Y_i)(\mathrm{Var}(Z_i)+\mathbb{E}[Z_i]^2) \tag{$\mathbb{E}[Y_i] = 0$ due to definition of $b(s)$} \\
            &= \BigO\left( \mathrm{Var}(Y_i)\mathrm{Var}(Z_i) +  \mathrm{Var}(Y_i)\mathbb{E}[Z_i]^2 \right)  \tag{by assumption, $\mathrm{Cov}(Y_i^2,Z_i^2) \leq c \mathrm{Var}(Y_i)  \mathrm{Var}(Z_i)$ for $c > 0$} \\
            &= \BigO\left( \mathrm{Var}(Y_i)\mathrm{Var}(Z_i) \right)  \tag{by assumption, $\mathrm{Var}(Z_i) \geq \mathbb{E}[Z_i]^2$} \\
            &= \BigO\left( \sigma_Q^2 \sigma_{\partial}(i)^2 \right) \,.
\end{align*}
From this computation, it follows that
\begin{align*}
\sqrt{\Tr(\Sigma_{X})}  &= \sqrt{\sum_{i=1}^{d} \mathrm{Var}_{\tilde{\nu}'}(X_i)} \\
                        &= \BigO\left(  \sqrt{\sum_{i=1}^{d} \sigma_Q^2 \sigma_{\partial}(i)^2} \right) \\ 
                    &= \BigO\left(  \norm{\sigma_Q \sigma_{\partial}(:)}{2} \right)  \tag{vectorisation over $i \in [d]$, definition of $\ell_2$ norm} \\
                    &= \BigO\left(  d^{\xi(p)} \norm{\sigma_Q \sigma_{\partial}(:)}{p} \right)  \tag{H\"older's inequality} \\
                    &= \BigO\left( d^{\xi(p)} \sigma_Q \sigma_{{\nabla}_p} \right) \tag{definition of $\sigma_{{\nabla}_p}$} \,. \\
\end{align*}
Following steps analogous to a), the QEstimator algorithm returns an $\epsilon$-correct estimate for $\mathbb{E}[X]$ within
\begin{align*}
n &= \BigO\left( \frac{d^{\xi(p)} \sigma_Q \sigma_{{\nabla}_p} \log(d/\sqrt{\delta})}{(1-\gamma)\epsilon} \right) \\
  &= \tildeO \left( \frac{d^{\xi(p)} \sigma_Q \sigma_{{\nabla}_p}}{(1-\gamma)\epsilon} \right) 
\end{align*}
$\BigO(T)$ time steps of interactions with the environment.
\qed

\section{CQRAC quadratic query complexity improvement}
\label{app: quadratic}
For the classical algorithm, one may use any sub-Gaussian multivariate mean estimator \cite{Lugosi2019} as these yield the same query complexity as the computationally efficient estimator of Hopkins \cite{Hopkins2020}, i.e. an $\ell_2$ error of 
\begin{align*}
\epsilon_2 \leq C \left(\sqrt{\Tr(\Sigma_X)/n} + \sqrt{\norm{\Sigma_X}{} \log(1/\delta) /n} \right)    
\end{align*}
with probability at least $1-\delta$ for some constant $C>0$. Applied to our $\ell_{\infty}$ setting, we obtain
\begin{align*}
n &= \BigO\left(\frac{\Tr(\Sigma_X) + \norm{\Sigma_X}{} \log(1/\delta)}{\epsilon_2^2}\right) \tag{$(x+y)^2 = \BigO(x^2 + y^2)$}\\
  &= \BigO\left(\frac{d^{2\xi(p)}\Delta_Q^2 B_p + \norm{\Sigma_X}{} \log(1/\delta)}{\epsilon_2^2}\right)  \tag{from upper bound on $\sqrt{\Tr(\Sigma_X)}$ in Appendix~\ref{app: actor-critic query complexity}\textbf{a}}\\
  &= \BigO\left(\frac{d^{2\xi(p)}\Delta_Q^2 B_p + \norm{\Sigma_X}{} \log(1/\delta)}{\epsilon^2}\right)  \tag{$\norm{X}{\infty} \leq \norm{X}{2}$ implies $\epsilon \leq \epsilon_2$}\\
  &= \tildeO\left(\frac{d^{2\xi(p)}\Delta_Q^2 B_p^2 + \norm{\Sigma_X}{}}{\epsilon^2}\right)  \tag{removing logarithmic factors}\,.
\end{align*}
Correcting for the discount factor, we obtain the first equation in Corollary~6.1. Analogous computations with upper bound $\Tr(\Sigma_X) \leq d^{2\xi(p)}\sigma_Q^2 \sigma_{{\nabla}_p}^2$ yield the second equation in Corollary~6.1. Comparing both results to Theorem~6.2\textbf{a} and \textbf{b}, respectively, proves the quadratic query complexity speedup. \qed

\section{CQRAC total query complexity}
\subsection{Tabular averaging critic}
\label{app: total complexity}
Let $\delta_1,\delta_2 \in (0,1)$ such that $\delta = \delta_1 + \delta_2$ represent failure probability upper bounds for the policy gradient and critic error, respectively. We will prove the query complexity for both failure probabilities separately and then note that both query complexities must hold with probability at least $1-\delta$. 

For the (classical) Compatible RKHS Actor-critic, the same samples are used for the critic and policy gradient estimates. Therefore, taking the worst-case of the query complexities for the policy gradient and the critic yields the desired result. For the critic, the number of queries $n_2$ relates to $m$, the total number of state-action samples, as $n_2 = \frac{m}{T}$. Then apply the classical multivariate Monte Carlo (see Appendix~\ref{app: classical MC}) based on the $\vert \mathcal{S} \times \mathcal{A} \vert$-dimensional vector of $Q$-values. Since each state-action pair is sampled independently, at least $m \vert \mathcal{S} \times \mathcal{A} \vert$ samples are required to ensure $m$ samples per state-action pair. However, we can drop the factor $\vert \mathcal{S} \times \mathcal{A} \vert$ from the big O notation due to the limited tabular state-action space. Consequently, with probability at least $1 - \delta_2$
\begin{align}
\label{eq: n2}
m  &= \BigO\left(\frac{V_{\text{max}}^2 \log(1/\delta_2)}{\epsilon'^2}\right)  \nonumber \\
  &= \tildeO\left(\frac{V_{\text{max}}^2}{\epsilon'^2}\right)  \nonumber \\
  &= \tildeO\left(\frac{d^{\xi(p)} T \Delta_Q B_p}{(1-\gamma)\epsilon}\right)   \nonumber \\
n_2 &= \tildeO\left(\frac{d^{\xi(p)} \Delta_Q B_p}{(1-\gamma)\epsilon} \right)  \,,
\end{align} 
where the third upper bound is due to $\epsilon' \geq \sqrt{\frac{(1-\gamma) \epsilon}{d^{\xi(p)}T\Delta_Q B_p}}V_{\text{max}}$.

For the policy gradient, with probability at least $1 - \delta_1$, the number of queries is bounded by Corollary~6.1:
\begin{align*}
n_1 &= \tildeO\left(\frac{d^{2\xi(p)}\Delta_Q^2 B_p^2 + \norm{\Sigma_X}{}}{(1-\gamma)^2 \epsilon^2}\right)\,.
\end{align*}
Note that $n_1 > n_2$ such that with probability $1 - \delta$, the number of total queries is bounded by 
\begin{align*}
n &= \tildeO\left(\frac{d^{2\xi(p)}\Delta_Q^2 B_p^2 + \norm{\Sigma_X}{}}{(1-\gamma)^2 \epsilon^2}\right) \,.
\end{align*}

For Compatible Quantum RKHS Actor-critic (see Algorithm~2), $n_1$ queries to a $T$-step implementation of $U_{X,\mathcal{S}\times\mathcal{A}}$ are used for quantum policy gradient estimates, and $n_2$ queries to a $T'$-step implementation of $U_P$ and $U_R$ are used for the critic estimate. Noting that $T' \leq 2T-1$, both represent $\BigO(T)$ steps of environment interaction. The result for $n_2$ follows directly from Eq.~\ref{eq: n2}. The result for $n_1$ is given by (see Lemma~6.2)
\begin{align*}
n_1 &= \tildeO\left(\frac{d^{\xi(p)}\Delta_Q B_p}{(1-\gamma) \epsilon}\right)\,.
\end{align*}
Summing the two equivalent query complexities yields the total query complexity, such that with probability $1-\delta$
\begin{align*}
n &= \tildeO\left(\frac{d^{\xi(p)}\Delta_Q B_p}{(1-\gamma) \epsilon} \right) \,.
\end{align*}
\qed 

\subsection{Kernel ridge regression critic}
\label{app: KRR}
From Lemma~3.4, it follows that the error converges in probability at a rate
\begin{align*}
\norm{\hat{Q} - Q}{L_2} &= \BigO_P\left(m^{-\frac{l}{2l+d}} \right) \,.
\end{align*}
It follows that
\begin{align*}
m  = \BigO_P\left( \epsilon'^{-\frac{2l+d}{l}} \right) = \BigO_P\left( \epsilon'^{-4} \right) \,,
\end{align*}
where the last equality follows from $l > d/2$ as in the preconditions of Lemma~3.4.
Due to setting $\epsilon' \geq \left(\frac{(1-\gamma)\epsilon}{T d^{\xi(p)} \Delta_Q B_p} \right)^{1/4}$, it follows that
\begin{align*}
m  = \BigO_P\left( \frac{T d^{\xi(p)} \Delta_Q B_p}{(1-\gamma)\epsilon}\right)
\end{align*}
and 
\begin{align*}
n_2  &= \BigO_P\left( \frac{d^{\xi(p)} \Delta_Q B_p}{(1-\gamma)\epsilon}\right) \,.
\end{align*}
Consequently, with high probability $1-\delta_2$ for some $\delta_2 > 0$, the desired $\epsilon'$ bound can be obtained  within 
\begin{align*}
n_2 = \tildeO\left(\frac{d^{\xi(p)} \Delta_Q B_p}{(1-\gamma)\epsilon}\right)
\end{align*} 
queries. The remainder of the proof is completely analogous to Appendix~\ref{app: total complexity}.
\qed

\section{DCQRAC query complexity}
\label{app: dpg query complexity}
\textbf{a)} We apply QBounded (algorithm in Theorem 3.3 of \cite{Cornelissen2022}) to a state occupancy oracle $U_{\tilde{X},\mathcal{S}}$ based on the normalised random variable $\tilde{X}(s) = \frac{\kappa(s,\cdot)  \nabla_{a} \hat{Q}(s,a) \vert_{a = \mu(s)}}{Z}$, where $Z=d^{\xi(p)}\kappa_p^{\text{max}}C_p$, which will be measured after $T$ time steps of interactions with the environment. 

Note that we can express $X(s)$ as an $NA$-dimensional vector 
\begin{align*}
X(s) = \left(\kappa(s,c_1) \partial_{a_1}  \hat{Q}(s,a) \vert_{a = \mu(s)},\kappa(s,c_2) \partial_{a_1}  \hat{Q}(s,a) \vert_{a = \mu(s)},\dots,\kappa(s,c_N) \partial_{a_A}  \hat{Q}(s,a) \vert_{a = \mu(s)}\right) \in \mathbb{R}^{NA} \,.
\end{align*}
It follows for any $s \in \mathcal{S}$ that 
\begin{align*}
\norm{\tilde{X}(s)}{2} &\leq \frac{1}{Z} \norm{(\kappa(s,c_1) \partial_{a_1}  \hat{Q}(s,a) \vert_{a = \mu(s)},\dots,\kappa(s,c_N) \partial_{a_A}  \hat{Q}(s,a) \vert_{a = \mu(s)})}{2} \\
                    &\leq  \frac{(NA)^{\xi(p)}}{d^{\xi(p)}\kappa_p^{\text{max}}C_p}  \norm{(\kappa(s,c_1)\partial_{a_1}  \hat{Q}(s,a) \vert_{a = \mu(s)},\dots,\kappa(s,c_N) \partial_{a_A}  \hat{Q}(s,a) \vert_{a = \mu(s)})}{p} \tag{H\"older's inequality} \nonumber \\
                    &=  \frac{(NA)^{\xi(p)}}{d^{\xi(p)}\kappa_p^{\text{max}}C_p}  \left(\sum_{i=1}^N \sum_{j=1}^A \vert \kappa(s,c_i) \partial_{a_j}  \hat{Q}(s,a) \vert_{a = \mu(s)}\vert^{p}\right)^{1/p} \tag{definition of $p$-norm} \nonumber \\
                    &=   \frac{(NA)^{\xi(p)}}{d^{\xi(p)}\kappa_p^{\text{max}}C_p}  \left(\sum_{i=1}^N \sum_{j=1}^A \vert \kappa(s,c_i) \vert^p \vert \partial_{a_j}  \hat{Q}(s,a) \vert_{a = \mu(s)}\vert^{p}\right)^{1/p} \tag{$\vert \sum_i x_i y_i \vert^p \leq \sum_i \vert  x_i y_i \vert^p \leq \sum_i \vert x_i \vert^p \vert y_i \vert^p $} \nonumber \\
                    &=  \frac{(NA)^{\xi(p)}}{d^{\xi(p)}\kappa_p^{\text{max}}C_p}  \left(\left(\sum_{i=1}^N \vert \kappa(s,c_i) \vert^p \right) \left(\sum_{j=1}^A  \vert \partial_{a_j}\hat{Q}(s,a) \vert_{a = \mu(s)}\vert^{p} \right)\right)^{1/p} \tag{rearranging} \nonumber \\
                     &= \frac{(NA)^{\xi(p)}}{d^{\xi(p)} \kappa_p^{\text{max}} C_p}  \norm{\kappa(s,:)}{p} \norm{\nabla_a  \hat{Q}(s,a) \vert_{a = \mu(s)}}{p} \tag{$p$-norm definition} \nonumber \\
                     &\leq \frac{(NA)^{\xi(p)}}{d^{\xi(p)} \kappa_p^{\text{max}} C_p}  \kappa_p^{\text{max}} C_p \tag{definition of $\kappa_p^{\text{max}}$ and $C_p$} \nonumber\\
                     &= 1 \tag{$d=NA$ for kernel policy} \nonumber \,.
\end{align*}

Applying Theorem 3.3 of \cite{Cornelissen2022} to $\tilde{X}$, QBounded returns an $\frac{\epsilon}{Z}$-precise estimate of $\mathbb{E}\left[\tilde{X}\right]$, such that 
\begin{align*}
\norm{\tilde{X} - \mathbb{E}\left[\tilde{X}\right]}{\infty} \leq \frac{\epsilon}{d^{\xi(p)} \kappa_p^{\text{max}}C_p}
\end{align*}
within 
\begin{align*}
n \leq \frac{d^{\xi(p)} \kappa_p^{\text{max}} C_p\log(d/\delta)}{\epsilon}
\end{align*}
oracle queries. Therefore, after renormalisation, an $\epsilon$-precise estimate $\bar{X}$ is obtained within the same number of oracle queries.

Following the reasoning of Lemma~6.3 with state occupancy oracle, we post-process $\bar{X} \gets \frac{\bar{X} - \gamma^T X(\mathbf{0})}{(1-\gamma)}$ to convert the state occupancy distribution to the occupancy measure, $\nu_{\pi}(s) = \frac{1}{1-\gamma} \tilde{\nu}_{\pi}(s)$. It follows that 
\begin{align*}
n &\leq \frac{d^{\xi(p)} \kappa_p^{\text{max}}C_p \log(d/\delta)}{(1 - \gamma)\epsilon} \\
  &= \tildeO\left(\frac{d^{\xi(p)} \kappa_p^{\text{max}}C_p}{(1-\gamma)\epsilon}\right) \,. 
\end{align*}
\qed

\textbf{b)} Follow the same procedure as \textbf{a)} but define the normalisation constant $Z = d^{\xi(p)} E_p C_p$, and express $X(s)$ as a $d$-dimensional vector 
\begin{align*}
X(s) = \nabla_{\theta} \mu_{\theta}(s)  \nabla_{a} \hat{Q}(s,a) \vert_{a =  \mu_{\theta}(s)} =  \left(\partial_{\theta_1} \mu_{\theta}(s)^{\intercal} \nabla_{a} \hat{Q}(s,a) \vert_{a =  \mu_{\theta}(s)}, \dots, \partial_{\theta_d} \mu_{\theta}(s)^{\intercal} \nabla_{a} \hat{Q}(s,a) \vert_{a =  \mu_{\theta}(s)}\right) \in \mathbb{R}^{d}  \,.
\end{align*}
Following analogous computations,  we obtain 
\begin{align*}
\norm{\tilde{X}(s)}{2} & \leq \frac{d^{\xi(p)}}{d^{\xi(p)}E_p C_p}  \left(\sum_{i=1}^{d} \vert \partial_{\theta_i} \mu_{\theta}(s)^{\intercal} \nabla_{a} \hat{Q}(s,a) \vert_{a =  \mu_{\theta}(s)}\vert^{p}\right)^{1/p} \tag{definition of $p$-norm} \nonumber \\
                    &= \frac{d^{\xi(p)}}{d^{\xi(p)}E_p C_p}  \left(\sum_{i=1}^{d} \vert \sum_{j=1}^{A} \partial_{\theta_i} \mu_{\theta}(s)_j \partial_{a_j} \hat{Q}(s,a) \vert_{a =  \mu_{\theta}(s)}\vert^{p}\right)^{1/p} \tag{inner product} \nonumber \\
                    & \leq \frac{d^{\xi(p)}}{d^{\xi(p)}E_p C_p}  \left(\sum_{i=1}^{d} \sum_{j=1}^{A} \vert \partial_{\theta_i} \mu_{\theta}(s)_j\vert^p \vert \partial_{a_j}  \hat{Q}(s,a) \vert_{a =  \mu_{\theta}(s)}\vert^{p}\right)^{1/p} \tag{$\vert \sum_i x_i y_i \vert^p \leq \sum_i \vert  x_i y_i \vert^p \leq \sum_i \vert x_i \vert^p \vert y_i \vert^p $} \nonumber \\
                   & \leq \frac{d^{\xi(p)}}{d^{\xi(p)}E_p C_p}  \left(\left(\sum_{i=1}^{d} \max_{j'} \vert \partial_{\theta_i} \mu_{\theta}(s)_{j'}\vert^p \right)\left(\sum_{j=1}^{A}  \vert \partial_{a_j}  \hat{Q}(s,a) \vert_{a =  \mu_{\theta}(s)}\vert^{p}\right)\right)^{1/p} \tag{max over action dimensions, rearrange} \nonumber \\
                    &\leq \frac{d^{\xi(p)}}{d^{\xi(p)}E_p C_p}  \norm{\nabla_{\theta} \mu_{\theta}(s)}{p,\text{max}}  \norm{\nabla_{\theta} \hat{Q}(s,a) \vert_{a =  \mu_{\theta}(s)}}{p}  \tag{$p$- and $p,\text{max}$-norm definitions} \nonumber \\
                     &\leq \frac{d^{\xi(p)}}{d^{\xi(p)}E_p C_p} E_p  C_p   \tag{definition of $E_p$ and $C_p$} \nonumber \\
                     &= 1  \nonumber \,.
\end{align*}
and $n = \tildeO\left(\frac{d^{\xi(p)} E_p C_p}{(1-\gamma)\epsilon}\right)$. \qed

\section{DCQRAC quadratic query complexity speedup}
\label{app: dpg quadratic}
Note that
\begin{align*}
 \norm{X}{\infty} &\leq \max_{s,a}  \norm{\kappa(s,\cdot)  \nabla_{a} \hat{Q}(s,a) \vert_{a = \mu(s)}}{\infty} \\ 
                  &\leq \max_{s,a} \norm{\kappa(s,\cdot)  \nabla_{a} \hat{Q}(s,a) \vert_{a = \mu(s)}}{p}  \tag{$\norm{x}{\infty} \leq \norm{x}{p})$}\\
                  &\leq \kappa_{p}^{\text{max}} C_{p} \tag{see Appendix~\ref{app: dpg query complexity}} \,.
\end{align*}
Correcting for the discount factor and applying classical multivariate Monte Carlo (see Appendix~\ref{app: classical MC}) to the range $[-B,B]$, where $B = \frac{ \kappa_{p}^{\text{max}} C_{p}}{1-\gamma}$ yields
\begin{align*}
 n &=  \tildeO\left( \frac{(\kappa_{p}^{\text{max}}  C_{p})^2}{(1-\gamma)^2\epsilon^2}  \right) \,.
\end{align*}
For the quantum algorithm, Lemma~6.3\textbf{a)} with $d^{\xi(p)}=1$ for $p \in [1,2]$ gives
\begin{align*}
n &= \tildeO\left(\frac{\kappa_{p}^{\text{max}} C_p}{(1-\gamma)\epsilon}\right) \,,
\end{align*}
 demonstrating the quadratic query complexity speedup. \qed

\section{DCQRAC total query complexity}
\subsection{Tabular averaging critic}
\label{app: total complexity deterministic}
Note that both $U_{X,\mathcal{S}}$ (for the policy gradient) as well as $U_P$ and $U_R$ (for the critic) require $T=\BigO(T)$ time steps of environment interactions per call. The proof follows a similar reasoning as in Appendix~\ref{app: total complexity}. Denote $n_1$ as the number of queries for computing the policy gradient and $n_2$ as the number of queries for computing the critic. Note that 
\begin{align*}
m &= \tildeO\left(\frac{V_{\text{max}}^2}{\epsilon'^2}\right) \\
    &= \tildeO\left(\frac{T \kappa_{p}^{\text{max}} C_p}{(1-\gamma) \epsilon} \right)  \tag{since $\epsilon' \geq \sqrt{\frac{(1-\gamma) \epsilon}{T d^{\xi(p)} \kappa_{p}^{\text{max}} C_p}}V_{\text{max}}$}
\end{align*}
and that based on Lemma~6.3\textbf{a)} and Algorithm~3,
\begin{align*}
n_2   &= n_1 = \tildeO\left(\frac{d^{\xi(p)} \kappa_{p}^{\text{max}} C_p}{(1-\gamma) \epsilon} \right) \,.
\end{align*}
The classical algorithm has the same error requirement and the same classical critic so has the same setting for $n_2$.

Sum $n_1$ and $n_2$ to obtain the total query complexity for DCQRAC,
\begin{align*}
n &= \tildeO\left(\frac{d^{\xi(p)} \kappa_{p}^{\text{max}} C_p}{(1-\gamma) \epsilon} \right) + \tildeO\left(\frac{d^{\xi(p)} \kappa_{p}^{\text{max}} C_p}{(1-\gamma) \epsilon}\right) = \tildeO\left(\frac{d^{\xi(p)} \kappa_{p}^{\text{max}} C_p}{(1-\gamma) \epsilon} \right) \,.
\end{align*}
For (classical) Deterministic Compatible RKHS Actor-Critic, note that the $n_1$ term dominates over $n_2$ as defined above, giving a query complexity of
\begin{align*}
n &= \tildeO\left(\frac{(\kappa_{p}^{\text{max}} C_p)^2}{(1-\gamma)^2\epsilon^2} \right) \,.
\end{align*}
The quantum algorithm therefore yields a quadratic improvement for $p \in [1,2]$, which gives $d^{\xi(p)} = d^{\max\{0,1/2-1/p\}} = 1$. 
\qed

\subsection{Kernel ridge regression critic}
\label{app: KRR-deterministic}
The start of the proof follows the reasoning of the proof of Appendix~\ref{app: KRR}. Due to setting $\epsilon'\geq \left( \frac{(1-\gamma)\epsilon}{T d^{\xi(p)}\kappa_{p}^{\text{max}} C_p}  \right)^{1/4}$ as a precondition, we require
\begin{align*}
m &= \BigO_P\left(\epsilon'^{-4}\right)  \tag{see Appendix~\ref{app: KRR} for the derivation} \\
  &= \BigO_P\left( \frac{T d^{\xi(p)}\kappa_{p}^{\text{max}} C_p}{(1-\gamma)\epsilon} \right)
\end{align*}
state-action samples and since $n_2 = \frac{m}{T}$,
\begin{align*}
n_2 &= \BigO_P\left( \frac{d^{\xi(p)} \kappa_{p}^{\text{max}} C_p}{(1-\gamma)\epsilon} \right) 
\end{align*}
queries. Therefore, with high probability $1-\delta_2$ for some $\delta_2 > 0$, the desired $\epsilon'$ bound can be obtained  within $n_2 = \tildeO\left( \frac{d^{\xi(p)} \kappa_{p}^{\text{max}} C_p}{(1-\gamma)\epsilon} \right)$ queries. The remainder of the proof is completely analogous to Appendix~\ref{app: total complexity deterministic}. 
\qed

\section{CQRNAC query complexity}
\label{app: CQRNAC query complexity}
We apply QBounded to a state-action occupancy oracle $U_{\nabla L}$ based on the normalised random variable $\tilde{X}(s,a) = \frac{\left(R(\tau \vert s,a) - \hat{Q}(s,a)\right) \phi(s,a)}{Z}$, where $Z=d^{\xi(p)}E B_p$, which will be measured after $2T - 1 = \BigO(T)$ time steps of interactions with the environment. 

The random variable is an $d$-dimensional vector and it follows that
\begin{align*}
\norm{\tilde{X}(s,a)}{2} & \leq \frac{1}{Z} \norm{\left(R(\tau \vert s,a) - \hat{Q}(s,a)\right) \phi(s,a)}{2} \nonumber \\
                    & \leq  \frac{d^{\xi(p)}}{d^{\xi(p)}E B_p} \norm{\left(R(\tau \vert s,a) - \hat{Q}(s,a)\right) \phi(s,a)}{p} \tag{H\"older's inequality} \nonumber \\
                    & \leq  \frac{1}{E B_p} \vert R(\tau \vert s,a) - \hat{Q}(s,a) \vert \norm{\phi(s,a)}{p}   \nonumber \\
                   & \leq 1  \nonumber \,.
\end{align*}

Applying Theorem 3.3 of \cite{Cornelissen2022} to $\tilde{X}$, QBounded returns an $\frac{\epsilon}{Z}$-precise estimate of $\mathbb{E}\left[\tilde{X}\right]$, such that 
\begin{align*}
\norm{\tilde{X} - \mathbb{E}\left[\tilde{X}\right]}{\infty} \leq \frac{\epsilon}{d^{\xi(p)}E B_p}
\end{align*}
within 
\begin{align*}
n \leq \frac{d^{\xi(p)} E B_p \log(d/\delta)}{\epsilon}
\end{align*}
oracle queries. Therefore, after renormalisation, an $\epsilon$-precise estimate $\bar{X}$ is obtained within the same number of oracle queries.

Following Lemma~6.3, the $\ell_{\infty}$ norm of the expectation of $U_{\nabla L}$ is given by 
\begin{align*}
\norm{\langle X \rangle}{\infty} &=  \norm{\sum_{(s,a) \in \mathcal{S}\times\mathcal{A}} \left( \sum_{t=0}^{T-1} (1-\gamma)\gamma^t  \mathbb{P}_t(s',a' \vert \pi) \right)\left(Q(s,a) - \hat{Q}(s,a)\right)\phi(s,a) + \gamma^T \left(Q(\mathbf{0},\mathbf{0}) - \hat{Q}(\mathbf{0},\mathbf{0})\right)\phi(\mathbf{0},\mathbf{0})}{\infty} \\
                &\leq \norm{\sum_{(s,a) \in \mathcal{S}\times\mathcal{A}} (1 - \gamma) \nu(s,a) \left(Q(s,a) - \hat{Q}(s,a)\right)\phi(s,a)}{\infty} + \norm{\gamma^T \left(Q(\mathbf{0},\mathbf{0}) - \hat{Q}(\mathbf{0},\mathbf{0})\right)\phi(\mathbf{0},\mathbf{0})}{\infty} \\
				  &= \norm{\sum_{(s,a) \in \mathcal{S}\times\mathcal{A}} (1 - \gamma) \nu(s,a) \left(Q(s,a) - \hat{Q}(s,a)\right)\phi(s,a)}{\infty} + \norm{\gamma^T \BigO\left(\frac{(1-\gamma)\epsilon}{\gamma^{T} \norm{\phi(\mathbf{0},\mathbf{0})}{\infty}}\right) \phi(\mathbf{0},\mathbf{0})}{\infty} \\
				  &= (1-\gamma) \nabla_{w} L(\hat{Q}) + \BigO((1-\gamma) \epsilon) \,,
\end{align*}
where the second step follows triangle inequality, the third step follows from the theorem preconditions, and the final step follows from the definition of $L(\hat{Q})$ and Cauchy-Schwarz inequality.
 
If we post-process $\bar{X} \gets \frac{\bar{X}}{(1-\gamma)}$, the occupancy distribution is converted to the occupancy measure, and moreover, $\mathbb{E}[\bar{X}] = \nabla_{w} L(\hat{Q}) + \BigO(\epsilon)$, which gives the desired expectation within an additional error $\BigO(\epsilon)$. Combining the estimation error with the $\BigO(\epsilon)$ term, the number of samples remains the same and we have guaranteed a $\BigO(\epsilon)$ error, concluding the proof:
\begin{align*}
n &\leq \frac{d^{\xi(p)} E B_p \log(d/\delta)}{(1 - \gamma)\epsilon} \\
  &= \tildeO\left(\frac{d^{\xi(p)} E B_p}{(1-\gamma)\epsilon}\right) \,. 
\end{align*}
\qed

\bibliographystyle{alpha}
\bibliography{library}